\documentclass[prd,singlecolumn,amsmath,nobibnotes,nofootinbib,superscriptaddress,preprintnumbers,floatfix]{revtex4}

\RequirePackage{epsfig}
\usepackage{graphicx}
\newcommand{\be}{\begin{equation}}
\newcommand{\ee}{\end{equation}}
\newcommand{\ba}{\begin{eqnarray}}
\newcommand{\ea}{\end{eqnarray}}
\newcommand{\bi}{\begin{itemize}}
\newcommand{\ei}{\end{itemize}}

\newcommand{\bfi}{\begin{figure}
\epsfxsize=9cm
\epsffile}
\newcommand{\bfinew}{\begin{figure}
\begin{center}
\includegraphics}
\newcommand{\efi}{\end{figure}}
\newcommand{\efinew}{
\end{center}
\end{figure}}
\newcommand{\no}{\nonumber}

\graphicspath{ {./figures/} }

\usepackage{epstopdf}
\usepackage{subfigure}
\usepackage[usenames, dvipsnames]{color}
\usepackage{bm} 
\usepackage[linktocpage]{hyperref}
\usepackage{caption}
\usepackage{mathtools}
\def\be{\begin{equation}}
\def\ee{\end{equation}}
\def\ba{\begin{eqnarray}}
\def\ea{\end{eqnarray}}
\def\nn{\nonumber}

\def\n{{\widehat{\bf n}}}

\def\Var{\mbox{Var}}

\def\n{{\widehat{\mathbf n}}}

\def\fsky{f_{\rm sky}}

\def\chimin{\chi_{\rm min}}
\def\chimax{\chi_{\rm max}}

\newcommand{\wj}[6]{\left(
                           \begin{array}{ccc}
        \! #1\! & #2\!  & #3\!  \\
        \! #4\! & #5\!  & #6\!
                           \end{array}
                   \right)}

\begin{document}

\title{Reconstruction of the remote dipole and quadrupole fields from the kinetic Sunyaev Zel'dovich and polarized Sunyaev Zel'dovich effects}

\author{Anne-Sylvie Deutsch}
\email{asdeutsch@psu.edu}
\affiliation{Institute for Gravitation and the Cosmos and Physics Department, The Pennsylvania State University, University Park, PA 16802, USA}

\author{Emanuela Dimastrogiovanni}
\email{exd191@case.edu}
\affiliation{Department of Physics/CERCA/Institute for the Science of Origins, Case Western Reserve University, Cleveland, OH 44106, USA}

\author{Matthew C. Johnson}
\email{mjohnson@perimeterinstitute.ca}
\affiliation{Department of Physics and Astronomy, York University, Toronto, Ontario, M3J 1P3, Canada}
\affiliation{Perimeter Institute for Theoretical Physics, Waterloo, Ontario N2L 2Y5, Canada}

\author{Moritz M{\"u}nchmeyer}
\email{mmunchmeyer@perimeterinstitute.ca}
\affiliation{Perimeter Institute for Theoretical Physics, Waterloo, Ontario N2L 2Y5, Canada}

\author{Alexandra Terrana}
\email{aterrana@perimeterinstitute.ca}
\affiliation{Department of Physics and Astronomy, York University, Toronto, Ontario, M3J 1P3, Canada}
\affiliation{Perimeter Institute for Theoretical Physics, Waterloo, Ontario N2L 2Y5, Canada}

\preprint{}

\date{\today}

\begin{abstract}
The kinetic Sunyaev Zel'dovich (kSZ) and polarized Sunyaev Zel'dovich (pSZ) effects are temperature and polarization anisotropies induced by the scattering of CMB photons from structure in the post-reionization Universe. In the case of the kSZ effect, small angular scale anisotropies in the optical depth are modulated by the cosmic microwave background (CMB) dipole field, i.e. the CMB dipole observed at each spacetime point, which is sourced by the primordial dipole and especially the local peculiar velocity. In the case of the pSZ effect, similar small-scale anisotropies are modulated by the CMB quadrupole field, which receives contributions from both scalar and tensor modes. Statistical anisotropies in the cross correlations of CMB temperature and polarization with tracers of the inhomogeneous distribution of electrons provide a means of isolating and reconstructing the dipole and quadrupole fields. In this paper, we present a set of unbiased minimum variance quadratic estimators for the reconstruction of the dipole and quadrupole fields, and forecast the ability of future CMB experiments and large scale structure surveys to perform this reconstruction. Consistent with previous work, we find that a high fidelity reconstruction of the dipole and quadrupole fields over a variety of scales is indeed possible, and demonstrate the sensitivity of the pSZ effect to primordial tensor modes. Using a principle component analysis, we estimate how many independent modes could be accessed in such a reconstruction. We also comment on a few first applications of a detection of the dipole and quadrupole fields, including a reconstruction of the primordial contribution to our locally observed CMB dipole, a test of statistical homogeneity on large scales from the first modes of the quadrupole field, and a reconstruction technique for the primordial potential on the largest scales. 
\end{abstract}

\maketitle

\section{Introduction}

The secondary Cosmic Microwave Background (CMB), temperature anisotropies induced by the scattering of CMB photons by mass or free charges, is becoming an important new frontier in observational cosmology. Unlike the primary CMB temperature anisotropies, which are mostly sourced by inhomogeneities near the time of last scattering, CMB secondaries are induced by inhomogeneities through much of the volume of the observable Universe. Therefore, CMB secondaries can in principle be a far more powerful direct probe of the large-scale homogeneity of the Universe than the primary CMB (see e.g.~\cite{Kamionkowski1997,Caldwell:2007yu,GarciaBellido:2008gd,Zhang11b,Clifton:2011sn,Maartens2011,2011CQGra..28p4005Z,Bull:2011wi,Yoo:2010ad}), and possibly shed light on a number of outstanding low-significance large-scale anomalies in the primary CMB (see e.g.~\cite{Schwarz:2015cma} for a recent review). The rapid development of detector technology in CMB experiments and the deployment of redshift surveys of increasing size and quality has in recent years driven the first detections of a variety of CMB secondaries (directly and in cross correlation) such as CMB lensing~\cite{Smith:2007rg,Hirata:2008cb,2011PhRvL.107b1301D,2012ApJ...756..142V,Ade:2015zua} (the scattering of CMB photons from mass~\cite{1987A&A...184....1B,1989MNRAS.239..195C,Lewis:2006aa}), the thermal Sunyaev Zel'dovich effect~\cite{Staniszewski:2008ma,2011A&A...536A...8P,Hasselfield:2013wf,2013ApJ...763..127R,Aghanim:2015eva} (spectral distortions of the CMB from scattering CMB photons off hot gas~\cite{1970Ap&SS...7....3S}), and the kinetic Sunyaev Zel'dovich effect~\cite{Hand12,DeBernardis:2016pdv,Soergel:2016mce,2016A&A...586A.140P,2016PhRvD..93h2002S,George14} (the scattering of CMB photons from the bulk motion of free electrons~\cite{SZ80}). Excitingly, next-generation CMB experiments (e.g. CMB Stage 4~\cite{CMBS42016}) in cross correlation with next-generation redshift surveys (e.g. LSST~\cite{LSSTScienceCollaboration2009}) will yield a drastic improvement in the measurement of these and other secondary effects.

In this paper, we focus on two CMB secondaries: the kinetic Sunyaev Zel'dovich (kSZ) effect and the polarized Sunyaev Zel'dovich (pSZ) effect. The kSZ effect is a temperature anisotropy induced by the bulk motion of free electrons during and after reionization relative to the CMB rest frame. The contribution from a given location at a given redshift is proportional to the locally observed CMB dipole. The pSZ effect is a polarization anisotropy induced by the scattering of CMB photons in the presence of a quadrupolar radiation field. The contribution from a given location at a given redshift is proportional to the locally observed CMB quadrupole. In the CMB, both of these effects are given by line-of-sight integrals. However, the redshift dependence can be extracted by direct cross correlation with tracers of large scale structure. This technique is known as kSZ~\cite{Ho09,Shao11b, Zhang11b, Zhang01,Munshi:2015anr,2016PhRvD..93h2002S,Ferraro:2016ymw,Hill:2016dta,Zhang10d,Zhang:2015uta,Terrana2016} and pSZ tomography~\cite{2012PhRvD..85l3540A,Deutsch:2017cja}. In both kSZ and pSZ tomography, the remote dipole and remote quadrupole fields, respectively, are encoded in characteristic statistical anisotropies in the cross correlation.  

In the case of the pSZ effect, previous work has assessed the ability of polarization measurements in the direction of clusters to reconstruct the quadrupole field and underlying primordial potential~\cite{Kamionkowski1997,Portsmouth2004,Seto:2000uc,Seto2005,Bunn2006,Abramo:2006gp,Liu2016,Louis:2017hoh}. In other previous work~\cite{Terrana2016,Deutsch:2017cja}, the authors introduced a set of estimators to assess the promise of kSZ and pSZ tomography to measure the remote dipole and quadrupole fields. These estimators did not include the full correlation structure of the signal, and therefore missed some of the available signal-to-noise. 

In this paper, we extend this previous work by defining a set of quadratic estimators for the remote dipole and quadrupole fields. Similar quadratic estimators have been defined for the lensing potential~\cite{Hu:2001kj,Okamoto:2002ik,Okamoto:2003zw}, where they played a key role in the first detections of CMB lensing. Other applications have included patchy screening and patchy reionization~\cite{Dvorkin:2008tf,Dvorkin:2009ah,Smith:2016lnt}. A set of quadratic estimators for the pSZ effect first appeared in Ref.~\cite{2012PhRvD..85l3540A}, where the primary goal was to assess how useful pSZ tomography is for the detection of primordial gravitational waves. A set of optimal estimators for kSZ tomography was developed for several specific scenarios including the detection of large bulk flows~\cite{Zhang10d} and cosmic bubble collisions~\cite{Zhang:2015uta}. The main contribution of this work is to present a completely general set of quadratic estimators for the remote dipole and quadrupole fields within a unified framework including scalar and tensor perturbations, to discuss the information content of the reconstructed fields using a principal component analysis, and to discuss a few first applications of measurements of the remote dipole and quadrupole fields to fundamental questions in cosmology. This paper also lays the foundation for a complete forecast of parameter constraints from measurements of the remote dipole and quadrupole fields, which we present elsewhere.

The plan of the paper is as follows. In Sec. \ref{sec:velocity} we first review the cross correlation between CMB and large scale structure induced by the kSZ effect. Then we derive the quadratic estimator for the dipole field. Finally we forecast the signal-to-noise for a representative set of experimental configurations. In Sec. \ref{sec:quadrupole} we apply the same methodology to the pSZ effect, deriving optimal estimators for the remote quadrupole field based on CMB $E$-modes and $B$-modes and forecasting the signal-to-noise. In Sec. \ref{sec:homogeneity} we give an initial discussion of cosmological applications of these estimators, including the reconstruction of the primordial potential on large scales. We conclude in Sec. \ref{sec:conclusions}. A set of appendices summarize the properties of the dipole and quadrupole fields.

\section{Quadratic estimator for the remote dipole field}
\label{sec:velocity}

The kSZ effect is determined by the locally observed CMB dipole, and can therefore be used to measure the remote CMB dipole field, e.g. the locally observed CMB dipole as a function of space and time. While the total kSZ signal is dominated by the Doppler effect from local velocity perturbations, on large scales it can reveal bulk flows, statistical anisotropies, and information about the local primordial CMB dipole. The kSZ is therefore an interesting probe to get more information about the largest scales in the universe. We have studied the different small-scale and large-scale contributions to the kSZ in detail in~\cite{Terrana2016}. Here we adapt the well-known quadratic estimator methodology to the reconstruction of the remote CMB dipole field via the kSZ effect. The quadratic estimator is optimal (in the sense that it is unbiased and gives the minimum variance reconstruction) and better suited for data analysis than the estimator presented in~\cite{Terrana2016}.

\subsection{Cross correlation induced by the kSZ effect}

The temperature perturbation due to the kSZ effect along the line of sight $\n$ is given by
\be
T(\n) \big|_{kSZ}  = - \sigma_T \int d\chi \ a \ n_e(\n,\chi) v_{\rm eff}(\n,\chi),
\ee
where $v_{\rm eff}(\n,\chi)$ is the CMB dipole projected along the line of sight, $\chi$ is comoving distance, and $n_e(\n,\chi)$ is the electron number density. We bin this equation in red shift shells (bins) $\alpha$ as
\be
\label{eq:bins1}
T(\n) \big|_{kSZ}  = - \sigma_T \sum_\alpha \int_{\chimin^{\alpha}}^{\chimax^\alpha} d\chi_\alpha \ a \ n_e(\n,\chi_\alpha) v_{\rm eff}(\n,\chi_\alpha),
\ee

We split up the dipole field into its mean over each bin $\bar{v}^\alpha_{\rm eff}(\n)$ and the small-scale dipole field $\delta v_{\rm eff} (\n,\chi_\alpha)$, which varies over the bin
\be
v_{\rm eff}(\n,\chi_\alpha) = \bar{v}^\alpha_{\rm eff}(\n) \left(1 + \delta v_{\rm eff} (\n,\chi_\alpha) \right)
\ee
where
\be
\bar{v}_{\rm eff}^\alpha(\n) = \frac{1}{\Delta \chi_\alpha} \int_{\chimin^{\alpha}}^{\chimax^\alpha} d\chi_\alpha v_{\rm eff}(\n,\chi_\alpha) 
\ee
We discuss the bin averaged dipole field and its power spectrum in detail in Appendix~\ref{sec:redshiftcorr}. We also write the electron density $n_e(\n,\chi)$ as its angular average plus a fluctuation term:
\be
n_e(\n,\chi) = \bar n_e(\chi) \left( 1 + \delta_e(\n,\chi) \right)
\ee
 
From Eq.~\eqref{eq:bins1} it follows that there is a contribution to the kSZ temperature anisotropies due to the mean field $\bar{v}_{\rm eff}$ given by 
\be
T(\n) \big|_{kSZ, \bar{v}_{\rm eff}} = \sum_\alpha \tau^\alpha(\n) \bar{v}^{\alpha}_{\rm eff}(\n).
\ee
where we defined anisotropies in the optical depth of the redshift bin by
\be
\tau^\alpha(\n) = -\sigma_T \int_{\chimin^{\alpha}}^{\chimax^\alpha} d\chi \ a \ \bar n_e(\chi) \left(1 + \delta n_e(\n,\chi)\right). 
\ee
There is an additional contribution from the isotropic optical depth which does not contribute significantly to the cross correlation we study below, and hence forward neglect.

Our estimator is based on the cross-correlation of the CMB temperature multipoles $a_{\ell m}^T$ with the red shift binned galaxy distribution
\be
\delta_g^\alpha(\n) = \int d\chi \ W^\alpha(\chi) \ \delta_g(\n,\chi)
\ee
where the normalized window fuction $W^\alpha(\chi)$ selects the bin range $(\chimin^{\alpha},\chimax^\alpha)$ and can take into account observational effects like a varying number density within the bin.
Within a redshift bin, the average dipole field induces the cross correlation, since the small-scale variations in the dipole field will cancel along the line of sight. The kSZ induced cross correlation of the CMB temperature and the binned galaxy distribution is
\ba\label{eq:Tdeltacross}
\left\langle a_{\ell_1m_1}^T  \delta_{g,\ell_2 m_2}^\alpha \right\rangle 
&=&  \left< \left( \int d\n_1 \ Y^*_{\ell_1 m_1}(\n_1) \sum_\beta \sum_{L_1 M_1} \bar{v}^{\beta}_{{\rm eff},L_1 M_1} Y_{L_1 M_1}(\n_1) \sum_{L_2 M_2}  \tau^\beta_{L_2 M_2}  Y_{L_2 M_2}(\n_1)\right) \delta_{g,\ell_2 m_2}^\alpha \right> \nn\\
&=& \sum_{\beta,L_1,M_1,L_2,M_2} (-1)^{m_1} \sqrt{\frac{(2\ell_1+1)(2L_1+1)(2L_2+1)}{4\pi}} \wj{\ell_1}{L_1}{L_2}{0}{0}{0} \wj{\ell_1}{L_1}{L_2}{-m_1}{M_1}{M_2}\\
& & \times \left< \tau^\beta_{L_2 M_2} \ \delta_{g,\ell_2 m_2}^\alpha \right>   \bar{v}^{\beta}_{{\rm eff},L_1 M_1}\nn 
\ea
We are interested in broad redshift bins, so that to good approximation $\left< \tau^\beta_{\ell_1 m_1} \ \delta_{g,\ell_2 m_2}^\alpha \right> = (-1)^{m_2} \ C^{\tau \delta_g}_{\alpha \ell_1} \ \delta_{\alpha\beta}  \delta_{\ell_1 \ell_2}  \delta_{m_1 -m_2}$. With this simplification we obtain
\ba
\left\langle a_{\ell_1m_1}^T  \delta_{g,\ell_2 m_2}^\alpha \right\rangle 
&=& \sum_{\ell,m} (-1)^{m_1+m_2} \Gamma^{\rm kSZ}_{\ell_1\ell_2 \ell \alpha}  \wj{\ell_1}{\ell_2}{\ell}{-m_1}{-m_2}{m}  \  \bar{v}^{\alpha}_{{\rm eff},\ell m}.
\ea
where we defined the coupling
\ba
\Gamma^{\rm kSZ}_{\ell_1\ell_2 \ell \alpha} = \sqrt{\frac{(2\ell_1+1)(2\ell_2+1)(2\ell+1)}{4\pi}} \wj{\ell_1}{\ell_2}{\ell}{0}{0}{0} \ C^{\tau \delta_g}_{\alpha,\ell_2}. 
\ea
Therefore, from Eq.~\eqref{eq:Tdeltacross}, the statistics of the small-scale field are modulated by the bin-averaged dipole field.

\subsection{Estimator and variance of the dipole field}

The induced cross correlation of CMB temperature and matter allows a quadratic estimator of the mean dipole field $\bar{v}_{{\rm eff}, \ell m}^{\alpha}$ of form 
\be
 \widehat{v}_{{\rm eff}, \ell m}^{\alpha} = \sum_{\ell_1m_1\ell_2m_2} W_{\ell m\ell_1 m_1\ell_2 m_2 \alpha} \ a_{\ell_1m_1}^T   \delta^\alpha_{g,\ell_2m_2}.
\ee
To find the optimal weights, we need to minimize the variance with respect to the weights, subject to the constraint  $\left<\widehat{v}^\alpha_{{\rm eff}, \ell m} \right>= \bar{v}^\alpha_{{\rm eff}, \ell m}$. The variance is
\ba
\Var(\widehat{v}_{{\rm eff}, \ell m}^{\alpha}) &=& \sum_{\ell_1m_1\ell_2m_2} W_{\ell m\ell_1m_1\ell_2m_2 \alpha} W^*_{\ell m \ell_1m_1\ell_2m_2 \alpha} \tilde{C}^{TT}_{\ell_1} \tilde{C}^{\delta_g \delta_g}_{\alpha,\ell_2}, 
\ea
where we considered the leading Gaussian contribution to the variance, where $\tilde{C}_\ell^{TT}$ includes the relevant contributions to CMB temperature, foregrounds, and instrumental noise, and where $\tilde{C}^{\delta_g \delta_g}_{\alpha,\ell}$ includes the galaxy power spectrum and shot noise. 
This optimization can be implemented using a Lagrange multiplier and gives the optimal estimator
\ba
\label{eq:estimator1}
\widehat{v}_{{\rm eff}, \ell m}^{\alpha} &=& {N^{\bar{v}\bar{v}}_{\alpha \ell}}
\sum_{\ell_1m_1\ell_2m_2} (-1)^m \
\Gamma^{\rm kSZ}_{\ell_1\ell_2\ell \alpha}
\wj{\ell_1}{\ell_2}{\ell}{m_1}{m_2}{-m} 
\frac{a^{T}_{\ell_1m_1} \delta^{\alpha}_{g,\ell_2m_2}}{\tilde{C}^{TT}_{\ell_1} \tilde{C}^{\delta_g \delta_g}_{\alpha \ell_2} },  
\ea
where
\ba
\label{eq:estimator1noise}
\frac{1}{N^{\bar{v}\bar{v}}_{\alpha \ell}} &=& \frac{1}{(2\ell+1)} \sum_{\ell_1\ell_2}
\frac{\Gamma^{\rm kSZ}_{\ell_1\ell_2\ell \alpha} \ \Gamma^{\rm kSZ}_{\ell_1\ell_2\ell \alpha}}{\tilde{C}^{TT}_{\ell_1} \tilde{C}^{\delta_g \delta_g}_{\alpha \ell_2} }. 
\ea

The signal and noise in our analysis are defined as follows (see~\cite{Dvorkin:2008tf} for a similar discussion in the case of the patchy-$\tau$ estimator). The two-point function of $\widehat{v}_{{\rm eff}, \ell m}$ is the sum of two terms:
\ba
\label{eq:vpower}
\left\langle \widehat{v}^{*\alpha}_{{\rm eff}, \ell m} \widehat{v}^{\beta}_{{\rm eff}, \ell'm'} \right\rangle &=& 
   \left\langle \widehat{v}^{*\alpha}_{{\rm eff}, \ell m} \widehat{v}^{\beta}_{{\rm eff}, \ell'm'} \right\rangle_{\rm noise} +
   \left\langle \widehat{v}^{*\alpha}_{{\rm eff}, \ell m} \widehat{v}^{\beta}_{{\rm eff}, \ell'm'} \right\rangle_{\rm signal}  \nn \\
\left\langle \widehat{v}^{*\alpha}_{{\rm eff}, \ell m} \widehat{v}^{\beta}_{{\rm eff}, \ell'm'} \right\rangle_{\rm noise} &=& N_{\alpha \ell}^{\bar{v}\bar{v}} \delta_{\alpha\beta} \delta_{\ell\ell'} \delta_{mm'} \nn \\
\left\langle \widehat{v}^{*\alpha}_{{\rm eff}, \ell m} \widehat{v}^{\beta}_{{\rm eff}, \ell'm'} \right\rangle_{\rm signal} &=& C_{\alpha \beta \ell}^{\bar{v}\bar{v}} \delta_{\ell\ell'} \delta_{mm'} \label{eq:signal_noise_split}
\ea
The first term $\left\langle \widehat{v}^{*\alpha}_{{\rm eff}, \ell m} \widehat{v}^{\beta}_{{\rm eff}, \ell'm'} \right\rangle_{\rm noise} $ is obtained by summing the Gaussian terms in the estimator variance. It is the power spectrum of the estimator in the absence of any large scale dipole field and is diagonal in red shift space. The second term $\left\langle \widehat{v}^{*\alpha}_{{\rm eff}, \ell m} \widehat{v}^{\beta}_{{\rm eff}, \ell'm'} \right\rangle_{\rm signal}$ is due to the expectation value $\left<\widehat{v}^{\alpha}_{{\rm eff}, \ell m} \right>= \bar{v} ^{\alpha}_{{\rm eff}, \ell m}$ and is our signal. It is an excess power in the reconstruction due to the presence of $\widehat{v}_{{\rm eff}}$ fluctuations. This part is not diagonal in redshift because large scale modes of the dipole field are correlated. Assuming a large-scale dipole field power spectrum $C_{\alpha \beta \ell}^{\bar{v}\bar{v}}$, the expected signal-to-noise per mode is
\be
\label{eq:vSN}
(S/N)_{\ell m} = \left[ \frac{\fsky}{2} \left( \frac{C_{\alpha \alpha \ell}^{\bar{v}\bar{v}}}{N_{\alpha \ell}^{\bar{v}\bar{v}}} \right)^2 \right]^{1/2} 
\ee
This is the expected ``number of sigmas'' for a  detection of a dipole field modulation of this mode. 
To make a map (as opposed to an overall detection), one roughly needs a mode-by-mode signal-to-noise bigger than unity. To get the signal-to-noise in all modes combined, one has to take into account the signal correlation in redshift space (see Appendix~\ref{sec:redshiftcorr}).

\subsection{Signal-to-noise forecast}

We now forecast the signal-to-noise of the estimator Eq.~\eqref{eq:estimator1} for a cosmic variance limited experiment that traces $\Delta\tau_{\ell m}^\alpha$ and $a^T_{\ell m}$ up to $\ell_{\rm max}$ for the signal and noise models specified below. 

\subsubsection{Dipole field signal model}

We use the expression for the $\Lambda$CDM dipole field from~\cite{Terrana2016}, which we recall here briefly. The primordial potential $\Psi_i ({\bf k})$ is related to the multipole moments of the dipole field as a function of comoving distance by
\begin{equation} \label{eq:almv}
	v_{{\rm eff}, \ell m} (\chi) = \int \frac{d^3 k}{(2 \pi)^3}\ \Delta_\ell^{v}(k,\chi) \Psi_i(\mathbf{k})Y^*_{\ell m}({\bf \widehat{k}}).
\end{equation}
An explicit expression for the transfer function $\Delta_\ell^{v}(k,\chi)$ is given in Appendix~\ref{sec:dipoletransfer}, and an extended discussion can be found in~\cite{Terrana2016}. The mean dipole field signal power spectrum $C_\ell^{\bar{v}\bar{v}}$ is then given by
\be\label{eq:vvcovmat}
C_{\alpha \beta \ell}^{\bar{v}\bar{v}} = \int \frac{d k}{(2 \pi)^3} k^2 P(k) \bar{\Delta}_{\alpha \ell }^{v}(k) \bar{\Delta}_{\beta \ell}^{v}(k)
\ee
where we have defined the bin averaged transfer functions 
\be
\bar{\Delta}_{\alpha \ell}^{v}(k) =  \frac{1}{\Delta \chi_\alpha} \int_{\chimin^{\alpha}}^{\chimax^\alpha} d\chi_\alpha \ \Delta_\ell^{v}(k,\chi)
\ee
In Appendix~\ref{sec:redshiftcorr}, we study the effect of binning by comparing the averaged and un-averaged dipole field power spectra. For the lowest multipoles, where mostly long-wavelength modes are already extracted from the spherical harmonic transform, the correlation length of the un-averaged dipole field is significant (see Fig.~\ref{fig:corrfunc}), and there is little difference between the power in the averaged and un-averaged dipole fields even for wide redshift bins (see the left panel of Fig.~\ref{fig:compvvbar}). For higher multipoles, where the correlation length of the un-averaged dipole field is not as long, the signal can be significantly affected by binning, unless a fine binning is chosen (see the right panel of Fig.~\ref{fig:compvvbar}). 

\subsubsection{Noise model}

The expression for the noise power spectrum $N_\ell^{\bar{v}\bar{v}}$ has been given in Eq.~\eqref{eq:estimator1noise}. It depends on the measured CMB and galaxy power spectra $\tilde{C}^{TT}_{\ell}$ and $\tilde{C}^{\delta_g \delta_g}_{\ell}$ as well as the cross-power $C^{\tau \delta_g}_{\alpha \ell}$. The CMB power spectrum includes primary CMB, lensing and small scale kSZ. The mass binned $\delta_g$ power spectrum is given by: 
\begin{equation}
C^{\delta_g \delta_g}_{\alpha \ell} = \int d\chi_1 \ W^\alpha(\chi_1) \ \int  d\chi_2 \ W^\alpha(\chi_2) \ C_\ell^{\delta_g\delta_g}(\chi_1,\chi_2)
\end{equation}
Using the Limber approximation we find
\begin{equation}
C^{\delta_g \delta_g}_{\alpha \ell} = \int \frac{dk}{\ell+\frac{1}{2}} (W^\alpha(\chi))^2 P_{gg}(k,\chi) \Big|_{\chi \rightarrow (\ell+1/2)/k}
\end{equation}
where the window function selects the redshift bin $(\chimin^{\alpha},\chimax^\alpha)$. Here $P_{gg}$ is the galaxy power spectrum including shot noise.

We also need the cross power spectrum of $\tau$ and $\delta_g$ given by
\begin{equation}
C^{\tau \delta_g}_{\alpha \ell} = \sigma_T \int_{\chimin^{\alpha}}^{\chimax^\alpha} d\chi_1 \ a(\chi_1) \ \bar{n}_e(\chi_1) \int d\chi_2 \ W^\alpha(\chi_2)\ C_\ell^{\delta_e \delta_g}(\chi_1,\chi_2).
\end{equation}
Using the Limber approximation we find
\begin{equation}
C^{\tau \delta_g}_{\alpha \ell} = \sigma_T \int \frac{dk}{\ell+\frac{1}{2}} \left(a(\chi) \bar{n}_e(\chi) \right) W^\alpha(\chi) P_{ge}(k,\chi) \Big|_{\chi \rightarrow (\ell+1/2)/k}.
\end{equation}
We assmune that $P_{gg} = b^2 P_{mm}$ and $P_{ge} = b P_{mm}$ where $P_{mm}$ is the nonlinear matter power spectrum and $b$ is the galaxy bias. This approach neglects baryonic feedback effects (which cause the electron distribution to differ from the dark matter distribution) and scale dependence of the galaxy bias.  A more precise calculation that incorporates these features using the halo model is deferred to future work.

\subsubsection{Forecast} \label{sec:forecastvv}

Armed with expressions for the signal and noise power spectra, we can calculate the signal-to-noise per mode from Eq.~\eqref{eq:vSN} in each redshift bin. In this work, following~\cite{Terrana2016}, we consider six redshift bins of equal width in $\chi$ covering the range $0 < z < 6$ as given in table~\ref{tab:bins}. 
For the CMB experiment we choose experimental properties in the range of the proposed CMB S4 mission. We assume the noise is Gaussian and given by
\be
N_\ell = N_T \ \mathrm{exp}\left(\frac{\ell(\ell+1)\theta_{\mathrm{FWHM}}^2}{8\mathrm{ln}2}\right).
\ee
with a beam $\theta_{\mathrm{FWHM}}=1.0$ arcminutes and an effective white noise level of $N_T=1.0 \mu K$-arcmin. The total measured CMB signal is then $C_\ell^{tot} = C_\ell^{\rm{CMB,lensed}} + C_\ell^{\rm{kSZ}} + N_\ell$. We compute the lensed CMB power spectrum using CAMB and the kSZ power spectrum using the results of Ref.~\cite{Terrana2016}. The measurement of the galaxy density is limited by shot noise due to discrete sampling of galaxies $N_{\alpha \ell}^{\text{gg}} = 1/N_{\alpha g}$, where $N_{\alpha g}$ is the number of galaxies per square radian in the redshift bin $\alpha$. We assume a predicted galaxy sample for Large Synoptic Survey Telescope (LSST) (\cite{LSSTScienceCollaboration2009}), where the number density $n$ per arcmin$^2$ is described by
\be
n(z) =  n_{\rm gal} \ \frac{1}{2 z_0} \left(\frac{z}{z_0}\right)^2 \exp(-z/z_0)  \label{eq:lsstnr}
\ee
with $z_0 = 0.3$ and $n_{\textrm{gal}}=40 \ {\rm arcmin}^{-2}$. The bias for the LSST data set is predicted to be
\be
b(z) = 0.95/D(z)
\ee
with the growth factor normalized as $D(z=0) = 1$. We assume full sky coverage in this forecast for simplicity. A more realistic overlap between CMB S4 and LSST might be $f_{\rm sky} \sim 0.5$. Finally, we use a range in $\ell$ for the reconstruction noise of $100 \leq \ell \leq 10^4$, which saturates the signal to noise for the experimental parameters considered here.

\begin{figure}[t!]
\resizebox{1.0\hsize}{!}{	
  \includegraphics{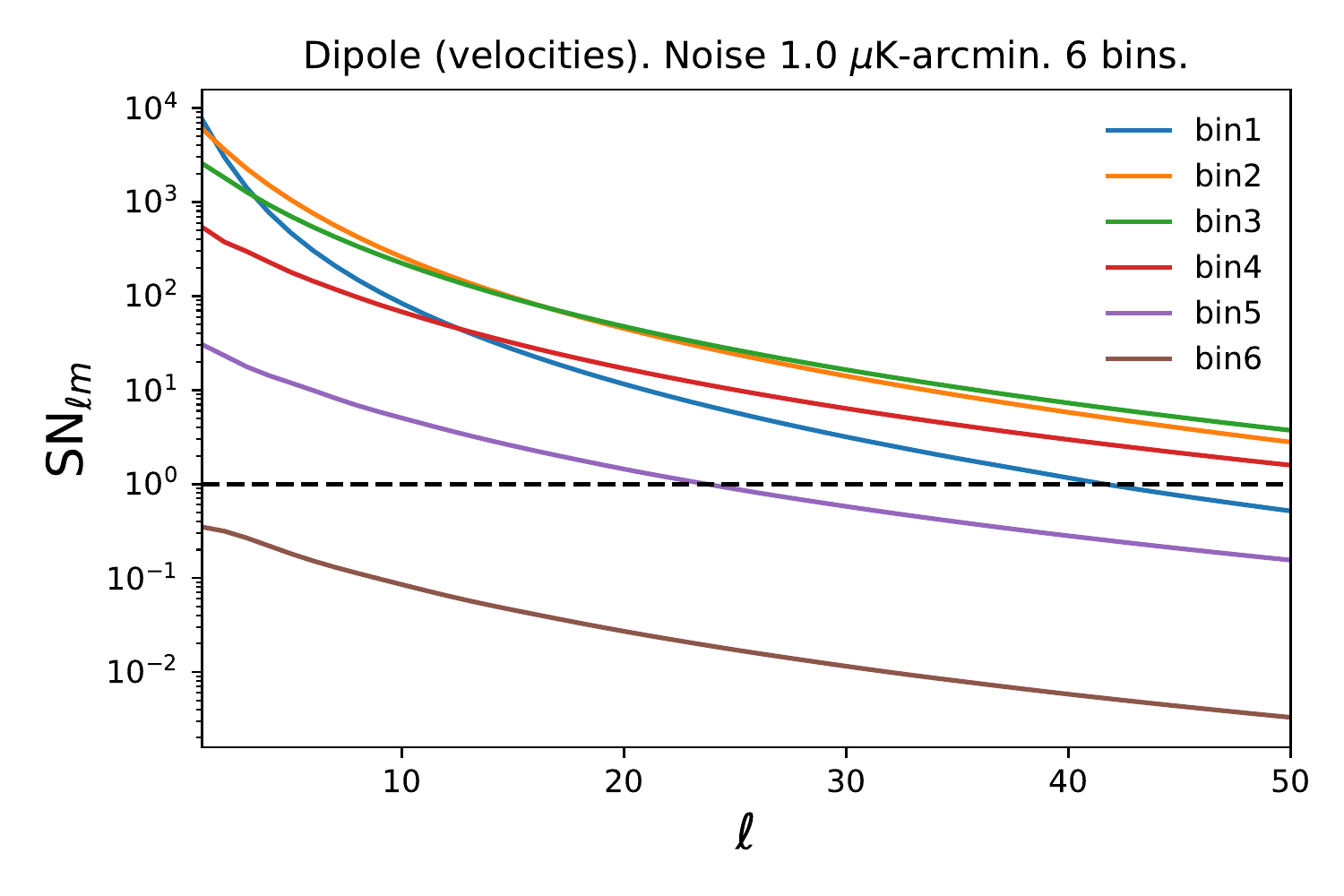}\includegraphics{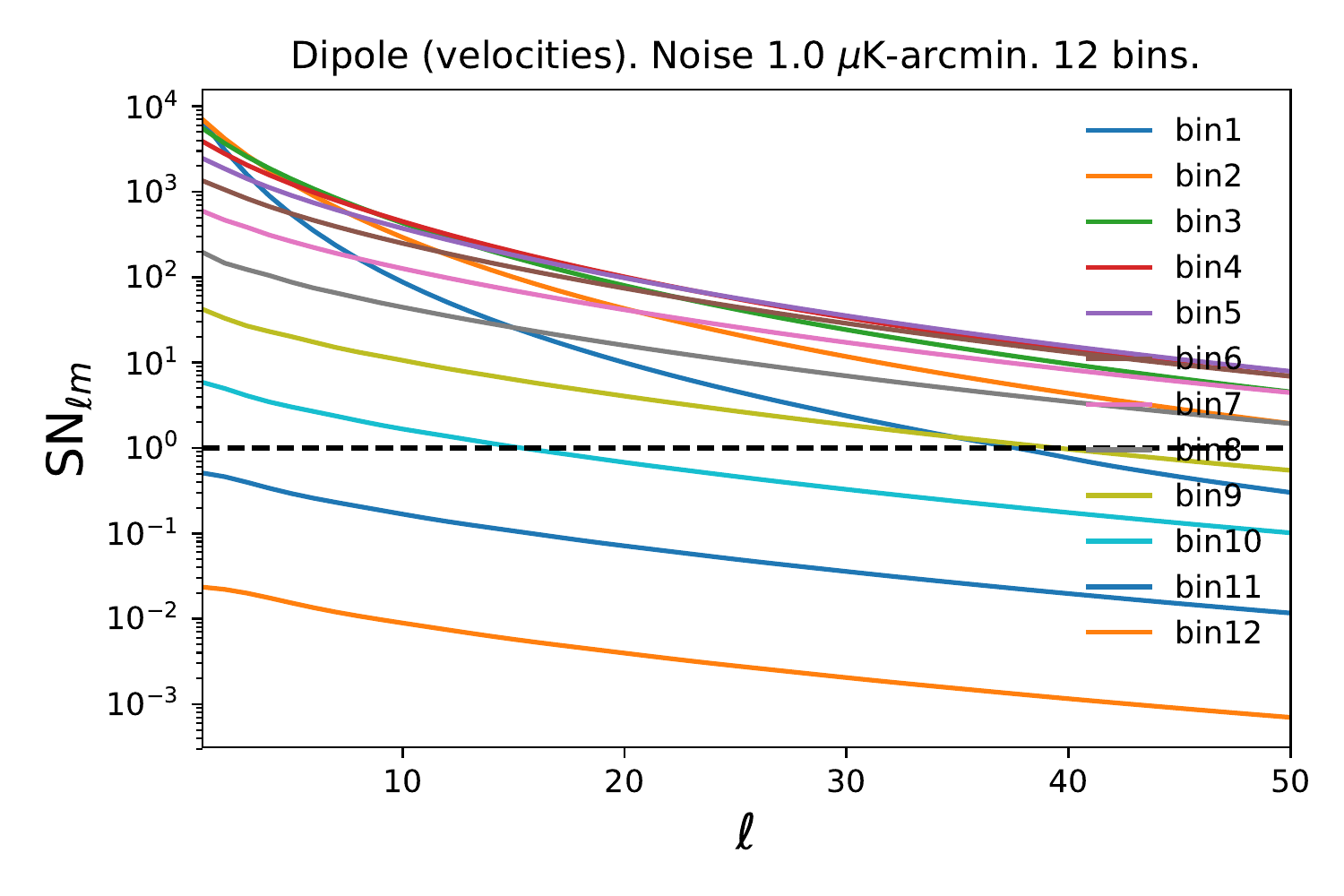}
}
	\caption{Left: Signal to noise per dipole field mode $\bar{v}_{{\rm eff}, \ell m}$ for a full sky experiment with experimental sensitivity oriented at CMB S4 and LSST (see main text), using the redshift binning given in table~\ref{tab:bins}. Right: Same with 12 bins, dividing each previous bin in two.}
	\label{fig:SN_vv}
\end{figure}

With these parameters, we obtain the cosmic variance limited signal to noise per mode in Fig.~\ref{fig:SN_vv}.  We conclude that the large-scale dipole field could be reconstructed with extremely high fidelity using our technique, with next generation experimental data, over most of the red-shift range. At high red shifts, the dropping number density of the LSST sample makes the dipole reconstruction impossible, but a deeper galaxy survey would recover these modes. By choosing more redshift bins in the same range $0 < z < 6$, it is possible to gain access to information about the remote dipole field on smaller scales. We therefore also show a 12 bin configuration in Fig.~\ref{fig:SN_vv} with the same experimental parameters.

\begin{table}[h!]
\centering
\begin{tabular}{ccc}
\hline
\multicolumn{1}{c}{$\alpha$\ \ \ \ } &\multicolumn{1}{c}{$(\chimin^{\alpha},\chimax^\alpha)$\ \ \ } & \multicolumn{1}{c}{\ \ $(z_\text{min}^{\alpha},z_\text{max}^\alpha)$} \\ \hline
1\ \ \ \ & $(0.00, 0.32)$ \ \ & \ \ $(0.00, 0.35)$ \\ 
2\ \ \ \ & $(0.32, 0.64)$ \ \ & \ \ $(0.35, 0.78)$ \\ 
3\ \ \ \ & $(0.64, 0.96)$ \ \ & \ \ $(0.78, 1.37)$ \\ 
4\ \ \ \ & $(0.96, 1.28)$ \ \ & \ \ $(1.37, 2.22)$ \\ 
5\ \ \ \ & $(1.28, 1.60)$ \ \ & \ \ $(2.22, 3.59)$ \\ 
6\ \ \ \ & $(1.60, 1.92)$ \ \ & \ \ $(3.59, 6.00)$ \\ 
\hline
\end{tabular}
\caption{We adopt the redshift binning of~\cite{Terrana2016} with six equally spaced redshift bins from the observer ($z=0$) to reionization ($z=6$). The comoving distance is given in units of $H_0^{-1}$.}
\label{tab:bins}
\end{table}

\section{Quadratic estimator for the quadrupole field}
\label{sec:quadrupole}

The same methodology that we used above for kSZ tomography can be used to reconstruct the CMB quadrupole field from the polarized SZ effect. The pSZ effect induces a correlation of the $E$-mode and $B$-mode CMB signal with the matter distribution. The CMB quadrupole field receives contributions from both scalar and tensor modes, which we forecast separately. The estimator described here has been previously presented in the context of detecting gravitational waves in~\cite{2012PhRvD..85l3540A}.

\subsection{Polarization from the CMB quadrupole field}

Polarization is generated by the line-of-sight integral of the local CMB temperature quadrupole, i.e.
\ba
(Q\pm iU)(\n) &=& -\frac{\sqrt{6}}{10} \int_0^\infty d\chi\, \dot\tau e^{-\tau(\n,\chi)} \tilde{q}_{\rm eff}^\pm(\n,\chi) \label{eq:line_of_sight_qu} \\
\tilde{q}_{\rm eff}^\pm(\n,\chi) &=& \sum_m \left[ q^{m, S}_{\rm eff} ({\bf \widehat{n}}, \chi) + q^{m, T}_{\rm eff} ({\bf \widehat{n}}, \chi)  \right]  {}_{\pm 2}Y_{2m}(\n) , \nn
\ea
where $q^{m, S}_{\rm eff}$ and $q^{m, T}_{\rm eff}$ are the scalar and tensor contributions to the components of the temperature quadrupole moment at each position in space, respectively. 
If we expand $(Q\pm iU)(\n)$~\cite{Dvorkin:2008tf}, in complete analogy with what was done above for the kSZ effect, we find that the CMB polarization due to the pSZ effect from the inhomogeneous distribution of electrons is
\ba
\label{eq:qu1_integral}
(Q\pm iU)(\n) \big|_{\rm pSZ} &=& -\frac{\sqrt{6}}{10} \sigma_T \int d\chi \ a \ \Delta n_e(\n,\chi) \tilde{q}_{\rm eff}^\pm(\n,\chi).
\ea
Now we redshift bin the equation in the same way as above to obtain
\be
(Q\pm iU)(\n) \big|_{\rm pSZ, \mathcal{L}} =  -\frac{\sqrt{6}}{10}  \sum_\alpha \Delta\tau^\alpha(\n) \tilde{q}_{\rm eff}^{\pm \alpha}(\n)  \label{eq:sec5_tqu}
\ee
Formally, $q^{\pm \alpha}(\n)$ should be interpreted as the average components of the quadrupole in each bin. However, since the correlation length for the quadrupole field is larger than any reasonable bin choice (see Appendix~\ref{sec:redshiftcorr}), unlike in the kSZ case here we do not discriminate between the averaged and un-averaged field. The binned equation Eq.~\eqref{eq:sec5_tqu} is the starting point for the quadratic estimator of the quadrupole field $q^\pm(\n,\chi)$. 

\subsection{CMB to matter cross power spectrum due to pSZ}

From Eq.~(\ref{eq:sec5_tqu}), it follows that the pSZ contribution to the CMB polarization is given by
\begin{align}
_{\pm2}a^{\rm pZS}_{\ell m} = &  -\frac{\sqrt{6}}{10}  \sum_\alpha \sum_{L_1,M_1,L_2,M_2} (-1)^m \ a^{q,\pm\alpha}_{L_1 M_1} \Delta \tau^\alpha_{L_2 M_2} \sqrt{\frac{(2\ell+1)(2L_1+1)(2L_2+1)}{4\pi}} \no \\
 & \times \wj{\ell}{L_1}{L_2}{\mp 2}{\pm 2}{0} \wj{\ell}{L_1}{L_2}{-m}{M_1}{M_2}
\end{align}
We decompose the CMB perturbations from pSZ in $E$- and $B$-modes,
\ba
a_{\ell m}^E = -\frac{1}{2} ( _{2}a_{\ell m} + _{-2}a_{\ell m})\\
a_{\ell m}^B = -\frac{1}{2i} ( _{2}a_{\ell m} - _{-2}a_{\ell m})
\ea
and equivalently for the remote quadrupole field
\ba
a^{q, E \alpha}_{\ell m} = -\frac{1}{2} ( a^{q, + \alpha}_{\ell m} + a^{q, - \alpha}_{\ell m} )\\
a^{q, B \alpha}_{\ell m} = -\frac{1}{2i} (a^{q, + \alpha}_{\ell m} - a^{q, - \alpha}_{\ell m} )
\ea
Note that if the remote quadrupole field is sourced only by scalar perturbations then $a^{q, + \alpha}_{\ell m} = a^{q, - \alpha}_{\ell m}$ and $a^{q, B \alpha}_{\ell m}=0$, but still $a^B_{\ell m} \neq 0$. That is, a purely $E$-mode type quadrupole field gives rise to both $E$-mode and $B$-mode power due to the spatial variations in optical depth~\cite{Dvorkin:2009ah}. The PSZ contribution to the CMB is then~\cite{2012PhRvD..85l3540A}
\ba
a_{\ell m}^E &=&  -\frac{\sqrt{6}}{10}   \sum_\alpha \sum_{L_1,M_1,L_2,M_2} (-1)^m \wj{\ell}{L_1}{L_2}{-m}{M_1}{M_2} F_{\ell,L_1,L_2} \nonumber \\ 
&& \times \Big( a^{q, E \alpha}_{L_1 M_1} \alpha_{\ell,L_1,L_2} - a^{q, B \alpha} _{L_1 M_1}  \gamma_{\ell,L_1,L_2}  \Big) \Delta \tau^\alpha_{L_2 M_2} \\
a_{\ell m}^B &=&  -\frac{\sqrt{6}}{10}   \sum_\alpha \sum_{L_1,M_1,L_2,M_2} (-1)^m  \wj{\ell}{L_1}{L_2}{-m}{M_1}{M_2} F_{\ell,L_1,L_2} \nonumber \\
&& \times \Big( a^{q, E \alpha}_{L_1 M_1} \gamma_{\ell,L_1,L_2} + a^{q, B \alpha} _{L_1 M_1} \alpha_{\ell,L_1,L_2} \Big) \Delta \tau^\alpha_{L_2 M_2} 
\ea
where we defined 
\ba
F_{\ell,L_1,L_2} &=& \sqrt{\frac{(2\ell+1)(2L_1+1)(2L_2+1)}{4\pi}} \wj{\ell}{L_1}{L_2}{2}{-2}{0}\\
\alpha_{\ell,L_1,L_2} &=& \frac{1}{2} (1 + (-1)^{\ell+L_1+L_2}) \\
\gamma_{\ell,L_1,L_2} &=& \frac{1}{2i} (1 - (-1)^{\ell+L_1+L_2})
\ea

We can now calculate the cross correlation of CMB $E$- and $B$-modes with the $\tau$ field as
\ba
\left\langle a^E_{\ell_1m_1}  \Delta\tau_{\ell_2 m_2}^\alpha \right\rangle &=&  -\frac{\sqrt{6}}{10} \sum_{\beta,L_1,M_1,L_2,M_2} 
(-1)^{m_1}  \wj{\ell_1}{L_1}{L_2}{-m_1}{M_1}{M_2} F_{\ell,L_1,L_2} \nonumber \\
&&\times \Big( a^{q,E\beta}_{L_1 M_1} \alpha_{\ell,L_1,L_2} - a^{q,B\beta} _{L_1 M_1}  \gamma_{\ell,L_1,L_2}  \Big) \left< \Delta \tau^\beta_{L_2 M_2} \ \Delta \tau_{\ell_2 m_2}^\alpha \right> \\
\left\langle a^B_{\ell_1m_1}  \Delta \tau_{\ell_2 m_2}^\alpha \right\rangle &=&  -\frac{\sqrt{6}}{10}   \sum_{\beta,L_1,M_1,L_2,M_2} 
(-1)^{m_1}  \wj{\ell_1}{L_1}{L_2}{-m_1}{M_1}{M_2} F_{\ell,L_1,L_2} \nonumber \\
&& \times \Big( a^{q,E\beta}_{L_1 M_1} \gamma_{\ell,L_1,L_2} + a^{q,B\beta} _{L_1 M_1}  \alpha_{\ell,L_1,L_2}  \Big) \left< \Delta \tau^\beta_{L_2 M_2} \ \Delta \tau_{\ell_2 m_2}^\alpha \right> 
\ea
We pulled $a^{q,E}_{L_1 M_1}$ and $a^{q,B}_{L_1 M_1}$ out of the correlators because they are the background fields that we want to estimate, not statistical quantities. Again, since we are considering large red shift bins, we can assume that
\be
\left< \Delta \tau^\alpha_{\ell_1 m_1} \ \Delta \tau_{\ell_2 m_2}^\beta \right> = (-1)^{m_2} C^{\Delta \tau \Delta \tau}_{\alpha,\ell_1} \delta_{\alpha \beta} \delta_{\ell_1 \ell_2} \delta_{m_1 -m_2}
\ee
and obtain
\ba
\label{eq:our_twopoint2}
\left\langle a^E_{\ell_1m_1}  \Delta\tau_{\ell_2 m_2}^\alpha \right\rangle &=& \sum_{L_1,M_1} 
 (-1)^{m_1+m_2} \wj{\ell_1}{L_1}{\ell_2}{-m_1}{M_1}{-m_2} \Gamma^{\rm pSZ }_{\ell_1 L_1 \ell_2 \alpha} \Big( a^{q,E\alpha}_{L_1 M_1} \alpha_{\ell,L_1,\ell_2} - a^{q,B\alpha} _{L_1 M_1}  \gamma_{\ell,L_1,\ell_2}  \Big) \\
\label{eq:our_twopoint3}
\left\langle a^B_{\ell_1m_1}  \Delta\tau_{\ell_2 m_2}^\alpha \right\rangle &=& \sum_{L_1,M_1} 
 (-1)^{m_1+m_2} \wj{\ell_1}{L_1}{\ell_2}{-m_1}{M_1}{-m_2} \Gamma^{\rm pSZ}_{\ell_1 L_1 \ell_2 \alpha} \Big( a^{q,E\alpha}_{L_1 M_1} \gamma_{\ell,L_1,\ell_2} + a^{q,B\alpha} _{L_1 M_1}  \alpha_{\ell,L_1,\ell_2}  \Big) 
\ea
where we defined the coupling
\ba
\Gamma^{\rm pSZ}_{\ell_1 L_1 \ell_2 \alpha} = - \frac{\sqrt{6}}{10} F_{\ell,L_1,\ell_2} \ C^{\Delta\tau \Delta\tau}_{\alpha,\ell_2}.
\ea
The sum in $L_1,M_1$ is dominated by very low $\ell$, since the quadrupole field has only low multipole contributions.

\subsection{Estimator and variance of the quadrupole field}

Based on Eq.\eqref{eq:our_twopoint2} and \eqref{eq:our_twopoint3}, we can construct an estimator of form
\be
 \widehat{a}^{q,X \alpha}_{\ell m} = \sum_{\ell_1m_1\ell_2m_2} \Big(  W^{X,E}_{\ell m\ell_1 m_1\ell_2 m_2} \ a^E_{\ell_1m_1} +  W^{X,B}_{\ell m\ell_1 m_1\ell_2 m_2} \ a^B_{\ell_1m_1}  \Big)  \Delta\tau^\alpha_{\ell_2m_2},
\ee
where $X = \{E,B\}$. In analogy with the case above we find
\ba
\label{eq:estimator1q}
\widehat{a}^{q,E \alpha}_{\ell m}&=& N^{E \alpha}_{\ell}
\sum_{\ell_1m_1\ell_2m_2} (-1)^m \
\Gamma^{\rm pSZ}_{\ell \ell_1\ell_2 \alpha}
\wj{\ell_1}{\ell_2}{\ell}{m_1}{m_2}{-m} 
\frac{   \Big( \alpha_{\ell,\ell_1,\ell_2} a^E_{\ell_1m_1}  - \gamma_{\ell,\ell_1,\ell_2}  a^B_{\ell_1m_1}  \Big) \Delta\tau^{*\alpha}_{\ell_2m_2}}   { \Big( |\alpha_{\ell,\ell_1,\ell_2}|^2 \tilde{C}^{EE}_{\ell_1} + |\gamma_{\ell,\ell_1,\ell_2}|^2 \tilde{C}^{BB}_{\ell_1} \Big)  \tilde{C}^{\Delta\tau \Delta\tau}_{\alpha \ell_2} } \\
\widehat{a}^{q,B \alpha}_{\ell m}&=& N^{B \alpha}_{\ell}
\sum_{\ell_1m_1\ell_2m_2} (-1)^m\ 
\Gamma^{\rm pSZ}_{\ell \ell_1\ell_2 \alpha}
\wj{\ell_1}{\ell_2}{\ell}{m_1}{m_2}{-m} 
\frac{   \Big( \gamma_{\ell,\ell_1,\ell_2} a^E_{\ell_1m_1}  + \alpha_{\ell,\ell_1,\ell_2}  a^B_{\ell_1m_1}  \Big) \Delta\tau^{*\alpha}_{\ell_2m_2}}   { \Big( |\gamma_{\ell,\ell_1,\ell_2}|^2 \tilde{C}^{EE}_{\ell_1} + |\alpha_{\ell,\ell_1,\ell_2}|^2 \tilde{C}^{BB}_{\ell_1} \Big)  \tilde{C}^{\Delta\tau \Delta\tau}_{\alpha \ell_2} },  
\ea
where
\ba
\label{eq:estimator1noiseq}
\frac{1}{N^{E \alpha}_{\ell}} &=& \frac{1}{(2\ell+1)} \sum_{\ell_1\ell_2}
\frac{\Gamma^{\rm pSZ}_{\ell \ell_1\ell_2 \alpha} \ \Gamma^{\rm pSZ}_{\ell \ell_1\ell_2 \alpha}}{\Big( |\alpha_{\ell,\ell_1,\ell_2}|^2 \tilde{C}^{EE}_{\ell_1} + |\gamma_{\ell,\ell_1,\ell_2}|^2 \tilde{C}^{BB}_{\ell_1} \Big)  \tilde{C}^{\Delta\tau \Delta\tau}_{\alpha \ell_2} }\\
\frac{1}{N^{B \alpha}_{\ell}} &=& \frac{1}{(2\ell+1)} \sum_{\ell_1\ell_2}
\frac{\Gamma^{\rm pSZ}_{\ell \ell_1\ell_2 \alpha} \ \Gamma^{\rm pSZ}_{\ell \ell_1\ell_2 \alpha}}{\Big( |\gamma_{\ell,\ell_1,\ell_2}|^2 \tilde{C}^{EE}_{\ell_1} + |\alpha_{\ell,\ell_1,\ell_2}|^2 \tilde{C}^{BB}_{\ell_1} \Big)  \tilde{C}^{\Delta\tau \Delta\tau}_{\alpha \ell_2} }.
\ea
The definition of signal and noise is the same as in the case of kSZ tomography discussed above, i.e. $N^{E,B}_\ell$ is the contribution we would get from statistical fluctuations if there were no remote quadrupole field. The expected signal-to-noise per quadrupole field mode is
\be
\label{eq:qSN}
(S/N)_{\ell m} = \left[ \frac{\fsky}{2} \left( \frac{C_{\alpha \alpha \ell}^{E/B}}{N_{\alpha \ell}^{E/B}} \right)^2 \right]^{1/2}.
\ee
which corresponds to the expected number of "sigmas" of a detection.

\subsection{Signal-to-noise forecast}

\subsubsection{Quadrupole field signal and noise}\label{sec:quadsignal}

Our signal is the projected quadrupole field $a^{q,E}_{\ell m}(\chi)$ and $a^{q,B}_{\ell m}(\chi)$. In~\cite{Deutsch:2017cja}, we calculated $a^{q,E}_{\ell m}(\chi)$ from scalar perturbations in terms of the primordial potential $\Psi_i({\bf k})$:
\begin{equation} \label{eq:almqfinal}
	a^{q,E}_{\ell m}(\chi) = \int \frac{d^3k}{(2\pi)^3} \Delta_\ell^q(k,\chi)\ \Psi_i({\bf k})\ Y^*_{\ell m}({\bf \widehat{k}}) \ ,
\end{equation}
The transfer function $\Delta_\ell^q(k,\chi)$ is reviewed in Appendix~\ref{sec:scalarcontribution} and discussed in detail in~\cite{Deutsch:2017cja}; it is dominated by the Sachs-Wolfe effect. In this case, there is a pure $E$-mode component of the quadrupole field. In the presence of primordial gravitational waves, there will also be a contribution to both the $E$-mode and $B$-mode components of the quadrupole field. This was first explored in Ref.~\cite{2012PhRvD..85l3540A}. We present a derivation of these contributions in Appendix~\ref{sec:tensorcontribution}, where the final result is:
\begin{align}
	a_{\ell m}^{q,E}(\chi) = & \int \frac{d^3 k}{(2\pi)^3}\ 5 i^\ell B_{\ell}(k,\chi)\left\{ \mathcal{G}^q_{T,+}(k,\chi) \left[{}_{2}Y^{*}_{\ell m}({\bf \widehat{k}}) + {}_{-2}Y^{*}_{\ell m}({\bf \widehat{k}}) \right]  \right.  + i \left. \mathcal{G}^q_{T,\times}(k,\chi) \left[{}_{2}Y^{*}_{\ell m}({\bf \widehat{k}}) - {}_{-2}Y^{*}_{\ell m}({\bf \widehat{k}}) \right] \right\}, \label{eq:talmE}\\
	a_{\ell m}^{q,B}(\chi) = & \int \frac{d^3 k}{(2\pi)^3}\ 5 i^\ell A_{\ell}(k,\chi)\left\{ -\mathcal{G}^q_{T,+}(k,\chi) \left[{}_{2}Y^{*}_{\ell m}({\bf \widehat{k}}) - {}_{-2}Y^{*}_{\ell m}({\bf \widehat{k}}) \right]  \right.  - i \left. \mathcal{G}^q_{T,\times}(k,\chi) \left[{}_{2}Y^{*}_{\ell m}({\bf \widehat{k}}) + {}_{-2}Y^{*}_{\ell m}({\bf \widehat{k}}) \right] \right\}. \label{eq:talmB}
\end{align}
Here, $A_{\ell}(k,\chi)$ and $B_{\ell}(k,\chi)$ encode projection effects, and have the limiting values of $A_{\ell}(k,\chi\rightarrow 0)\rightarrow 0$ and $B_{\ell}(k,\chi\rightarrow 0)\rightarrow -1/5$. The functions $\mathcal{G}^q_{T,(+,\times)}(k,\chi)$ fix the amplitude of the effect from the two gravitational wave polarization states. Definitions of these functions can be found in Appendix~\ref{sec:tensorcontribution}.

From the multipoles, assuming equal amplitudes for the two gravitational wave polarization states we obtain the signal power spectra:
\ba
\left\langle a^{E \alpha*}_{\ell m} a^{E \beta*}_{\ell'm'} \right\rangle_{\rm signal} &=& \left( C_{\alpha \beta \ell}^{S,E} + C_{\alpha \beta \ell}^{T,E} \right) \delta_{\ell\ell'} \delta_{mm'}, \\
\left\langle a^{B \alpha*}_{\ell m} a^{B \beta*}_{\ell'm'} \right\rangle_{\rm signal} &=& \left( C_{\alpha \beta \ell}^{T,B} \right) \delta_{\ell\ell'} \delta_{mm'}, \\
\left\langle a^{E \alpha*}_{\ell m} a^{B \beta*}_{\ell'm'} \right\rangle_{\rm signal} &=& 0.
\ea
where
\begin{eqnarray}
C_{\alpha \beta \ell}^{S,E}  &=& \int_0^{k_\text{max}} \frac{k^2dk}{(2\pi)^3} P_\Psi(k)  \Delta^q_{L}(k, \chi_\alpha)  \Delta^q_{L}(k, \chi_\beta)^* \ , \label{eq:clq} \\
C_{\alpha \beta \ell}^{T,E} &=& 2\int\frac{k^2\ dk}{(2\pi)^{3}}  \ 50\  P_h (k)\   \mathcal{I}^q_T(k,\chi_\alpha ) \mathcal{I}^q_T(k,\chi_\beta)\ B_{\ell}(k,\chi_\alpha)B_{\ell}(k,\chi_\beta), \\
C_{\alpha \beta \ell}^{T,B} &=& 2\int\frac{k^2\ dk}{(2\pi)^{3}}  \ 50\  P_h (k)\   \mathcal{I}^q_T(k,\chi_\alpha ) \mathcal{I}^q_T(k,\chi_\beta)\ A_{\ell}(k,\chi_\alpha)A_{\ell}(k,\chi_\beta).
\end{eqnarray}
The $\tau$ field is given in terms of its power spectrum $\tilde{C}^{\Delta\tau \Delta\tau}_\ell$ as in the previous section. The polarization field has the CMB power spectra $\tilde{C}^{EE}_{\ell}$ and $\tilde{C}^{BB}_{\ell}$, which includes primary CMB and lensing.

\subsubsection{Forecast} \label{sec:forecastqq}

We forecast the remote quadrupole field from scalar perturbations, a guaranteed signal, as well as that of tensor perturbations assuming a tensor ratio of $r=0.1$, the upper limit of current experimental constraints. Scalar perturbations induce only an E-type remote quadrupole field, while tensor perturbations source both E-type and B-type. Our experimental configuration is as above in the kSZ case with the same 6-bin setup up to red-shift 6, using the LSST galaxy sample and CMB S4 oriented noise values, as well as a hypothetical experiment with 10 times reduced noise.

\begin{figure}[t!]
\resizebox{1.0\hsize}{!}{
\includegraphics{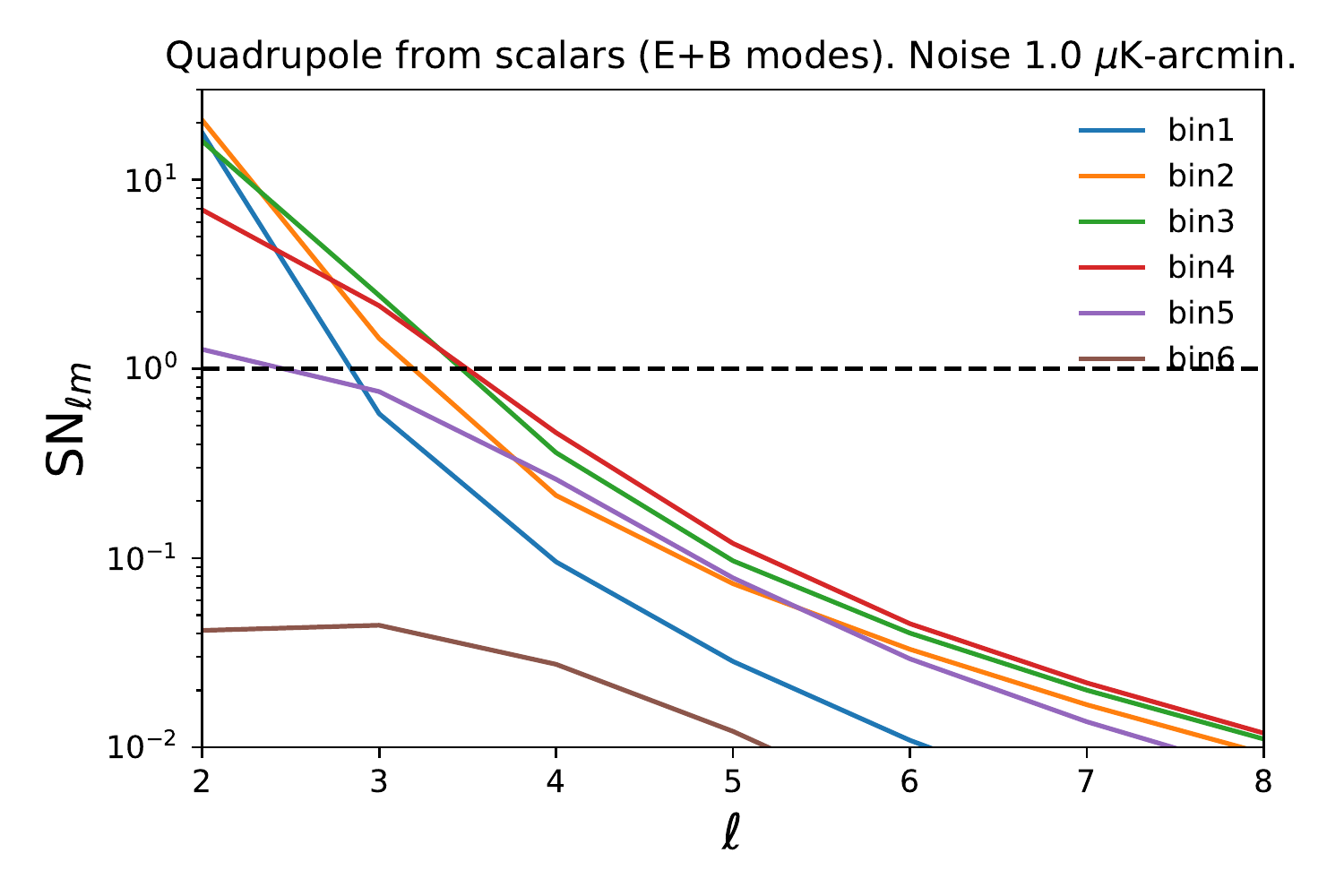}\includegraphics{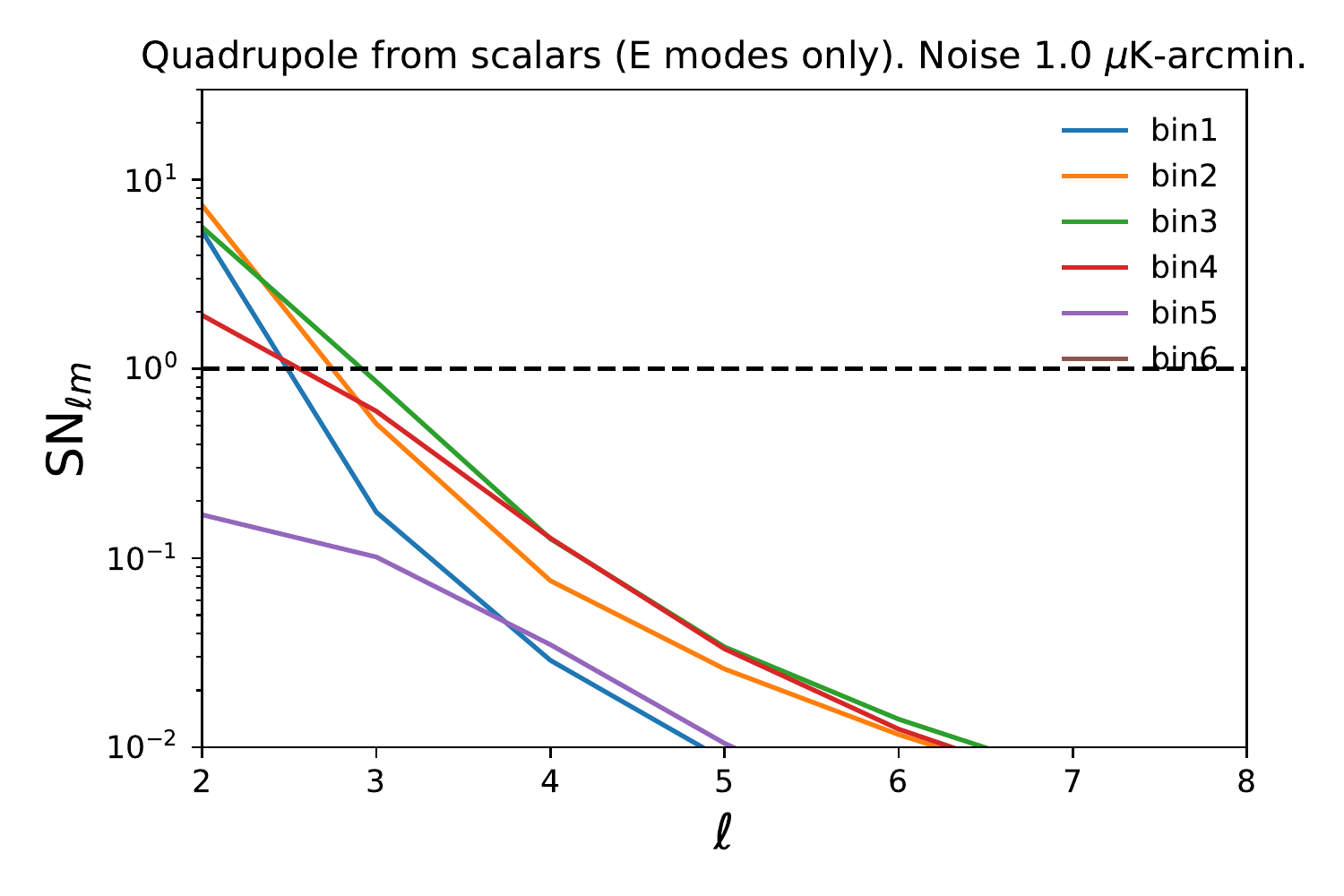}
}
	\caption{Signal to noise from scalar perturbations per quadrupole field mode $a^{q,E}_{\ell m}$ for a full sky experiment with experimental sensitivity oriented at CMB S4 and LSST, using the redshift binning in table~\ref{tab:bins}. Left: using $E$- and $B$-modes. Right: using only $E$-modes. This illustrates the power of $B$-modes for this application.}
	\label{fig:SN_SS}
\end{figure}

In Fig.~\ref{fig:SN_SS}, we show the per-mode signal to noise for the scalar sourced $E$-mode quadrupole field. In the left panel, we include information both from the $E$- and $B$-mode polarization anisotropies in the CMB. Here, it can be seen that a reconstruction of the quadrupole field can be made up to about $\ell=4$. The lower signal to noise in the high red shift bins is due to the falling galaxy density. For comparison, in the right panel we show the per-mode signal to noise including only (small scale) $E$-mode polarization anisotropies. The signal to noise is considerably smaller in this case due to the smaller amplitude of the lensed $B$-modes which enter the noise.

\begin{figure}[t!]
\resizebox{1.0\hsize}{!}{	
\includegraphics{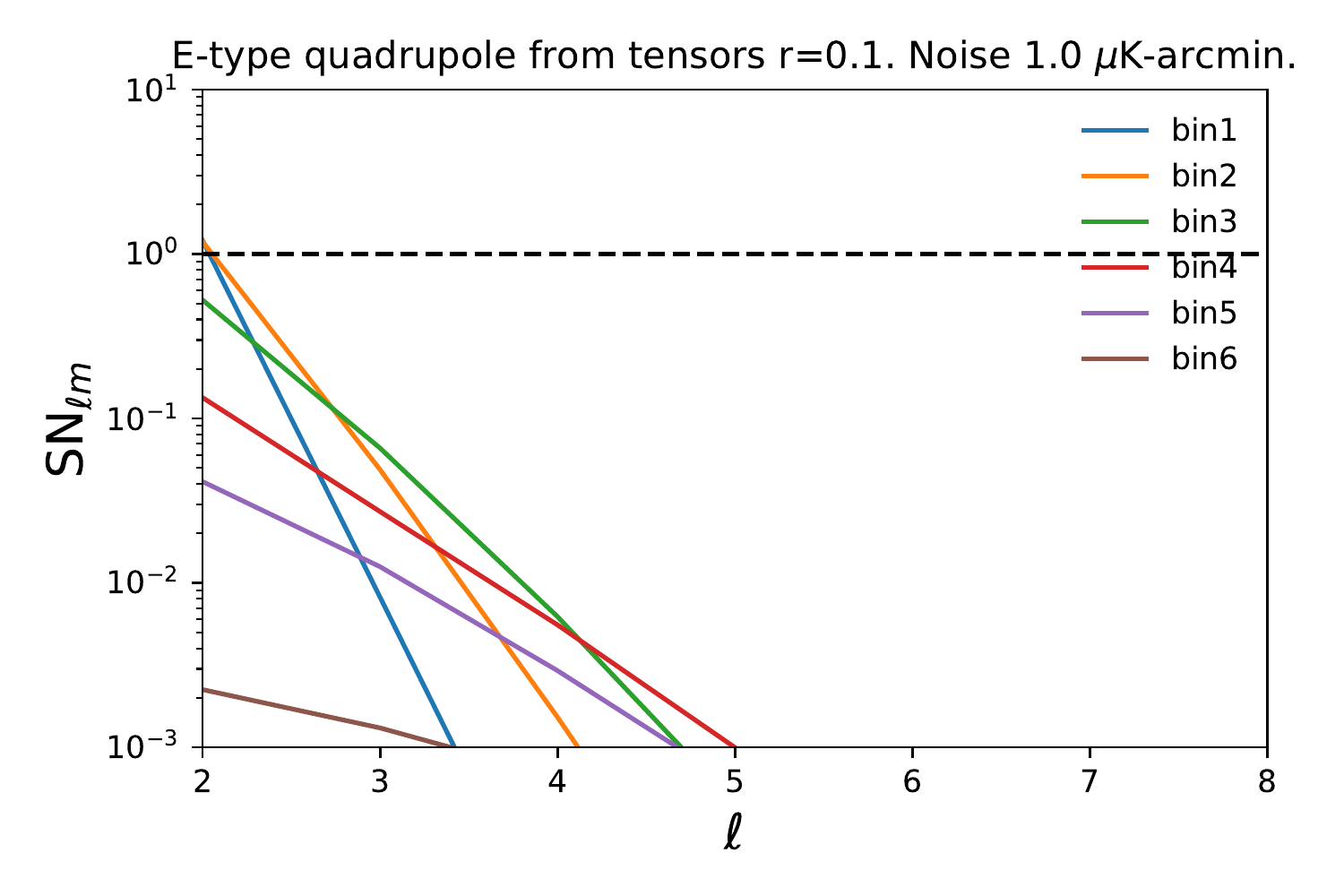}\includegraphics{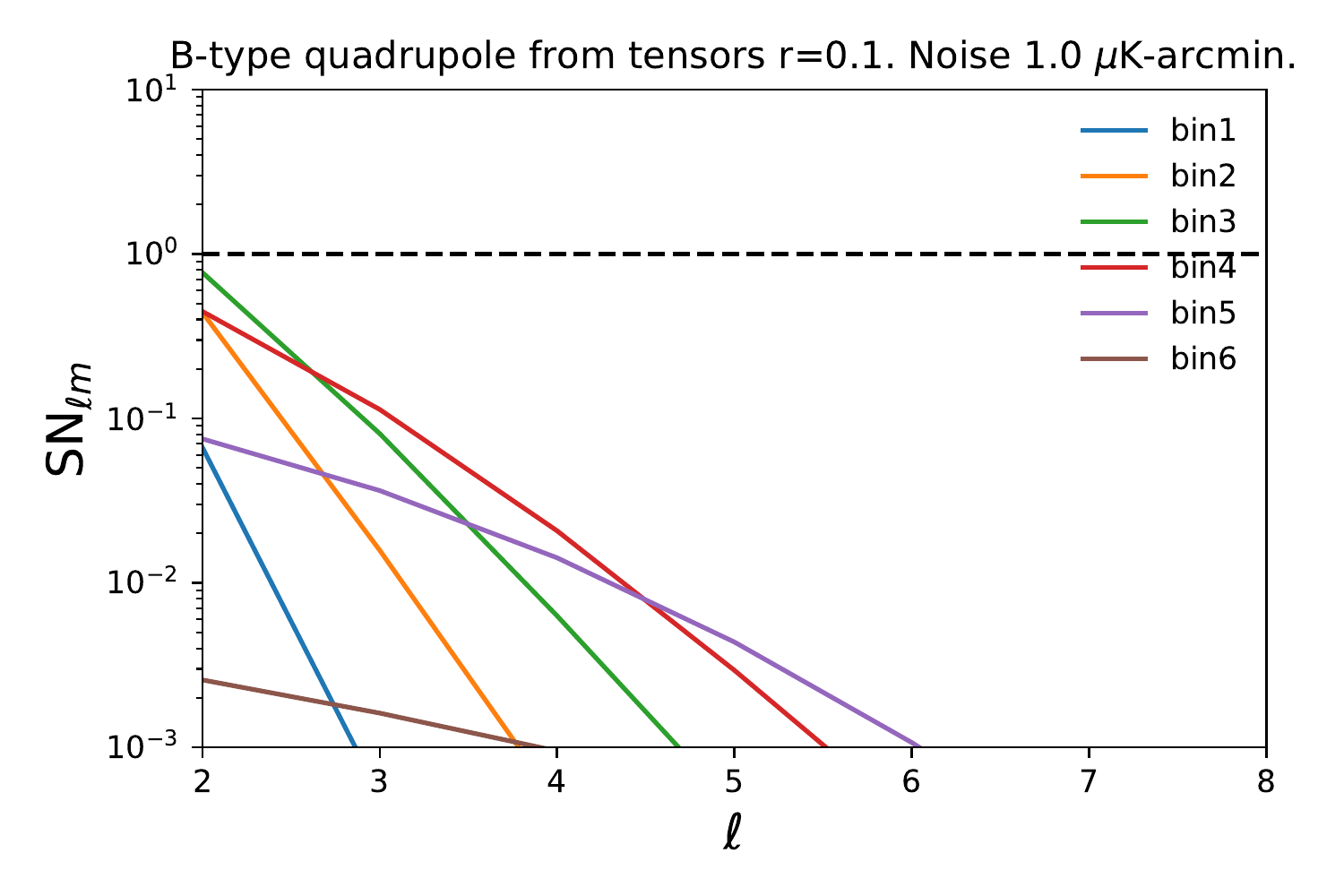}
}
\resizebox{1.0\hsize}{!}{	
\includegraphics{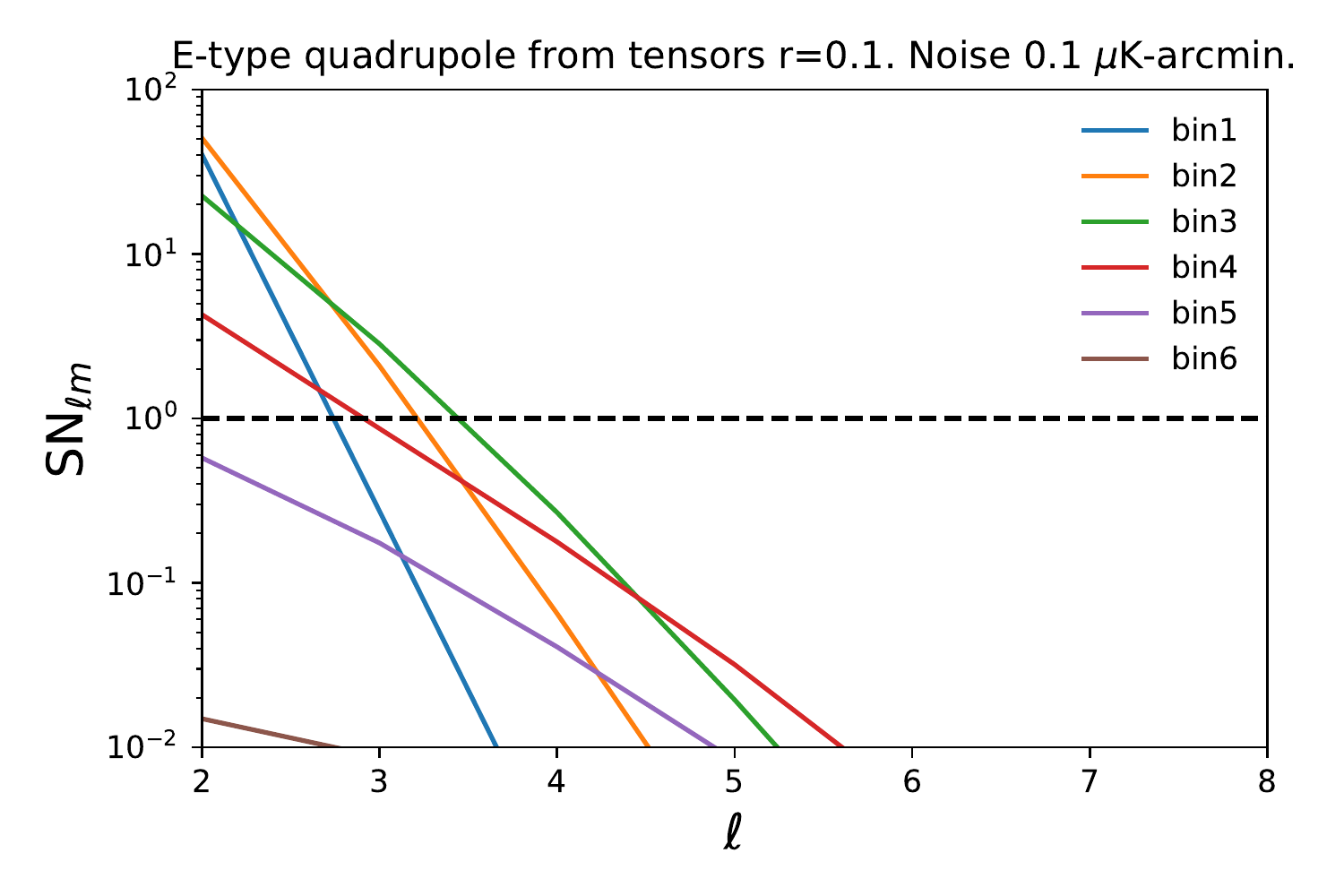}\includegraphics{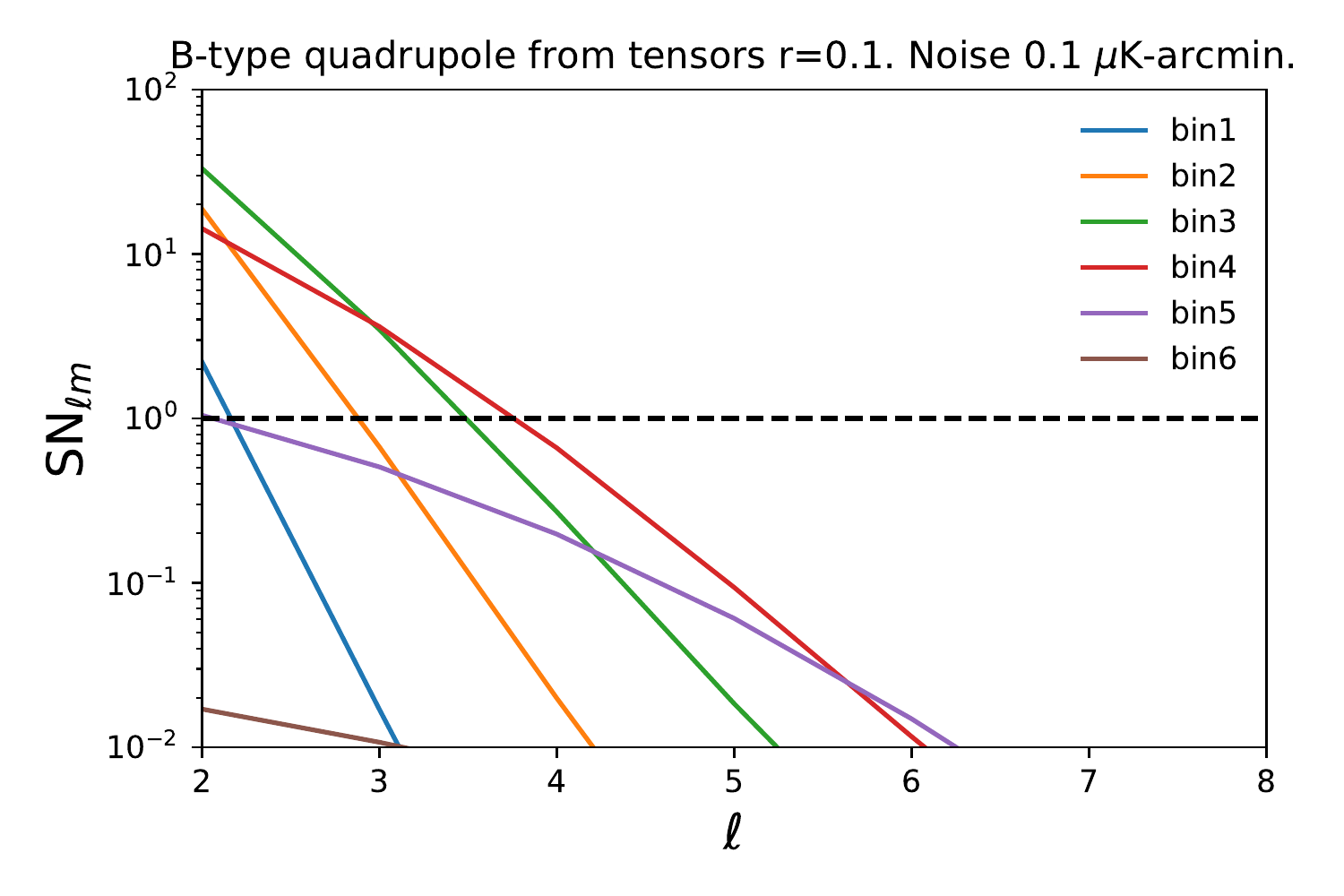}
}
	\caption{Top: Signal to noise for tensors with $r=0.1$ per quadrupole field mode $a^{q,E}_{\ell m}$ (left) and $a^{q,B}_{\ell m}$ (right) for a full sky experiment with experimental sensitivity oriented at CMB S4 and LSST, using the redshift binning in table~\ref{tab:bins}, using $E$- and $B$-modes. Bottom: Using the same configuration but reducing the CMB noise by a factor of 10.}
	\label{fig:SN_SStens}
\end{figure}

In Fig.~\ref{fig:SN_SStens}, we show the per-mode signal to noise for the tensor sourced $E$-mode (left) and $B$-mode (right) components of the quadrupole field for $N^{BB,EE} \simeq 1.0 \ \mu$K-arcmin (top) and $N^{BB,EE} \simeq 0.1 \ \mu$K-arcmin (bottom), both using the LSST galaxy density. Here we assumed a fiducial value $r=0.1$. The signal to noise ratio scales linearly with $r$. For the CMB S4 type noise a combined detection may be possible, but not a mode by mode reconstruction. On the other hand for the 10 times lowered noise, tensor modes would be detected easily and a noisy large scale map could be constructed. An interesting effect is that because the $B$-mode component of the quadrupole field vanishes at low redshift (see Eq.~\eqref{eq:talmB}), there is little signal in the first bin in the plots on the right. A measurement of the $B$-mode component of the quadrupole field therefore benefits from relatively high redshift surveys. 

\section{Principal component analysis}
\label{sec:PCA}

\begin{figure}[t!]
\resizebox{0.80\hsize}{!}{	
	\includegraphics{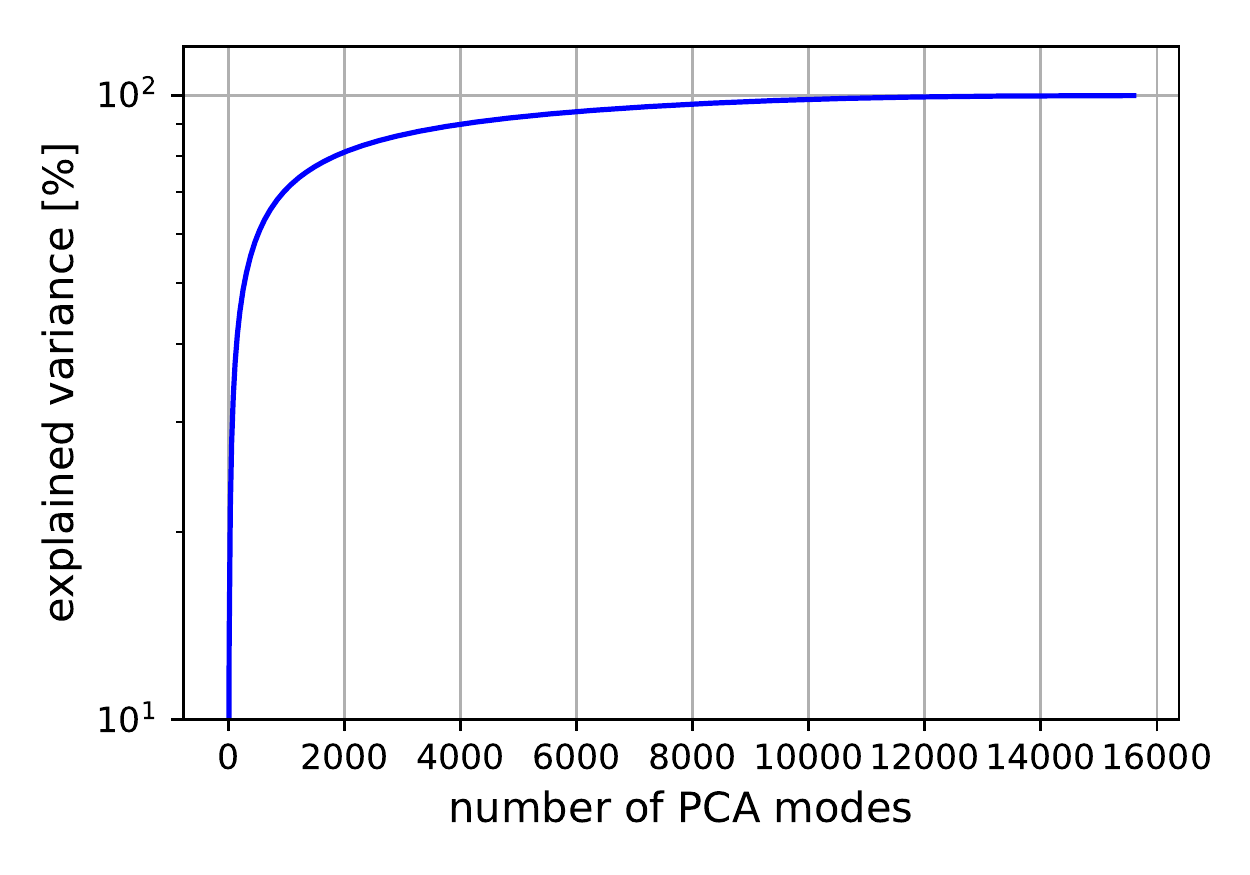}\includegraphics{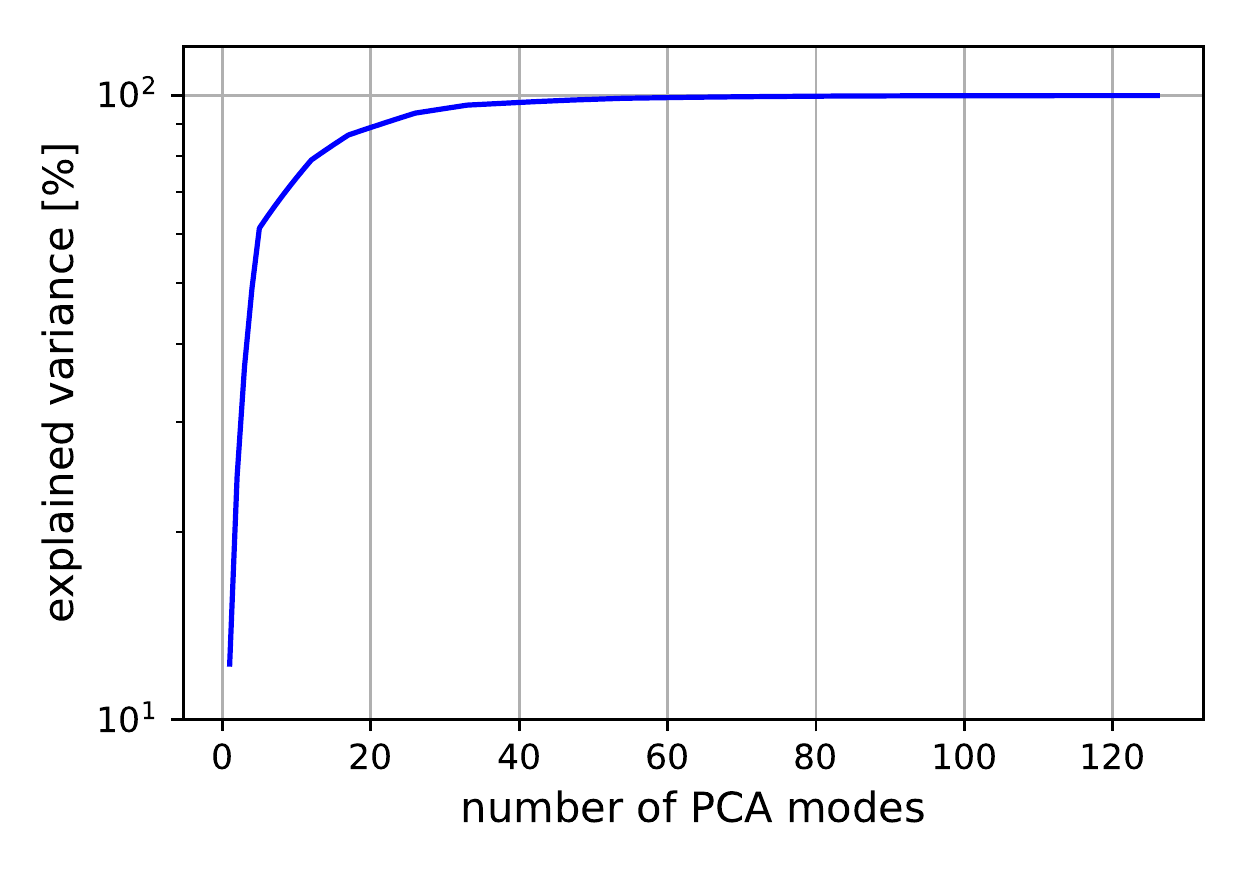}
}
	\caption{Explained variance as a function of the number of principal components. Left: dipole field $\bar{v}_{LM}$ with $\ell_{\rm max} = 50$ and 6 redshift bins. Right: quadrupole field $a^{q,E}_{LM}$ with $\ell_{\rm max} = 4$ and 6 redshift bins.}
	\label{fig:pca}
\end{figure}

Our signal, the effective dipole field and the remote quadrupole field, is correlated between different red shift bins. This correlation is in particular strong for the remote quadrupole field, i.e. there is a modest number of independent quadrupoles contained in the observable universe. To quantify how much degenerate information we obtain from our estimators, we performed a principal component analysis (PCA). The PCA for the dipole and quadrupole fields are based on the signal covariance matrices Eq.~\ref{eq:vvcovmat} and~\ref{eq:clq} respectively, which are $N\times N$ matrix where $N=N_{\rm bins}\sum_\ell (2\ell+1)$, of which most elements are zero. Here, we consider only scalar contributions to the quadrupole field. Based on the covariance matrix, we plot the ``explained variance'' as a function of the number of PCA components, which is the usual diagnostic for the number of components in a PCA. For the kSZ case we choose $\ell_{\rm max} = 50$ and for the pSZ case $\ell_{\rm max} =4$, motivated by the signal-to-noise forecasts above. The results are shown in Fig.~\ref{fig:pca}. As expected, in the case of the quadrupole field most of the structure is described by a very small number of modes, due to the large correlation length. This is consistent with previous observations to this effect~\cite{Kamionkowski1997,Bunn2006}. A much greater number of independent modes is available from the dipole field, which is sensitive to peculiar velocities with a small correlation length. We stress, however, that the modes probed by the remote dipole and quadrupole are different than those probed by the primary CMB on large angular scales, as they involve information from the volume enclosed by the past light cone, as opposed to a projection onto the past light cone.

\section{Probes of homogeneity}
\label{sec:homogeneity}

Above, we have established that the largest modes of the dipole and quadrupole fields can be measured with the highest signal to noise. This suggests that for a set of experiments with significant sky overlap, the largest scale modes will be the first to be detected. What might we learn from the measurement of these modes? Below, we explore two possible applications that might be addressed with a first measurement: reconstructing the primordial CMB dipole and testing statistical homogeneity of the observable Universe using the quadrupole field. Beyond the first detected modes, with more information it becomes possible to use observations of the remote dipole and quadrupole fields to reconstruct the primordial gravitational potential on large scales, as we also outline here.

\subsection{Reconstructing the primordial CMB dipole}

In contrast to the higher multipole moments of the CMB, the observed CMB dipole is determined almost entirely by structure on non-linear scales, e.g. the peculiar velocity induced by solar-system, galactic, group, and cluster scales. This is the kinematic CMB dipole. There is, however, a contribution from linear scales which we refer to here as the ``primordial dipole". Note that there exists some ambiguity in the precise definition of the kinematic and primordial dipole because the overall observed dipole can be boosted to any desired value, at the price of inducing aberration of the primary CMB anisotropies (among other effects such as the moving lens effect; see Ref.~\cite{Lewis:2006aa} for an extended discussion). Moreover, different observables will be sensitive to different components of the dipole. For example, the lensing reconstruction method introduced in Ref.~\cite{Meerburg:2017xga} and the spectral method described in~\cite{Yasini:2016pby,Yasini:2016dnd} are sensitive to the CMB dipole observed in a frame with vanishing aberration of the primary CMB anisotropies. In any case, the various primordial dipoles can be unambiguously related back to the underlying inhomogeneities in one's favourite gauge (see e.g.~\cite{Terrana2016}). 

The first modes we might hope to reconstruct using kSZ tomography, $\bar{v}^1_{\rm eff, 1m}$ at low redshift, provide a direct measurement the local primordial dipole. To see this, first decompose the dipole field $v_{\rm eff}$, defined by
\be\label{eq:veffdef}
v_{\rm eff}({\bf \widehat{n}}_e, \chi)=\frac{3}{4\pi}\int d^2{\bf \widehat{n}} \ \Theta({\bf \widehat{n}}_e, \chi, {\bf \widehat{n}}) \ ({\bf \widehat{n}}\cdot{\bf \widehat{n}}_e) ,
\ee
into a weighted sum of local dipole components
\begin{equation}
v_{\rm eff}({\bf \widehat{n}}_e, \chi) = \sum_{m=-1}^{1} v_{\rm eff}^m({\bf \widehat{n}}_e, \chi) Y_{1m} ({\bf \widehat{n}}_e), \ \ \ v_{\rm eff}^m({\bf \widehat{n}}_e, \chi) \equiv \int d^2 {\bf \widehat{n}} \ \Theta({\bf \widehat{n}}_e, \chi, {\bf \widehat{n}}) Y^*_{1m} ({\bf \widehat{n}})
\end{equation}
The $m=-1,0,1$ components of the locally observed CMB dipole are defied by $v_{\rm eff}^m(0) \equiv v_{\rm eff}^m({\bf \widehat{n}}_e, \chi \rightarrow 0)$. At $\chi \neq 0$, from this definition and Eq.~\ref{eq:almv} we see that $v_{\rm eff, 1m}(\chi)$ is an angular average of each component of the locally observed dipole at fixed redshift. The bin-averaged dipole field $\bar{v}^1_{\rm eff, 1m}$ introduces an additional smoothing of small-scale velocities due to cancelation along the line of sight. In a redshift bin large enough to encompass linear scales, the kinematic dipole components average out, and the estimator for the averaged dipole field is a measure of the average primordial dipole in the volume. Because the correlation length of the averaged field is, by definition, large on the scales of the redshift bin, a measurement of $\bar{v}^1_{\rm eff, 1m}$ can be used as an estimate of the components of the locally observed primordial CMB dipole. As we have seen in Sec.~\ref{sec:forecastvv}, it is possible to achieve a very high signal to noise detection of $\bar{v}^1_{\rm eff, 1m}$, and therefore a high signal to noise detection of the local primordial dipole.

\subsection{Testing statistical isotropy using the CMB quadrupole}

Using the \texttt{Commander} approach, the Planck 2015 data yield a central value for the power in the CMB quadrupole of $C_2^{\rm measured} = 253.6 \ \mu{\rm K}$~\cite{web:pla}. This can be compared with the theoretical prediction for the best-fit $\Lambda$CDM model~\cite{Ade:2015xua} of $C_2^{\Lambda{\rm CDM}} = 1123.6 \pm 355.393 \ \mu{\rm K}$, where we have included the theoretical cosmic variance error bar. The measured value is a factor of roughly 4 smaller than the theoretical prediction, and the degree to which the measured value for the CMB quadrupole is anomalously low has been debated since it's first measurement by the COBE satellite~\cite{1996ApJ...464L..17H}. Although it is still entirely consistent to take the position that we simply inhabit a rare realization~\cite{Efstathiou:2003wr}, the low quadrupole may be a hint of new physics on ultra-large scales. 

The first modes measured by pSZ tomography can be used to test the hypothesis that observers at other locations measure a quadrupole consistent with our own, and therefore that we do not inhabit a special location in the Universe. Because of the large correlation length of the quadrupole field (see appendix~\ref{sec:redshiftcorr}), a measurement of $a^{q,E1}_{2m}$, the $\ell=2$ moments of the $E$-mode type quadrupole field in the first redshift bin, should closely match our observed CMB quadrupole. However, the large correlation length is a consequence of the assumption of statistical homogeneity on large scales. Therefore, this first measurement will provide an check of this assumption. A more systematic check would involve a measurement of the bin-to-bin correlation of the $\ell=2$ moments of the $E$-mode type quadrupole field. Given the possibility of a relatively high signal to noise detection, as outlined in Sec.~\ref{sec:forecastqq}, this could be among the first applications of pSZ tomography.

\subsection{Primordial potential from the remote dipole and quadrupole fields}

Above, we estimated a map of the remote CMB dipole and quadrupole fields, we now show how such a measurement can be transformed into an estimate of the primordial potential at the same scales. This was first worked out in~\cite{Yadav:2005tf} which we review and generalize here. To reconstruct the primordial potential $\Psi$, the appropriate coordinates for $\Psi$ are spherical harmonics with the observation point at the origin, i.e. $\Psi(\widehat{n},\chi) = \sum_{\ell m} \psi_{\ell m}(\chi) Y_{\ell m}(\widehat{n})$. The dipole and quadrupole fields depend primarily on linear scales, and one can therefore find a linear transformation between $a_{\ell m}$ and the reconstructed underlying $\psi_{\ell m}$. Assuming statistical isotropy and homogeneity, the transformation $\mathcal{O}^{X,\alpha}$ is a function of $\ell$ and $\chi$ only, where $X=\bar{v},q$ the dipole and quadrupole fields, and $\alpha$ labels the redshift bin. The reconstructed primordial field is thus of form $\widehat{\psi}_{\ell m}(\chi) = \sum_{X,\alpha} \mathcal{O}^{X,\alpha}_\ell (\chi) a^{X,\alpha}_{\ell m}$.  One can find the transformation $\mathcal{O}^{X,\alpha}_\ell (\chi)$ analytically by minimizing the expected deviation of the reconstructed field from the true $\psi_{\ell m}(\chi)$ as
\begin{align}
\frac{\partial}{\partial \mathcal{O}^{X,\alpha}_\ell (\chi)} \left< | \sum_{X,\alpha} \mathcal{O}^{X,\alpha}_\ell (\chi) a^{X,\alpha}_{\ell m} - \psi_{\ell m}(\chi)|^2 \right> = 0
\end{align}
which is solved by
\begin{equation}
\begin{bmatrix}
    \mathcal{O}^{q,1}_\ell (\chi)\\
    \mathcal{O}^{v,1}_\ell (\chi) \\
    \vdots \\
 \mathcal{O}^{q,N_{\rm bins}}_\ell (\chi)\\
    \mathcal{O}^{v,N_{\rm bins}}_\ell (\chi)
\end{bmatrix}
=
\begin{bmatrix}
    C_{11\ell}^{\bar{v} \bar{v}} & C_{11\ell}^{\bar{v} q} & C_{12\ell}^{\bar{v} \bar{v}} & \dots  & C_{1N_{\rm bins}\ell}^{\bar{v} q}\\
    C_{11\ell}^{q\bar{v}} & C_{11\ell}^{q q} & C_{12\ell}^{q\bar{v}} & \dots  & C_{1N_{\rm bins}\ell}^{q q} \\
    \vdots & \vdots & \vdots & \ddots & \vdots \\
    C_{ N_{\rm bins} 1 \ell}^{q\bar{v}} & C_{N_{\rm bins}1\ell}^{q q} & C_{N_{\rm bins}2\ell}^{q \bar{v}} & \dots  & C_{N_{\rm bins}N_{\rm bins}\ell}^{q q}
\end{bmatrix}^{-1}
\begin{bmatrix}
    \beta^{q,1}_\ell (\chi)\\
    \beta^{v,1}_\ell (\chi) \\
    \vdots \\
 \beta^{q,N_{\rm bins}}_\ell (\chi)\\
    \beta^{v,N_{\rm bins}}_\ell (\chi)
\end{bmatrix}
\end{equation}
where the $C_{\alpha \beta \ell}^{XY}$ are the possible auto and cross power spectra of the binned dipole and quadrupole fields and
\begin{align}
\beta^{X,\alpha}_\ell (\chi) = \langle \psi_{\ell m} (\chi) {a^{X \alpha}_{\ell m}}^* \rangle = \frac{2}{\pi} \int k^2 dk P(k) \Delta^{X,\alpha}_\ell (k) j_{\ell}(k \chi)
\end{align}
where $\Delta^{X,\alpha}_\ell (k)$ are the transfer functions for $X=q,\bar{v}$ in redshift bin $\alpha$ and $P(k)$ is the primordial scalar power spectrum. Applying these equations, one can in principle assemble a 3d map of the primordial potential on large scales. One can incorporate the measurements of the low multipole temperature and polarization moments of the primary CMB with the information from the dipole and quadrupole fields to improve the reconstruction. We explore this reconstruction problem further in future work.

\section{Conclusions}
\label{sec:conclusions}

In this paper we presented a method to extract large-scale information about the universe from the kinetic and polarized Sunyaev Zel'dovich effects in a unified way. We constructed optimal quadratic estimators for the remote dipole and quadrupole fields using the cross correlation of CMB and large scale structure. For the calculation of the signal we used the results from our previous publications~\cite{Terrana2016, Deutsch:2017cja}, except for the tensor contribution to the quadrupole field which we added in appendix~\ref{sec:tensorcontribution}. This gives a complete description of these large-scale observables which we will explore further for cosmological applications in the future. In the case of the kSZ effect, we found that the dipole field can be reconstructed mode by mode over a variety of scales with high fidelity using next generation experiments such as CMB S4 and LSST. 

In the case of the remote quadrupole field an equivalent experiment in polarization can reconstruct the remote quadrupole field from scalar perturbations mode by mode on the largest scales with a signal to noise of order 1 to 10. Although the pSZ signal is far smaller, this is possible because the noise from the primary and lensing $B$-mode CMB polarization is correspondingly small.  For the remote quadrupole field from tensors we found that the same setup is not enough for a mode by mode reconstruction (but might be enough for a first detection), even with the optimistic assumption of $r=0.1$. However a ten times reduced CMB noise level would provide a mode by mode reconstruction of the remote B-mode quadrupole field from tensors for $\ell \lesssim 4$. 

We have also presented some of the first applications of a measurement of the remote dipole and quadrupole fields. In particular, it should be possible to measure the local CMB dipole sourced by linear scales (what we refer to as the "primordial dipole") and to test the statistical homogeneity of the Universe with the first modes detected at the lowest redshifts reconstructed using kSZ and pSZ tomography respectively. A high fidelity reconstruction of the dipole and quadrupole fields also opens the possibility of reconstructing a 3-D map of the primordial potential and the tensor field, and we presented a generalization of the reconstruction methods of Ref.~\cite{Yadav:2005tf} that would be well-suited to this task. 
 
The results of this paper illustrate the exciting opportunity to learn more about the largest observable scales from the statistics of small-scale fluctuations. The next generation of CMB experiments is well-poised to take advantage of the this opportunity, motivating a more realistic suite of forecasts tailored to these efforts, and further exploration of the range of early-Universe and large-scale physics that can be probed with new  observables. We hope that our results can contribute to a new sense of optimism for the prospects of learning a great deal more about the early Universe in the near future.

\acknowledgments
We thank Gil Holder, P. Daniel Meerburg, and Alexander Van Engelen for important conversations. MCJ is supported by the National Science and Engineering Research Council through a Discovery grant. AT acknowledges support from the Vanier Canada Graduate Scholarships program. AD is supported by NSF Award PHY-1417385. AD thanks the Perimeter Institute for Theoretical Physics for its hospitality. This research was supported in part by Perimeter Institute for Theoretical Physics. Research at Perimeter Institute is supported by the Government of Canada through the Department of Innovation, Science and Economic Development Canada and by the Province of Ontario through the Ministry of Research, Innovation and Science. ED acknowledges support by DOE grant DE-SC0009946. MCJ and ED thank NORDITA for their hospitality during the completion of this work.

\appendix

\section{Transfer function for the dipole field}\label{sec:dipoletransfer}

In this appendix, we summarize the contributions to the transfer function for the dipole field $\Delta_{\ell}^{v}(k,\chi)$ appearing in Eq.~\ref{eq:almv}. From Ref.~\cite{Terrana2016}, we have:
\be\label{eq:transfer}
\Delta^{v}_{\ell}(k, \chi) \equiv \frac{4\pi \ i^{\ell}}{2 \ell + 1} \mathcal{K}^v(k,\chi) \left[ \ell\ j_{\ell - 1} (k \chi) - (\ell + 1)  j_{\ell + 1} (k \chi) \right] T(k).
\ee
Here, we encorporate the transfer function $T(k)$ for the Newtonian potential (e.g. evaluated using the BBKS fitting function~\cite{Bardeen1986}), and the dipole field receives contributions from the Sachs Wolfe (SW), Integrated Sachs Wolfe (ISW), and Doppler effects through the kernel $\mathcal{K}^v(k,z\chi)$:
\begin{equation}
\mathcal{K}^v(k,\chi) \equiv \left[ \mathcal{K}_{\rm D} (k, \chi) + \mathcal{K}_{\rm SW} (k, \chi) + \mathcal{K}_{\rm ISW} (k, \chi)\right].
\end{equation} 
where
\ba\label{eq:kernel}
\mathcal{K}_{\rm D} (k, \chi) &\equiv& k D_v (\chi_\text{dec}) j_{0} (k \Delta \chi_\text{dec} ) - 2 k D_v (\chi_\text{dec}) j_{2} (k \Delta \chi_\text{dec} ) - k D_v (\chi),  \\ \label{eq:kernel2}
\mathcal{K}_{\rm SW} (k, \chi) &\equiv& 3\left( 2D_\Psi(\chi_\text{dec}) -\frac{3}{2} \right) j_{1} (k \Delta \chi_\text{dec}), \\ \label{eq:kernel3}
\mathcal{K}_{\rm ISW} (k, \chi) &\equiv& 6  \int_{a_{\rm dec}}^{a_e} da \frac{dD_\Psi}{da}  \ j_{1} (k \Delta \chi (a)) .
\ea
Here, $\chi_\text{dec}$ is the total distance to decoupling and $\Delta \chi_\text{dec} = \chi_\text{dec} - \chi$ is the distance from $\chi$ to decoupling. $D_v (\chi)$ is the velocity growth function, and $D_\Psi (\chi)$ is the potential growth function for long-wavelength modes, which are defined as:
\begin{eqnarray}
\Psi({\bf x},t) =  D_\Psi (t) \Psi_i ({\bf x}), \ \ \ \ {\bf v} = - D_v (t) {\bf \nabla} \Psi_i ({\bf x}).
\end{eqnarray}
Approximate forms for these growth functions can be found in Refs.~\cite{Zhang:2015uta,Terrana2016}. The leading order behavior of $\mathcal{K}^v(k,\chi)$ in the limit where $k \rightarrow 0$ is cubic in $k$, not linear as would be expected from the individual kernels Eq.~\ref{eq:kernel}-\ref{eq:kernel3}. This is a consequence of diffeomorphism invariance, as discussed in Ref.~\cite{Terrana2016}.

The dominant contribution to the remote dipole field on all but the very largest scales is the local doppler contribution in Eq.~\eqref{eq:kernel}. In this limit, the remote dipole field reduces to the line-of-sight peculiar velocity field. To illustrate this, we have calculated the remote dipole power spectrum including all the terms above, as well as only using only the peculiar velocity term (the third term in Eq.~\eqref{eq:kernel}). The ratio of these two power spectra is shown in Fig.~\ref{eq:kernel2}. We see that the peculiar velocity term constitutes the majority of the signal, with the additional components contributing at the $\sim10\%$ level on the largest scales. 

\begin{figure}[t!]
\resizebox{0.5\hsize}{!}{	
  \includegraphics{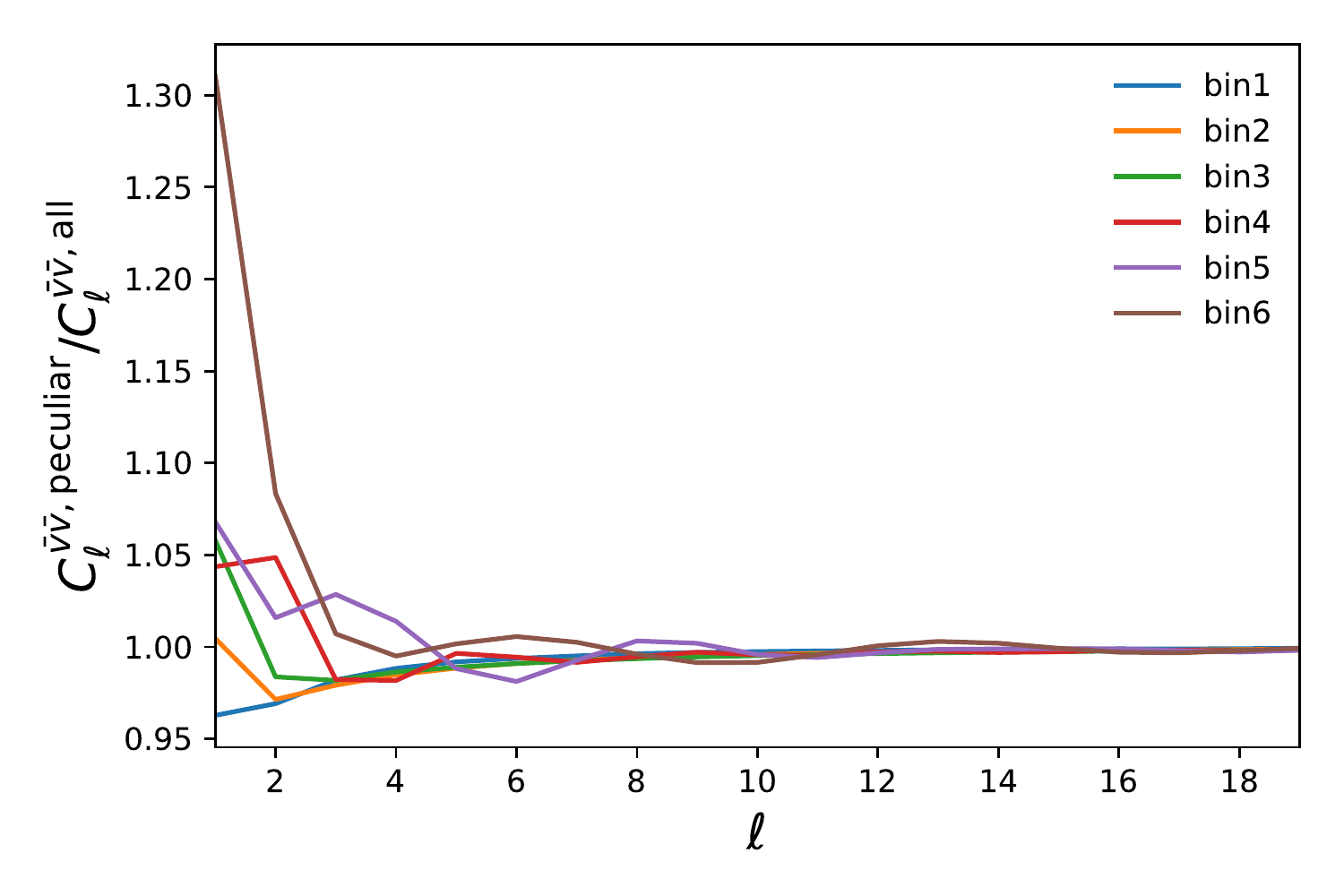}
}
	\caption{Ratio of the binned power spectrum $C_l^{\bar{v}\bar{v}}$ of the dipole field computed using only the peculiar velocity term (the third term in Eq.~\eqref{eq:kernel}) to the full remote dipole power spectrum. }
	\label{dipole_velocities}
\end{figure}

\section{Red shift correlations and bin averaging}
\label{sec:redshiftcorr}

\begin{figure}[t!]
	\includegraphics[width=12cm]{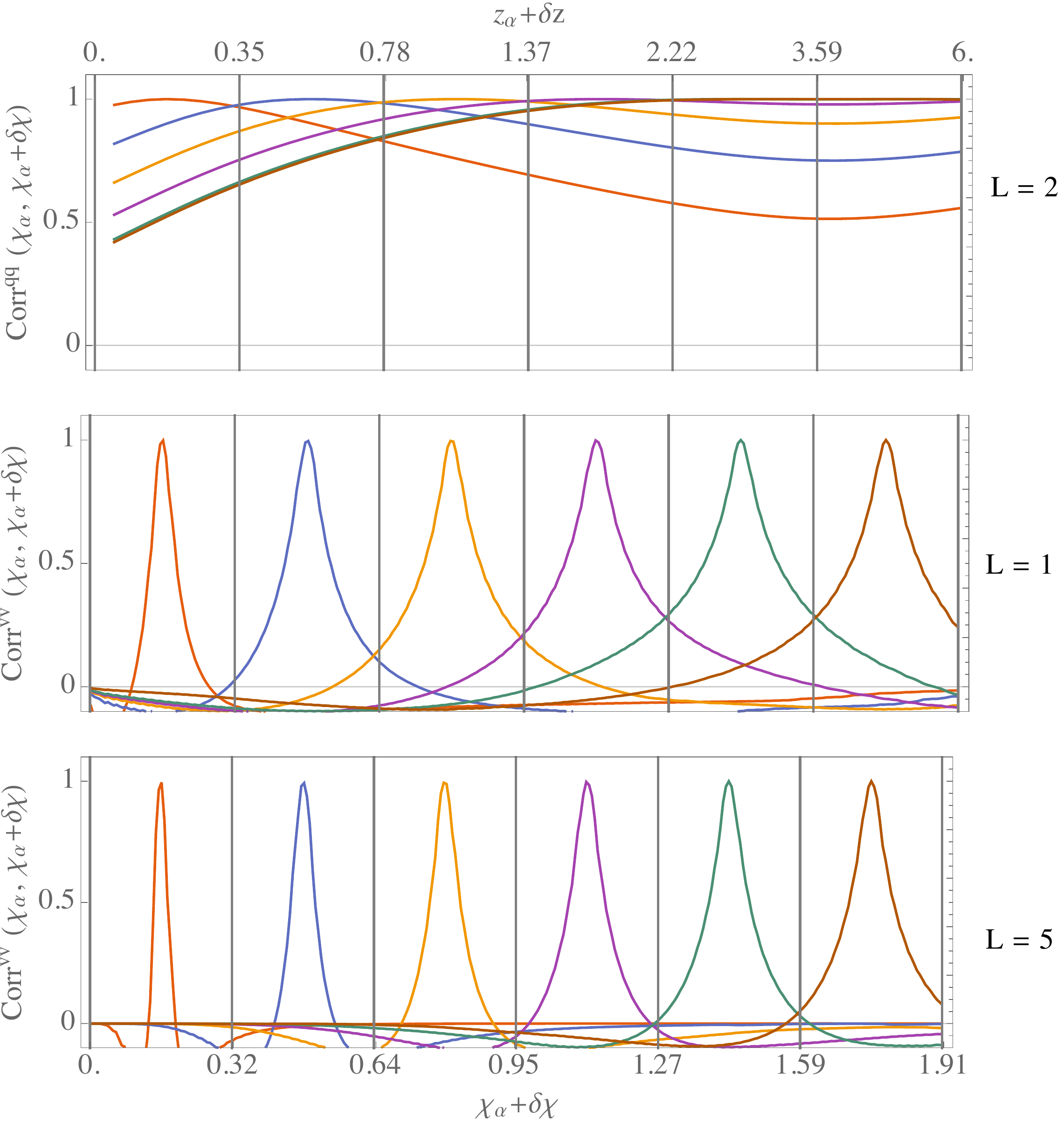}
	\caption{The correlation function of the dipole and quadrupole field is plotted at $(\chi_\alpha,\ {\chi}_\alpha + \delta\chi)$, where ${\chi_\alpha}$ is fixed to be the midpoint of a redshift bin. The redshift values are given across the top axis, and the corresponding comoving distance values along the bottom axis, consistent with the binning in table~\ref{tab:bins}. Specifically, the different curves from left to right fix ${\chi_\alpha} =$ 0.16, 0.48, 0.79, 1.11, 1.43, 1.75, in units of $H_0^{-1}$. Top: Correlation of the quadrupole field $q_{2M}(\chi)$, which has a very large correlation length. Middle: Correlation of the dipole field $v_{1M}(\chi)$. The correlations are smaller than in the quadrupole case because the effective velocities are dominated by the Doppler effect. Bottom: Correlation of the dipole field $v_{5M}(\chi)$. The correlation length is much smaller than the bin size, in particular for low redshift bins.}
	\label{fig:corrfunc}
\end{figure}

Our quadratic estimators are sensitive to the bin-averaged multipole moments of the dipole field and quadrupole field. To understand the physical meaning of these quantities and their power spectra, it is helpful to examine the redshift correlation function of the un-averaged fields; several examples are show in Fig.~\ref{fig:corrfunc}. The top plot shows the un-averaged correlation function of the $L=2$ mode of the remote quadrupole field from scalar perturbations. It is clear from this plot that the quadrupole field at $L=2$ within a redshift bin (here we used 6 bins) is essentially constant. This property holds over a variety of multipoles, as discussed in Ref.~\cite{Deutsch:2017cja}. Therefore the bin-average quadrupole field for all practical purposes can be identified with the un-averaged field. 

\begin{figure}[t!]
\resizebox{1.0\hsize}{!}{	
  \includegraphics{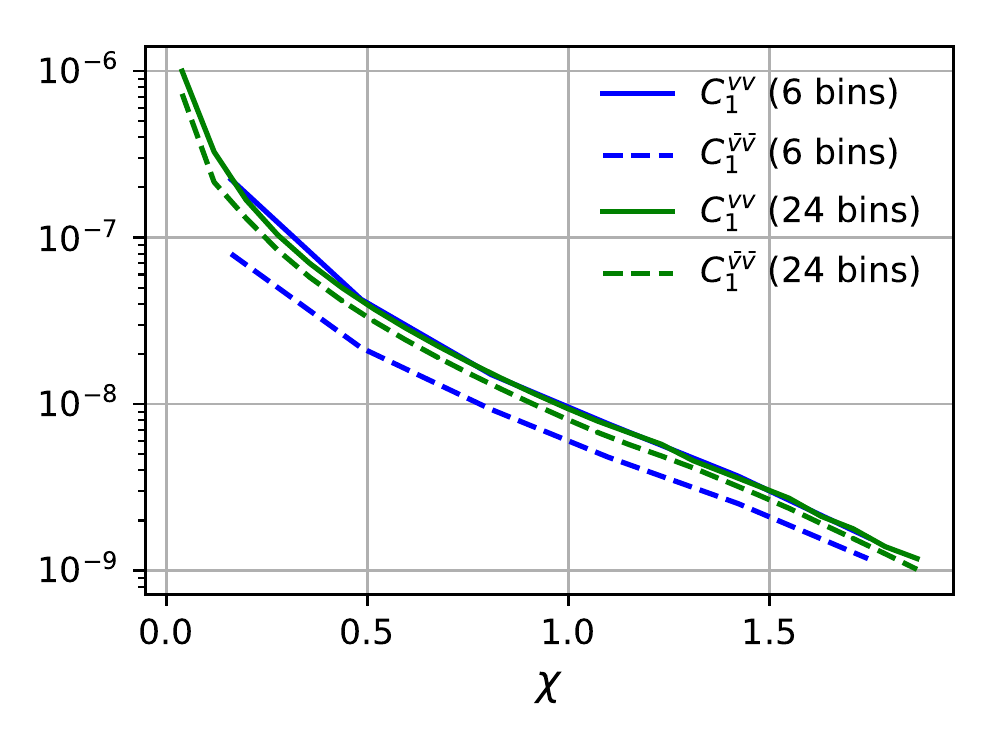}
  \includegraphics{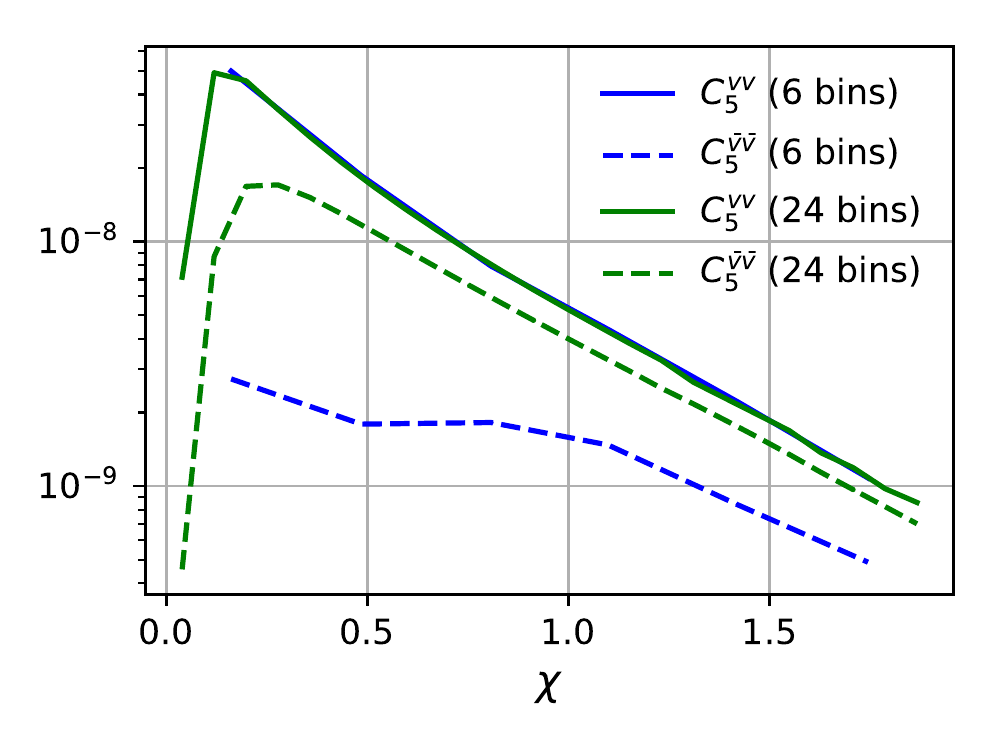}
}

	\caption{Power spectrum $C_l^{vv}$ of the dipole field as a function of red shift $\chi$. We compare $C_l^{vv}$, the power evaluated in the bin center, with $C_l^{\bar{v}\bar{v}}$, the power of the average field. At $L=1$ the average field is comparable in size to the unaveraged field, while for $L=5$ the cancellations are important. The effect of cancelations is expectedly larger for larger bins.}
	\label{fig:compvvbar}
\end{figure}

The second plot shows the un-averaged $L=1$ moment of the dipole field. Here the correlation length is much smaller, due to the Doppler contribution from small-scale velocities. If the correlation length is smaller than the bin size then, due to cancellations within the bin, the averaged field will have substantially smaller power than the un-averaged field. The bottom plot shows that for $L=5$, in particular for the low red shift bins, this condition is strongly violated, and we expect a much smaller signal in the mean field. We quantify this effect on the signal in Fig.~\ref{fig:compvvbar} by comparing the un-averaged and averaged power spectra. Here, it can be seen that at $L=1$, the power in the averaged and un-averaged dipole field is comparable for both a 6-bin and 24-bin configuration, while for $L=5$ there is a substantial difference, especially for the $6$-bin case. This effect was also seen in the simulations performed in Ref.~\cite{Terrana2016}. These results illustrate the importance of distinguishing the bin-averaged and un-averaged power when characterizing the dipole field signal.

\section{Scalar contributions to the remote quadrupole}\label{sec:scalarcontribution}

In this appendix, we review the contributions to the quadrupole field transfer function $\Delta_\ell^q(k,\chi)$ appearing in Eq.~\ref{eq:almqfinal}. From Ref.~\cite{Deutsch:2017cja}, we have
\begin{equation}\label{eq:Deltaq}
	\Delta_\ell^q(k,\chi) = -5 i^\ell \sqrt{\frac{3}{8}}\sqrt{\frac{(\ell+2)!}{(\ell-2)!}} \frac{j_\ell(k\chi)}{(k\chi)^2}\ T(k)\ \left[ \mathcal{G}_{\rm SW}(k,\chi) + \mathcal{G}_{\rm ISW}(k,\chi) + \mathcal{G}_{\rm Doppler }(k,\chi)\right], 
\end{equation}
Note that this expression is zero for $\ell =0$ and $\ell =1$. The various doppler kernels are given by
\begin{align} 
	\mathcal{G}_{\rm SW}(k,\chi) = & -4\pi \left( 2D_\Psi(\chi_{\rm dec}) -\frac{3}{2} \right) j_{2} (k \Delta \chi_\text{dec}), \no \\
	\mathcal{G}_{\rm ISW}(k,\chi) = & -8\pi \int_{a_{\rm dec}}^{a_e} da \frac{dD_\Psi}{da}  \ j_{2} (k \Delta \chi (a)), \no \\
	\mathcal{G}_{\rm Doppler }(k,\chi) = & \frac{4\pi}{5} k D_v(\chi_\text{dec}) \left[ 3j_3(k\Delta\chi_\text{dec}) - 2j_1(k\Delta\chi_\text{dec}) \right]. \label{eq:qkernel}
\end{align}
where the growth functions $D_\Psi$ and $D_v$ are as defined in Appendix~\ref{sec:dipoletransfer}.

\section{Tensor contributions to the remote quadrupole}\label{sec:tensorcontribution}

We present here the detailed derivation of the primordial tensor modes contributions to the CMB temperature quadrupole, used in Sec.~\ref{sec:quadsignal}. The CMB quadrupole as seen by an electron along our past light cone in the ${\bf \widehat{n}}_{e}$ direction at comoving distance $\chi$ is given by 
\begin{equation}
q^{m}_{\rm eff} ({\bf \widehat{n}}_e, \chi) = \int d^2 {\bf \widehat{n}} \ \left[ \Theta^S (\chi, {\bf \widehat{n}}_e, {\bf \widehat{n}}) + \Theta^T (\chi, {\bf \widehat{n}}_e, {\bf \widehat{n}}) \right] \ Y^*_{2 m} ({\bf \widehat{n}}).
\end{equation}
Let us focus on the anisotropy due to tensors, $\Theta^T (\chi, {\bf \widehat{n}}_e, {\bf \widehat{n}})$, and define
\begin{equation}
q^{m,T}_{\rm eff} ({\bf \widehat{n}}_e, \chi) = \int d^2 {\bf \widehat{n}} \  \Theta^T (\chi, {\bf \widehat{n}}_e, {\bf \widehat{n}}) \ Y^*_{2 m} ({\bf \widehat{n}}).
\label{eq:effective-quadrupole-def}
\end{equation}
It will be helpful to go to Fourier space,
\begin{equation}
	\Theta^T (\chi, {\bf \widehat{n}}_e, {\bf \widehat{n}}) = \int \frac{d^3k}{(2\pi)^3} \Theta^T(k,\widehat{\bf n}) e^{i\chi \widehat{\bf n}_e \cdot \widehat{\bf k}}, \label{eq:Theta_fourier}
\end{equation}
and then decompose into $+$ and $\times$ components 
\begin{equation}
	\Theta^T(k,\mu,\phi) = \Theta_+^T(k,\mu)(1-\mu^2)\cos{(2\phi)} +  \Theta_\times^T(k,\mu)(1-\mu^2)\sin{(2\phi)}, \label{eq:plus_cross_decomp}
\end{equation}
where $\mu \equiv {\bf \widehat{k} \cdot \widehat{n}}$. These two components can further be expanded into multipoles using
\begin{equation}
	\Theta_{(+,\times)}^T(k,\mu) = \sum_\ell (-i)^\ell (2\ell +1) \mathcal{P}_\ell(\mu)\Theta_{\ell,(+,\times)}^T(k),
\end{equation}
where $\mathcal{P}_\ell$ are the Legendre polynomials. The latter can be expressed in terms of spherical harmonics as
\begin{equation}
 	\mathcal{P}_\ell ({\bf \widehat{n}}\cdot{\bf \widehat{k}})=\frac{4\pi}{2\ell +1} \sum_{m=-\ell }^\ell Y^*_{\ell m}({\bf \widehat{n}})\ Y_{\ell m}({\bf \widehat{k}}).
 \end{equation}
For ${\bf \widehat{k}} = {\bf \widehat{z}}$, $\mu$ is simply equal to $\cos{\theta}$ with $(\theta,\phi)$ being the angles associated with the ${\bf \widehat{n}}$ vector. In this case, the decomposition~\eqref{eq:plus_cross_decomp} can be easily rewritten in terms of spherical harmonics using the following relations
\begin{align}
	(1-\mu^2)\cos{(2\phi)} = & \sin^2\theta\cos{(2\phi)} = 2 \sqrt{\frac{2\pi}{15}}[Y_{2,2}({\bf \widehat{n}}) + Y_{2,-2}({\bf \widehat{n}})], \label{eq:muphi1} \\
	(1-\mu^2)\sin{(2\phi)} = & \sin^2\theta\sin{(2\phi)} = \frac{2}{i} \sqrt{\frac{2\pi}{15}}[Y_{2,2}({\bf \widehat{n}}) - Y_{2,-2}({\bf \widehat{n}})].\label{eq:muphi2}
\end{align}
For a general wavenumber $k$, this result can be actively rotated using the Wigner rotation operator $\widehat{D}$. The rotation of spherical harmonics satisfies $\widehat{D}Y_{\ell m} = \sum_{m''}D^{\ell}_{m''m}Y_{\ell m''}$, where the Wigner rotation matrix is related to the spin-weighted spherical harmonics according to 
\begin{equation}
	D^\ell_{m s}(\phi_k,\theta_k,0) = (-1)^m\sqrt{\frac{4\pi}{2\ell+1}} {}_s Y_{\ell, -m}(\widehat{\bf k}). \label{eq:rotationD}
\end{equation}
The expressions in Eqs.~\eqref{eq:muphi1}-\eqref{eq:muphi2} generalize to
\begin{align}
	(1-\mu^2)\cos{(2\phi)} = & \frac{4\pi}{5}\sqrt{\frac{2}{3}} \sum_{m''} (-1)^{m''} [{}_2Y_{2,-m''}({\bf \widehat{k}}) + {}_{-2}Y_{2,-m''}({\bf \widehat{k}})] Y_{2m''}({\widehat{\bf n}}), \label{eq:muphi3} \\
	(1-\mu^2)\sin{(2\phi)} = & \frac{4\pi}{5i}\sqrt{\frac{2}{3}} \sum_{m''}  (-1)^{m''}[{}_2Y_{2,-m''}({\bf \widehat{k}}) - {}_{-2}Y_{2,-m''}({\bf \widehat{k}})] Y_{2m''}({\widehat{\bf n}}).\label{eq:muphi4}
\end{align}
This allows for $\Theta^T(k,\widehat{\bf n})$ to be expressed entirely in terms of spherical harmonics:
\begin{align}
	\Theta^T(k,\widehat{\bf n}) = & \frac{(4\pi)^2}{5} \sqrt{\frac{2}{3}} \sum_{\ell',m',m''} (-i)^{\ell'} (-1)^{m''} \ Y^*_{\ell' m'}({\bf \widehat{k}})\ Y_{\ell' m'}({\bf \widehat{n}})  \no \\
	 & \times \left[ \left({}_2Y_{2,-m''}({\bf \widehat{k}}) + {}_{-2}Y_{2,-m''}({\bf \widehat{k}})\right) Y_{2m''}({\widehat{\bf n}}) \ \Theta_{\ell',+}^T(k) \right. \no \\
	 & - \left. i \left( {}_2Y_{2,-m''}({\bf \widehat{k}}) - {}_{-2}Y_{2,-m''}({\bf \widehat{k}}) \right) Y_{2m''}({\widehat{\bf n}})\ \Theta_{\ell',\times}^T(k) \right]. \label{eq:thetaT_decomp_Y}
\end{align}
Our interest is determining the contribution of $\Theta^T_{(+,\times)}(k,\widehat{\bf n})$ to the effective quadrupole $q^{m, T}_{\rm eff} ({\bf \widehat{n}}_e, \chi)$. One can verify that when Eq.~\eqref{eq:thetaT_decomp_Y} is replaced into Eq.~\eqref{eq:Theta_fourier} and then into Eq.~\eqref{eq:effective-quadrupole-def}, there will be a product of 5 spherical harmonics: three with argument $\widehat{\bf n}$ and two with argument $\widehat{\bf{k}}$. The integral over $\widehat{\bf n}$ is performed using the triple product identity for the Wigner 3-$j$ symbols. This gives
\begin{equation}
\int d^2 {\bf \widehat{n}} \ Y_{\ell' m'} ({\bf \widehat{n}})Y_{2 m''} ({\bf \widehat{n}}) Y^*_{2 m} ({\bf \widehat{n}}) = (-1)^m\sqrt{ \frac{(2\ell' +1) (5)(5)}{4\pi} } \left(\begin{array}{ccc} \ell' & 2 & 2 \\ m' & m'' & -m \end{array} \right) \left(\begin{array}{ccc}  \ell' & 2 & 2 \\ 0 & 0 & 0 \end{array} \right).
\label{eq:YYYn}
\end{equation}
Note that the selection rules for the 3-$j$ symbols above imply that we need $\ell' = 0,2,4$ and $m' + m'' -m = 0$. The products of spherical harmonics with argument $\widehat{\bf{k}}$ can be simplified using the identity,
\begin{equation}
\begin{split}
	{}_{s_1}Y_{\ell_1 m_1} ({\bf \widehat{n}}) \ {}_{s_{2}} Y_{\ell_{2} m_{2}}({\bf \widehat{n}}) = \sum_{S,L,M} & (-1)^{\ell_1+\ell_{2}+L} \sqrt{ \frac{(2\ell_1+1) (2\ell_{2}+1)(2L+1)}{4\pi} } \\
	& \quad \times \left(\begin{array}{ccc} \ell_1 & \ell_{2} & L \\ m_1 & m_{2} & M \end{array} \right) \left(\begin{array}{ccc} \ell_1 & \ell_2 & L \\ s_1 & s_{2} & S \end{array} \right) {}_{S}Y_{LM}^{*}({\bf \widehat{n}}),
	\end{split} \label{eq:identity-two-spherical-harmonics}
\end{equation}
which when applied on the $+$ component results in, 
\begin{equation}
\begin{split}
	Y^*_{\ell' m'} ({\bf \widehat{k}}) \left[ {}_{2} Y_{2, -m''}({\bf \widehat{k}}) +  {}_{-2} Y_{2, -m''}({\bf \widehat{k}}) \right] = (-1)^{m'} \sum_{S,L,M} & (-1)^{\ell'+L} \sqrt{ \frac{(2\ell'+1) (5)(2L+1)}{4\pi} } \left(\begin{array}{ccc}  \ell' & 2 & L \\ -m' & -m'' & M \end{array} \right) \\
	& \quad \times \left[ \left(\begin{array}{ccc} \ell' & 2 & L \\ 0 & 2 & S \end{array} \right) + \left(\begin{array}{ccc} \ell' & 2 & L \\ 0 & -2 & S \end{array} \right) \right] {}_{S}Y_{LM}^{*}({\bf \widehat{k}}).
	\end{split} \label{eq:YYk}
\end{equation} 
Here the selection rules imply $S=\pm 2$. The sum over the first Wigner 3-$j$ symbol in Eq.~\eqref{eq:YYYn} and the first Wigner 3-$j$ symbol in Eq.~\eqref{eq:YYk} can be simplified using orthogonality relations, yielding
\begin{align}
	\sum_{m',m''} \left(\begin{array}{ccc} \ell' & 2 & 2 \\ m' & m'' & -m \end{array} \right) \left(\begin{array}{ccc} \ell' & 2 & L \\ -m' & -m'' & M \end{array} \right) = & (-1)^{\ell'} \frac{\delta_{L2}\delta_{Mm}}{5}. \label{eq:sumoverm''m'}
\end{align}
\\
Eqs.~\eqref{eq:YYYn}, \eqref{eq:YYk}, \eqref{eq:sumoverm''m'} lead to the following expression for the $+$ component of the tensor contribution to the effective quadrupole 
\begin{align}
q^{m, T}_{{\rm eff},+} ({\bf \widehat{n}}_e, \chi) = & \int \frac{d^3k}{(2\pi)^3}\ e^{i\chi \widehat{\bf n}_e \cdot \widehat{\bf k}}  \frac{(4\pi)^2}{5} \sqrt{\frac{2}{3}} \sum_{\ell'=0,2,4} (-i)^{\ell'} \frac{(2\ell' +1) (5)}{4\pi}\ \Theta_{\ell',+}^T(k)  \no \\ 
& \times \left(\begin{array}{ccc}  \ell' & 2 & 2 \\ 0 & 0 & 0 \end{array} \right)\left(\begin{array}{ccc} \ell' & 2 & L \\ 0 & 2 & -2 \end{array} \right)  \left[ {}_2 Y^*_{2m}(\widehat{\bf k}) + {}_{-2} Y^*_{2m}(\widehat{\bf k}) \right] \no \\
= &  \int \frac{d^3k}{(2\pi)^3}\ e^{i\chi \widehat{\bf n}_e \cdot \widehat{\bf k}} \ 4\pi \sqrt{6} \left[  \frac{1}{15} \Theta_{0,+}^T(k) + \frac{2}{21} \Theta_{2,+}^T(k) + \frac{1}{35} \Theta_{4,+}^T(k)\right]   \left[ {}_2 Y^*_{2m}(\widehat{\bf k}) + {}_{-2} Y^*_{2m}(\widehat{\bf k}) \right],\label{eq:qeff_Theta_024}
\end{align}
where we evaluated the non-zero 3-$j$ symbols at $\ell'=0,2,4$. The expression for the $\times$ component is obtained with the replacements $\Theta_{\ell,+}^T(k) \rightarrow \Theta_{\ell,\times}^T(k)$ and $  \left[ {}_2 Y^*_{2m}(\widehat{\bf k}) + {}_{-2} Y^*_{2m}(\widehat{\bf k}) \right] \rightarrow i  \left[ {}_2 Y^*_{2m}(\widehat{\bf k}) - {}_{-2} Y^*_{2m}(\widehat{\bf k}) \right]$.  \\
The next step is to express the $\ell$th moment due to tensor perturbations as 
\begin{equation}
	\Theta_{\ell,(+,\times)}^T(k) = -\frac{1}{2} \int_{a_\text{dec}}^{a_e} da \frac{d h_{(+,\times)}}{da} j_\ell (k\Delta\chi(a)), \label{eq:Theta_ell_integral}
\end{equation}
where $\Delta\chi(a) = -\int_{a_e}^{a} da' [H(a')a'^2]^{-1}$ and the  metric perturbations, $h_{(+,\times)}$, are defined as
\begin{equation}
	g_{ij} = a^2 \left(\begin{array}{ccc} 1+h_+ & h_\times & 0 \\ h_\times & 1-h_+ & 0 \\ 0 & 0 & 1 \end{array}\right).
\end{equation}
When the result in Eq.~\eqref{eq:Theta_ell_integral} is replaced into the quadrupole \eqref{eq:qeff_Theta_024}, the recursion relations for the spherical Bessel functions, $(2\ell+1)j_\ell(x) = xj_{\ell-1}(x) + xj_{\ell+1}(x)$, allow for a simplification. If applied twice, the recursion relations imply,
\begin{equation}
	\frac{j_{\ell+2}(x)}{(2\ell+1)(2\ell+3)} + 2 \frac{j_{\ell}(x)}{(2\ell+3)(2\ell-1)} + \frac{j_{\ell-2}(x)}{(2\ell+1)(2\ell-1)} = \frac{j_{\ell}(x)}{x^2},
\end{equation}
which yields a simple expression for the quadrupole,
\begin{align}
	q^{m, T}_{{\rm eff}} ({\bf \widehat{n}}_e, \chi) =\, & q^{m, T}_{{\rm eff},+} ({\bf \widehat{n}}_e, \chi) + q^{m, T}_{{\rm eff},\times} ({\bf \widehat{n}}_e, \chi) \no \\
	= &\int \frac{d^3k}{(2\pi)^3}\ e^{i\chi \widehat{\bf n}_e \cdot \widehat{\bf k}} \ \left\{ \mathcal{G}^q_{T,+}(k,\chi)\ \left[ {}_2 Y^*_{2m}(\widehat{\bf k}) + {}_{-2} Y^*_{2m}(\widehat{\bf k}) \right] \right. \no \\
	& + \left. i \mathcal{G}^q_{T,\times}(k,\chi)\ \left[ {}_2 Y^*_{2m}(\widehat{\bf k}) - {}_{-2} Y^*_{2m}(\widehat{\bf k}) \right] \right\}, \label{eq:qeff_tensors_+}
\end{align}
with the kernel defined as
\begin{equation}\label{eq:kernel_tensors}
	 \mathcal{G}^q_{T,(+,\times)}(k,\chi) \equiv 2\pi\sqrt{6}  \int_{a_e}^{a_\text{dec}} da \frac{d h_{(+,\times)}}{da} \ \frac{j_2 (k\Delta\chi(a))}{[k\Delta\chi(a)]^2}.
\end{equation}
It will be convenient to define a total effective quadrupole as the sum of the projections on the spin-weighted basis:
\begin{equation}\label{eq:qeffeff}
	\tilde{q}_\text{eff}^{\pm,T}({\bf \widehat{n}}_e,\chi) \equiv \sum_{m=-2}^2 q^{m, T}_{\rm eff} ({\bf \widehat{n}}_e, \chi) \left._{\pm 2}Y_{2 m}\right. ({\bf \widehat{n}}_e) \ .
\end{equation}
We also will make use of the multipolar expansion of this quantity
\begin{equation}\label{eq:qsumalm_new}
	\tilde{q}_\text{eff}^{\pm,T}({\bf \widehat{n}}_e,\chi) =\sum_{\ell m} \left( a_{\ell m}^{q,E}(\chi) \pm i a_{\ell m}^{q,B}(\chi) \right) \left._{\pm 2}Y_{\ell m}\right. ({\bf \widehat{n}}_e) \ .
\end{equation}
This relationship can be inverted to solve for the multipole coefficients,
\begin{equation}\label{eq:almqdef_new}
	a_{\ell m}^{q,E}(\chi) \pm i a_{\ell m}^{q,B}(\chi)  = \int  d^2{\bf \widehat{n}}_e\ \tilde{q}_\text{eff}^{\pm,T}({\bf \widehat{n}}_e,\chi) \ {}_{\pm 2}Y_{\ell m}^* ({\bf \widehat{n}}_e) \ .
\end{equation}
Using the result from Eqs.~(\ref{eq:qeff_tensors_+}) and (\ref{eq:qeffeff}), Eq.~(\ref{eq:almqdef_new}) becomes 
\begin{align}\label{altra}
	a_{\ell m}^{q,E}(\chi) \pm i a_{\ell m}^{q,B}(\chi) = & \int d^2 {\bf \widehat{n}}_e \sum_{m'=-2}^2 \int \frac{d^3 k}{(2\pi)^3}  e^{i \chi {\bf k} \cdot {\bf \widehat{n}}_e} \left\{ \mathcal{G}^q_{T,+}(k,\chi) \left[ {}_2 Y^*_{2m'}(\widehat{\bf k}) + {}_{-2} Y^*_{2m'}(\widehat{\bf k}) \right] \right. \no \\
	& +i \left. \mathcal{G}^q_{T,\times}(k,\chi) \left[ {}_2 Y^*_{2m'}(\widehat{\bf k}) - {}_{-2} Y^*_{2m'}(\widehat{\bf k}) \right]   \right\}\left._{\pm 2}Y_{2 m'}\right. ({\bf \widehat{n}}_e) \left._{\pm 2}Y^*_{\ell m}\right. ({\bf \widehat{n}}_e) .
\end{align}
Expanding the exponential with the identity,
\begin{equation} \label{eq:expidentity}
	e^{i \chi {\bf k} \cdot {\bf \widehat{n}_{e}} } = \sum_{L,M} 4\pi \ i^{L} \ j_{L} (k \chi) \ Y^*_{LM}({\bf \widehat{k}})  Y_{LM}({\bf \widehat{n}_{e}} ),
\end{equation}
there will be five spherical harmonics: three with argument ${\bf \widehat{n}}_e$  and two with argument ${\bf \widehat{k}}$. The former yield
\begin{equation} 
\begin{split}
\int d^2 {\bf \widehat{n}_{e}} \ \left._{\pm 2}Y_{2 m'}\right. ({\bf \widehat{n}}_e) \left._{\pm 2}Y^*_{\ell m}\right. ({\bf \widehat{n}}_e) Y_{LM}({\bf \widehat{n}}_e) & = (-1)^m\sqrt{ \frac{(5)(2 \ell+1)(2L+1)}{4\pi} } \\
& \quad \times \left(\begin{array}{ccc} \ell & 2 & L \\ -m & m' & M \end{array} \right) \left(\begin{array}{ccc} \ell & 2 & L \\ \pm2 & \mp2 & 0 \end{array} \right).
\end{split}\label{eq:YYY}
\end{equation}
For the remaining two spherical harmonics with argument $\widehat{\bf k}$, we can use the identity in Eq.~\eqref{eq:identity-two-spherical-harmonics} to express them as just one spherical harmonic:
\begin{align}
	\left[ {}_2 Y^*_{2m'}(\widehat{\bf k}) \pm {}_{-2} Y^*_{2m'}(\widehat{\bf k}) \right]  Y_{LM}^{*}({\bf \widehat{k}}) = & (-1)^{M+m^{\prime}} \sum_{L^{\prime}, M^{\prime}} \sqrt{ \frac{5(2L+1)(2L^{\prime}+1)}{4\pi} } \left[ {}_{2}Y^{*}_{L^{\prime}M^{\prime}}({\bf \widehat{k}}) \pm (-1)^{L+L^{\prime}} {}_{-2}Y^{*}_{L^{\prime}M^{\prime}}({\bf \widehat{k}}) \right]	\no \\
	& \quad \times \left( \begin{array}{ccc} L & 2 & L^{\prime} \\ -M & -m^{\prime} & M^{\prime} \end{array} \right) \left( \begin{array}{ccc} L & 2 & L^{\prime} \\ 0 & 2 & -2 \end{array} \right).
\label{eq:YY}
\end{align}
When the results of equations \eqref{eq:YYY} and \eqref{eq:YY} are combined in (\ref{altra}), the four Wigner 3-$j$ symbols can be simplified as follows:
\begin{align}
	\sum_{M,m^{\prime}} (-1)^{M+m^{\prime}+m} \left( \begin{array}{ccc} \ell & 2 & L \\ -m & m^{\prime} & M \end{array} \right) \left( \begin{array}{ccc} L & 2 & L^{\prime} \\ -M & -m^{\prime} & M^{\prime} \end{array} \right) & = \frac{\delta_{\ell L^{\prime}} \delta_{m M^{\prime}}}{2 \ell +1},
\end{align}
where we used the selection rule of the 3-$j$ symbols $M+m^{\prime}-m=0$, and
\begin{equation}
	\left(\begin{array}{ccc} \ell & 2 & L \\ \pm2 & \mp2 & 0 \end{array} \right)  \left( \begin{array}{ccc} L & 2 & \ell\\ 0 & 2 & -2 \end{array} \right) = (-1)^{\ell+L} \left(\begin{array}{ccc} \ell & 2 & L \\ \pm2 & \mp2 & 0 \end{array} \right)  \left( \begin{array}{ccc} \ell & 2 & L \\ -2 & 2 & 0 \end{array} \right) = \left(\begin{array}{ccc} \ell & 2 & L \\ \pm2 & \mp2 & 0 \end{array} \right)  \left( \begin{array}{ccc} \ell & 2 & L \\ 2 & -2 & 0 \end{array} \right).
\end{equation}
The result thus far reads,
\begin{align}
	a_{\ell m}^{q,E}(\chi) \pm i a_{\ell m}^{q,B}(\chi) = &\int \frac{d^3 k}{(2\pi)^3} \sum_L 5i^L(2L+1) \left(\begin{array}{ccc} \ell & 2 & L \\ 2 & -2 & 0 \end{array} \right) \left(\begin{array}{ccc} \ell & 2 & L \\ \pm2 & \mp2 & 0 \end{array} \right)  j_L(k\chi) \no \\
	& \times \bigg\{ \mathcal{G}^q_{T,+}(k,\chi) \left[ {}_{2}Y^{*}_{\ell m}({\bf \widehat{k}}) + (-1)^{L+\ell} {}_{-2}Y^{*}_{\ell m}({\bf \widehat{k}}) \right]	 \no \\
	& - i\mathcal{G}^q_{T,\times}(k,\chi) \left[  {}_{2}Y^{*}_{\ell m}({\bf \widehat{k}})- (-1)^{L+\ell} {}_{-2}Y^{*}_{\ell m}({\bf \widehat{k}}) \right]\bigg\}.
\end{align}
This expression can be further simplified with the selection rule $\mid\ell-2\mid \leq L \leq \ell + 2$. This implies that for all $\ell \geq 2$, only the terms $L=\ell-2,\ \ell-1,\ \ell,\ \ell+1,\ \ell+2$ will contribute. The 3-$j$ symbols can then be expressed in each case as:
\begin{align}
	i^L(2L+1) \left(\begin{array}{ccc} \ell & 2 & L \\ 2 & -2 & 0 \end{array} \right) \left(\begin{array}{ccc} \ell & 2 & L \\ \pm2 & \mp2 & 0 \end{array} \right)j_L(k\chi) \Bigg|_{L=\ell+2} = &  - i^\ell \frac{\ell(\ell-1)}{4(2\ell+1)(2\ell+3)} j_{\ell+2}(k\chi)\\
	i^L(2L+1) \left(\begin{array}{ccc} \ell & 2 & L \\ 2 & -2 & 0 \end{array} \right) \left(\begin{array}{ccc} \ell & 2 & L \\ \pm2 & \mp2 & 0 \end{array} \right)j_L(k\chi)\Bigg|_{L=\ell+1} = & \pm  i^{\ell+1} \frac{(\ell-1)}{2(2\ell+1)} j_{\ell+1}(k\chi)\\
	i^L(2L+1) \left(\begin{array}{ccc} \ell & 2 & L \\ 2 & -2 & 0 \end{array} \right) \left(\begin{array}{ccc} \ell & 2 & L \\ \pm2 & \mp2 & 0 \end{array} \right)j_L(k\chi) \Bigg|_{L=\ell} = & \ i^\ell \frac{3(\ell-1)(\ell+2)}{2(2\ell-1)(2\ell+3)} j_{\ell}(k\chi)\\
	i^L(2L+1) \left(\begin{array}{ccc} \ell & 2 & L \\ 2 & -2 & 0 \end{array} \right) \left(\begin{array}{ccc} \ell & 2 & L \\ \pm2 & \mp2 & 0 \end{array} \right)j_L(k\chi) \Bigg|_{L=\ell-1} = & \mp  i^{\ell+1} \frac{(\ell+2)}{2(2\ell+1)}j_{\ell-1}(k\chi) \\
	i^L(2L+1) \left(\begin{array}{ccc} \ell & 2 & L \\ 2 & -2 & 0 \end{array} \right) \left(\begin{array}{ccc} \ell & 2 & L \\ \pm2 & \mp2 & 0 \end{array} \right)j_L(k\chi) \Bigg|_{L=\ell-2} = &  - i^\ell \frac{(\ell+1)(\ell+2)}{4(2\ell+1)(2\ell-1)} j_{\ell-2}(k\chi)
\end{align}
Some recursion relations are now required:
\begin{align}
	j_\ell(x) = & \frac{x}{2\ell+1} \left[ j_{\ell-1}(x) + j_{\ell+1}(x) \right], \\
	j'_\ell(x) = & \frac{1}{2\ell+1} \left[\ell j_{\ell-1}(x) -(\ell+1) j_{\ell+1}(x) \right], \\
	\frac{j_{\ell}(x)}{x^2} = & \frac{j_{\ell-2}(x)}{(2\ell+1)(2\ell-1)} + 2 \frac{j_{\ell}(x)}{(2\ell+3)(2\ell-1)} + \frac{j_{\ell+2}(x)}{(2\ell+1)(2\ell+3)}\,, \\
	\frac{j'_\ell(x)}{x} = & \frac{\ell j_{\ell-2}(x)}{(2\ell+1)(2\ell-1)}  + \frac{j_\ell(x) }{(2\ell+3)(2\ell-1)}-\frac{(\ell+1)j_{\ell+2}(x)}{(2\ell+1)(2\ell+3)}\,,\\
	j''_\ell(x)= & \frac{\ell(\ell-1) j_{\ell-2}(x)}{(2\ell+1)(2\ell-1)}  - \frac{(2\ell^2+2\ell-1)j_\ell(x) }{(2\ell+3)(2\ell-1)}+\frac{(\ell+1)(\ell+2)j_{\ell+2}(x)}{(2\ell+1)(2\ell+3)}\,.
\end{align}
The first two relations allow for the $L=\ell\pm1$ terms to be written as 
\begin{equation}
	\mp i^{\ell+1} \frac{(\ell+2)}{2(2\ell+1)} j_{\ell-1}(k\chi)   \pm  i^{\ell+1} \frac{(\ell-1)}{2(2\ell+1)}j_{\ell+1}(k\chi) = \mp \frac{i^{\ell+1}}{2}\left[ \frac{2j_\ell(k\chi)}{k\chi} + \frac{1}{k}\frac{dj_\ell(k\chi)}{d\chi}  \right].
\end{equation}
A simplification occurs for the $L=\ell-2,\ \ell,\ \ \ell+2$ terms relying on the last three recursion relations,
\begin{eqnarray}\label{11}
&&-\frac{\ell (\ell-1)}{4 (2\ell+1) (2\ell+3)}j_{\ell+2}-\frac{(\ell+1) (\ell+2)}{4 (2 \ell+1) (2 \ell-1)}j_{\ell-2}(k\chi)+\frac{3 (\ell+2) (\ell-1)}{2 (2 \ell-1) (2 \ell+3)}j_{\ell}(k\chi) \\&&= -\frac{1}{4k^2}\frac{d^2 j_{\ell}(k\chi)}{d\chi^2}-\frac{1}{k^2\chi}\frac{d j_{\ell}(k\chi)}{d\chi}+j_{\ell}(k\chi)\left(\frac{1}{4}-\frac{1}{2(k\chi)^2}\right)  \,.\nonumber
\end{eqnarray}
Combing the above relations one finds:
\begin{align}
	a_{\ell m}^{q,E}(\chi) \pm i a_{\ell m}^{q,B}(\chi) = & \int \frac{d^3 k}{(2\pi)^3}\ 5 i^\ell\left\{ \mathcal{G}^q_{T,+}(k,\chi)\bigg[ \mp i A_{\ell}(k,\chi) \left[ {}_{2}Y^{*}_{\ell m}({\bf \widehat{k}}) - {}_{-2}Y^{*}_{\ell m}({\bf \widehat{k}}) \right]	+ B_{\ell}(k,\chi) \left[{}_{2}Y^{*}_{\ell m}({\bf \widehat{k}}) + {}_{-2}Y^{*}_{\ell m}({\bf \widehat{k}}) \right]  \bigg] \right. \no \\
	& + i \left. \mathcal{G}^q_{T,\times}(k,\chi)\bigg[ \mp i A_{\ell}(k,\chi) \left[ {}_{2}Y^{*}_{\ell m}({\bf \widehat{k}}) + {}_{-2}Y^{*}_{\ell m}({\bf \widehat{k}}) \right]	+ B_{\ell}(k,\chi) \left[{}_{2}Y^{*}_{\ell m}({\bf \widehat{k}}) - {}_{-2}Y^{*}_{\ell m}({\bf \widehat{k}}) \right]  \bigg] \right\}, \label{eq:almTfinal}
\end{align}
where
\begin{eqnarray} \label{eq:Al_def}
	A_{\ell}(k,\chi)&\equiv&\frac{1}{2}\left(\frac{2 j_{\ell}(k\chi_{e})}{k\chi_{e}}+\frac{1}{k}\frac{d}{d\chi_{e}}j_{\ell}(k\chi_{e})\right)  \,,\\ 
	B_{\ell}(k,\chi)&\equiv& -\frac{1}{4k^2}\frac{d^2 j_{\ell}(k\chi)}{d\chi^2}-\frac{1}{k^2\chi}\frac{d j_{\ell}(k\chi)}{d\chi}+j_{\ell}(k\chi)\left(\frac{1}{4}-\frac{1}{2(k\chi)^2}\right)  \,.  \label{eq:Bl_def}
\end{eqnarray}
From Eq.~(\ref{eq:almTfinal}) one easily infers the expressions for $a_{\ell m}^{q,E}$ and $ a_{\ell m}^{q,B}$ 
\begin{align}
	a_{\ell m}^{q,E}(\chi) = & \int \frac{d^3 k}{(2\pi)^3}\ 5 i^\ell B_{\ell}(k,\chi)\left\{ \mathcal{G}^q_{T,+}(k,\chi) \left[{}_{2}Y^{*}_{\ell m}({\bf \widehat{k}}) + {}_{-2}Y^{*}_{\ell m}({\bf \widehat{k}}) \right]  \right.  + i \left. \mathcal{G}^q_{T,\times}(k,\chi) \left[{}_{2}Y^{*}_{\ell m}({\bf \widehat{k}}) - {}_{-2}Y^{*}_{\ell m}({\bf \widehat{k}}) \right] \right\}, \\
	a_{\ell m}^{q,B}(\chi) = & \int \frac{d^3 k}{(2\pi)^3}\ 5 i^\ell A_{\ell}(k,\chi)\left\{ -\mathcal{G}^q_{T,+}(k,\chi) \left[{}_{2}Y^{*}_{\ell m}({\bf \widehat{k}}) - {}_{-2}Y^{*}_{\ell m}({\bf \widehat{k}}) \right]  \right.  - i \left. \mathcal{G}^q_{T,\times}(k,\chi) \left[{}_{2}Y^{*}_{\ell m}({\bf \widehat{k}}) + {}_{-2}Y^{*}_{\ell m}({\bf \widehat{k}}) \right] \right\}.
\end{align}
It follows from these definitions that
\begin{align}
	C_{\ell}^{T,E}(\chi,\chi') = & \langle a_{\ell m}^{q,E}(\chi) a_{\ell' m'}^{q,E,*}(\chi')  \rangle \ \delta_{\ell \ell'}\delta_{mm'} \no \\
	 = &\ 2\int\frac{k^2\ dk}{(2\pi)^{3}}  \ 50\  P_h (k)\   \mathcal{I}^q_T(k,\chi) \mathcal{I}^q_T(k,\chi')\ B_{\ell}(k,\chi)B_{\ell}(k,\chi')\  \delta_{\ell \ell'}\delta_{mm'} \\
	C_{\ell}^{T,B}(\chi,\chi') = & \langle a_{\ell m}^{q,B}(\chi) a_{\ell' m'}^{q,B,*}(\chi')  \rangle \ \delta_{\ell \ell'}\delta_{mm'} \no \\
	= & \ 2\int\frac{k^2\ dk}{(2\pi)^{3}}  \ 50\  P_h (k)\   \mathcal{I}^q_T(k,\chi) \mathcal{I}^q_T(k,\chi')\ A_{\ell}(k,\chi)A_{\ell}(k,\chi') \ \delta_{\ell \ell'}\delta_{mm'} \\
	C_{\ell}^{T,EB}(\chi,\chi') = & \langle a_{\ell m}^{q,E}(\chi) a_{\ell' m'}^{q,B,*}(\chi')  \rangle = \ 0.
\end{align}
\begin{figure}
  \includegraphics[width=.75\textwidth]{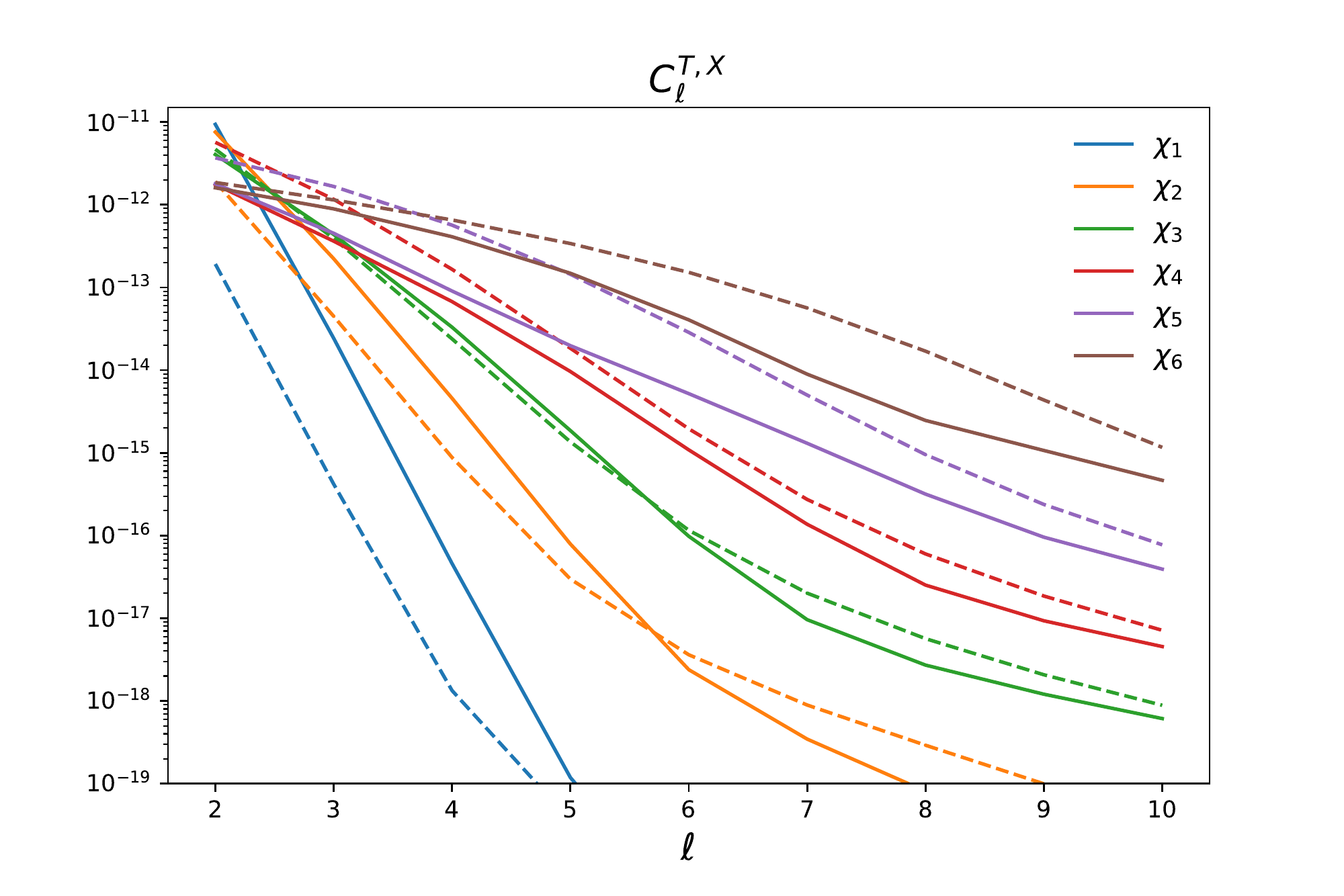}
  \caption{Power spectra of primordial tensor modes contributions to the CMB temperature quadrupole from $E$-modes ($C^{T, E}_{\ell}(\chi, \chi')$, plain lines) and $B$-modes ($C^{T, B}_{\ell}(\chi, \chi')$, dashed lines), evaluated at the same redshift, $\chi = \chi' = \chi_\alpha$, taken to be the midpoints of the bins given in table~\ref{tab:bins}. The power increases with redshift, and is roughly similar for $E$- and $B$-modes.}
  \label{fig:Clqq_EB}
\end{figure}
Fig.~\ref{fig:Clqq_EB} shows $C^{T, E}_{\ell}(\chi, \chi)$, in plain lines, and $C^{T, B}_{\ell}(\chi, \chi)$, in dashed lines, evaluated in the same redshift bin (see table~\ref{tab:bins}).
In deriving these expressions we have used the fact that the $+$ and $\times$ polarizations do not mix ($\langle \mathcal{G}^q_{T,+}(k,\chi) \ \mathcal{G}^q_{T,\times}(k',\chi') \rangle=0$), and we have rewritten the kernel explicitly in terms of the primordial tensor power spectrum $P_h$. To this end, one can define a tensor growth function, $D^T$, as 
\begin{equation}
	h_{(+,\times)}(k,a) \equiv D^T_{}(k,a) h_{i,(+,\times)}(k)
\end{equation}
where $h_{i,(+,\times)}(k) $ is the primordial tensor perturbation. This allows for the kernel in Eq.~\eqref{eq:kernel_tensors} to be rewritten as
\begin{equation}
		 \mathcal{G}^q_{T,(+,\times)}(k,\chi) = 2\pi\sqrt{6} \ h_{i,(+,\times)}(k) \int_{a_e}^{a_\text{dec}} da \frac{d D^T_{}(k,a)}{da} \ \frac{j_2 (k\Delta\chi(a))}{[k\Delta\chi(a)]^2} \equiv h_{i,(+,\times)}(k)\ \mathcal{I}^q_{T}(k,\chi), \label{eq:kernel_growthDT}
\end{equation}
hence one finds
\begin{equation}
	\langle \mathcal{G}^q_{T,(+,\times)}(k,\chi) \ \mathcal{G}^q_{T,(+,\times)}(k',\chi') \rangle = (2\pi)^3 \delta^{(3)}\ ({\bf k - k'}) \ P_h (k) \ \mathcal{I}^q_{T}(k,\chi) \ \mathcal{I}^q_{T}(k,\chi').
\end{equation}

\bibliography{psz}

\begin{thebibliography}{66}
\expandafter\ifx\csname natexlab\endcsname\relax\def\natexlab#1{#1}\fi
\expandafter\ifx\csname bibnamefont\endcsname\relax
  \def\bibnamefont#1{#1}\fi
\expandafter\ifx\csname bibfnamefont\endcsname\relax
  \def\bibfnamefont#1{#1}\fi
\expandafter\ifx\csname citenamefont\endcsname\relax
  \def\citenamefont#1{#1}\fi
\expandafter\ifx\csname url\endcsname\relax
  \def\url#1{\texttt{#1}}\fi
\expandafter\ifx\csname urlprefix\endcsname\relax\def\urlprefix{URL }\fi
\providecommand{\bibinfo}[2]{#2}
\providecommand{\eprint}[2][]{\url{#2}}

\bibitem[{\citenamefont{Kamionkowski and Loeb}(1997)}]{Kamionkowski1997}
\bibinfo{author}{\bibfnamefont{M.}~\bibnamefont{Kamionkowski}}
  \bibnamefont{and} \bibinfo{author}{\bibfnamefont{A.}~\bibnamefont{Loeb}},
  \bibinfo{journal}{Physical Review D} \textbf{\bibinfo{volume}{56}},
  \bibinfo{pages}{4511} (\bibinfo{year}{1997}).

\bibitem[{\citenamefont{Caldwell and Stebbins}(2008)}]{Caldwell:2007yu}
\bibinfo{author}{\bibfnamefont{R.~R.} \bibnamefont{Caldwell}} \bibnamefont{and}
  \bibinfo{author}{\bibfnamefont{A.}~\bibnamefont{Stebbins}},
  \bibinfo{journal}{Phys. Rev. Lett.} \textbf{\bibinfo{volume}{100}},
  \bibinfo{pages}{191302} (\bibinfo{year}{2008}).

\bibitem[{\citenamefont{Garcia-Bellido and
  Haugboelle}(2008)}]{GarciaBellido:2008gd}
\bibinfo{author}{\bibfnamefont{J.}~\bibnamefont{Garcia-Bellido}}
  \bibnamefont{and}
  \bibinfo{author}{\bibfnamefont{T.}~\bibnamefont{Haugboelle}},
  \bibinfo{journal}{JCAP} \textbf{\bibinfo{volume}{0809}}, \bibinfo{pages}{016}
  (\bibinfo{year}{2008}).

\bibitem[{\citenamefont{{Zhang} and {Stebbins}}(2011)}]{Zhang11b}
\bibinfo{author}{\bibfnamefont{P.}~\bibnamefont{{Zhang}}} \bibnamefont{and}
  \bibinfo{author}{\bibfnamefont{A.}~\bibnamefont{{Stebbins}}},
  \bibinfo{journal}{Physical Review Letters} \textbf{\bibinfo{volume}{107}},
  \bibinfo{pages}{041301} (\bibinfo{year}{2011}).

\bibitem[{\citenamefont{Clifton et~al.}(2012)\citenamefont{Clifton, Clarkson,
  and Bull}}]{Clifton:2011sn}
\bibinfo{author}{\bibfnamefont{T.}~\bibnamefont{Clifton}},
  \bibinfo{author}{\bibfnamefont{C.}~\bibnamefont{Clarkson}}, \bibnamefont{and}
  \bibinfo{author}{\bibfnamefont{P.}~\bibnamefont{Bull}},
  \bibinfo{journal}{Phys. Rev. Lett.} \textbf{\bibinfo{volume}{109}},
  \bibinfo{pages}{051303} (\bibinfo{year}{2012}).

\bibitem[{\citenamefont{Maartens}(2011)}]{Maartens2011}
\bibinfo{author}{\bibfnamefont{R.}~\bibnamefont{Maartens}},
  \bibinfo{journal}{Philosophical Transactions of the Royal Society A:
  Mathematical, Physical and Engineering Sciences}
  \textbf{\bibinfo{volume}{369}}, \bibinfo{pages}{5115} (\bibinfo{year}{2011}).

\bibitem[{\citenamefont{{Zibin} and {Moss}}(2011)}]{2011CQGra..28p4005Z}
\bibinfo{author}{\bibfnamefont{J.~P.} \bibnamefont{{Zibin}}} \bibnamefont{and}
  \bibinfo{author}{\bibfnamefont{A.}~\bibnamefont{{Moss}}},
  \bibinfo{journal}{Classical and Quantum Gravity}
  \textbf{\bibinfo{volume}{28}}, \bibinfo{pages}{164005}
  (\bibinfo{year}{2011}).

\bibitem[{\citenamefont{Bull et~al.}(2012)\citenamefont{Bull, Clifton, and
  Ferreira}}]{Bull:2011wi}
\bibinfo{author}{\bibfnamefont{P.}~\bibnamefont{Bull}},
  \bibinfo{author}{\bibfnamefont{T.}~\bibnamefont{Clifton}}, \bibnamefont{and}
  \bibinfo{author}{\bibfnamefont{P.~G.} \bibnamefont{Ferreira}},
  \bibinfo{journal}{Phys. Rev.} \textbf{\bibinfo{volume}{D85}},
  \bibinfo{pages}{024002} (\bibinfo{year}{2012}).

\bibitem[{\citenamefont{Yoo et~al.}(2010)\citenamefont{Yoo, Nakao, and
  Sasaki}}]{Yoo:2010ad}
\bibinfo{author}{\bibfnamefont{C.-M.} \bibnamefont{Yoo}},
  \bibinfo{author}{\bibfnamefont{K.-i.} \bibnamefont{Nakao}}, \bibnamefont{and}
  \bibinfo{author}{\bibfnamefont{M.}~\bibnamefont{Sasaki}},
  \bibinfo{journal}{JCAP} \textbf{\bibinfo{volume}{1010}}, \bibinfo{pages}{011}
  (\bibinfo{year}{2010}).

\bibitem[{\citenamefont{Schwarz et~al.}(2016)\citenamefont{Schwarz, Copi,
  Huterer, and Starkman}}]{Schwarz:2015cma}
\bibinfo{author}{\bibfnamefont{D.~J.} \bibnamefont{Schwarz}},
  \bibinfo{author}{\bibfnamefont{C.~J.} \bibnamefont{Copi}},
  \bibinfo{author}{\bibfnamefont{D.}~\bibnamefont{Huterer}}, \bibnamefont{and}
  \bibinfo{author}{\bibfnamefont{G.~D.} \bibnamefont{Starkman}},
  \bibinfo{journal}{Class. Quant. Grav.} \textbf{\bibinfo{volume}{33}},
  \bibinfo{pages}{184001} (\bibinfo{year}{2016}).

\bibitem[{\citenamefont{Smith et~al.}(2007)\citenamefont{Smith, Zahn, and
  Dore}}]{Smith:2007rg}
\bibinfo{author}{\bibfnamefont{K.~M.} \bibnamefont{Smith}},
  \bibinfo{author}{\bibfnamefont{O.}~\bibnamefont{Zahn}}, \bibnamefont{and}
  \bibinfo{author}{\bibfnamefont{O.}~\bibnamefont{Dore}},
  \bibinfo{journal}{Phys. Rev.} \textbf{\bibinfo{volume}{D76}},
  \bibinfo{pages}{043510} (\bibinfo{year}{2007}), \eprint{0705.3980}.

\bibitem[{\citenamefont{Hirata et~al.}(2008)\citenamefont{Hirata, Ho,
  Padmanabhan, Seljak, and Bahcall}}]{Hirata:2008cb}
\bibinfo{author}{\bibfnamefont{C.~M.} \bibnamefont{Hirata}},
  \bibinfo{author}{\bibfnamefont{S.}~\bibnamefont{Ho}},
  \bibinfo{author}{\bibfnamefont{N.}~\bibnamefont{Padmanabhan}},
  \bibinfo{author}{\bibfnamefont{U.}~\bibnamefont{Seljak}}, \bibnamefont{and}
  \bibinfo{author}{\bibfnamefont{N.~A.} \bibnamefont{Bahcall}},
  \bibinfo{journal}{Phys. Rev.} \textbf{\bibinfo{volume}{D78}},
  \bibinfo{pages}{043520} (\bibinfo{year}{2008}), \eprint{0801.0644}.

\bibitem[{\citenamefont{{Das} et~al.}(2011)\citenamefont{{Das}, {Sherwin},
  {Aguirre}, {Appel}, {Bond}, {Carvalho}, {Devlin}, {Dunkley}, {D{\"u}nner},
  {Essinger-Hileman} et~al.}}]{2011PhRvL.107b1301D}
\bibinfo{author}{\bibfnamefont{S.}~\bibnamefont{{Das}}},
  \bibinfo{author}{\bibfnamefont{B.~D.} \bibnamefont{{Sherwin}}},
  \bibinfo{author}{\bibfnamefont{P.}~\bibnamefont{{Aguirre}}},
  \bibinfo{author}{\bibfnamefont{J.~W.} \bibnamefont{{Appel}}},
  \bibinfo{author}{\bibfnamefont{J.~R.} \bibnamefont{{Bond}}},
  \bibinfo{author}{\bibfnamefont{C.~S.} \bibnamefont{{Carvalho}}},
  \bibinfo{author}{\bibfnamefont{M.~J.} \bibnamefont{{Devlin}}},
  \bibinfo{author}{\bibfnamefont{J.}~\bibnamefont{{Dunkley}}},
  \bibinfo{author}{\bibfnamefont{R.}~\bibnamefont{{D{\"u}nner}}},
  \bibinfo{author}{\bibfnamefont{T.}~\bibnamefont{{Essinger-Hileman}}},
  \bibnamefont{et~al.}, \bibinfo{journal}{Physical Review Letters}
  \textbf{\bibinfo{volume}{107}}, \bibinfo{eid}{021301} (\bibinfo{year}{2011}),
  \eprint{1103.2124}.

\bibitem[{\citenamefont{{van Engelen} et~al.}(2012)\citenamefont{{van Engelen},
  {Keisler}, {Zahn}, {Aird}, {Benson}, {Bleem}, {Carlstrom}, {Chang}, {Cho},
  {Crawford} et~al.}}]{2012ApJ...756..142V}
\bibinfo{author}{\bibfnamefont{A.}~\bibnamefont{{van Engelen}}},
  \bibinfo{author}{\bibfnamefont{R.}~\bibnamefont{{Keisler}}},
  \bibinfo{author}{\bibfnamefont{O.}~\bibnamefont{{Zahn}}},
  \bibinfo{author}{\bibfnamefont{K.~A.} \bibnamefont{{Aird}}},
  \bibinfo{author}{\bibfnamefont{B.~A.} \bibnamefont{{Benson}}},
  \bibinfo{author}{\bibfnamefont{L.~E.} \bibnamefont{{Bleem}}},
  \bibinfo{author}{\bibfnamefont{J.~E.} \bibnamefont{{Carlstrom}}},
  \bibinfo{author}{\bibfnamefont{C.~L.} \bibnamefont{{Chang}}},
  \bibinfo{author}{\bibfnamefont{H.~M.} \bibnamefont{{Cho}}},
  \bibinfo{author}{\bibfnamefont{T.~M.} \bibnamefont{{Crawford}}},
  \bibnamefont{et~al.}, \bibinfo{journal}{\apj} \textbf{\bibinfo{volume}{756}},
  \bibinfo{eid}{142} (\bibinfo{year}{2012}), \eprint{1202.0546}.

\bibitem[{\citenamefont{Ade et~al.}(2016{\natexlab{a}})}]{Ade:2015zua}
\bibinfo{author}{\bibfnamefont{P.~A.~R.} \bibnamefont{Ade}}
  \bibnamefont{et~al.} (\bibinfo{collaboration}{Planck}),
  \bibinfo{journal}{Astron. Astrophys.} \textbf{\bibinfo{volume}{594}},
  \bibinfo{pages}{A15} (\bibinfo{year}{2016}{\natexlab{a}}),
  \eprint{1502.01591}.

\bibitem[{\citenamefont{{Blanchard} and
  {Schneider}}(1987)}]{1987A&A...184....1B}
\bibinfo{author}{\bibfnamefont{A.}~\bibnamefont{{Blanchard}}} \bibnamefont{and}
  \bibinfo{author}{\bibfnamefont{J.}~\bibnamefont{{Schneider}}},
  \bibinfo{journal}{Astron. Astrophys.} \textbf{\bibinfo{volume}{184}},
  \bibinfo{pages}{1} (\bibinfo{year}{1987}).

\bibitem[{\citenamefont{{Cole} and {Efstathiou}}(1989)}]{1989MNRAS.239..195C}
\bibinfo{author}{\bibfnamefont{S.}~\bibnamefont{{Cole}}} \bibnamefont{and}
  \bibinfo{author}{\bibfnamefont{G.}~\bibnamefont{{Efstathiou}}},
  \bibinfo{journal}{Mon. Not. Roy. Astron. Soc.}
  \textbf{\bibinfo{volume}{239}}, \bibinfo{pages}{195} (\bibinfo{year}{1989}).

\bibitem[{\citenamefont{{Lewis} and {Challinor}}(2006)}]{Lewis:2006aa}
\bibinfo{author}{\bibfnamefont{A.}~\bibnamefont{{Lewis}}} \bibnamefont{and}
  \bibinfo{author}{\bibfnamefont{A.}~\bibnamefont{{Challinor}}},
  \bibinfo{journal}{\physrep} \textbf{\bibinfo{volume}{429}},
  \bibinfo{pages}{1} (\bibinfo{year}{2006}), \eprint{astro-ph/0601594}.

\bibitem[{\citenamefont{Staniszewski et~al.}(2009)}]{Staniszewski:2008ma}
\bibinfo{author}{\bibfnamefont{Z.}~\bibnamefont{Staniszewski}}
  \bibnamefont{et~al.}, \bibinfo{journal}{Astrophys. J.}
  \textbf{\bibinfo{volume}{701}}, \bibinfo{pages}{32} (\bibinfo{year}{2009}),
  \eprint{0810.1578}.

\bibitem[{\citenamefont{{Planck Collaboration}
  et~al.}(2011)\citenamefont{{Planck Collaboration}, {Ade}, {Aghanim},
  {Arnaud}, {Ashdown}, {Aumont}, {Baccigalupi}, {Balbi}, {Banday}, {Barreiro}
  et~al.}}]{2011A&A...536A...8P}
\bibinfo{author}{\bibnamefont{{Planck Collaboration}}},
  \bibinfo{author}{\bibfnamefont{P.~A.~R.} \bibnamefont{{Ade}}},
  \bibinfo{author}{\bibfnamefont{N.}~\bibnamefont{{Aghanim}}},
  \bibinfo{author}{\bibfnamefont{M.}~\bibnamefont{{Arnaud}}},
  \bibinfo{author}{\bibfnamefont{M.}~\bibnamefont{{Ashdown}}},
  \bibinfo{author}{\bibfnamefont{J.}~\bibnamefont{{Aumont}}},
  \bibinfo{author}{\bibfnamefont{C.}~\bibnamefont{{Baccigalupi}}},
  \bibinfo{author}{\bibfnamefont{A.}~\bibnamefont{{Balbi}}},
  \bibinfo{author}{\bibfnamefont{A.~J.} \bibnamefont{{Banday}}},
  \bibinfo{author}{\bibfnamefont{R.~B.} \bibnamefont{{Barreiro}}},
  \bibnamefont{et~al.}, \bibinfo{journal}{Astron. Astrophys.}
  \textbf{\bibinfo{volume}{536}}, \bibinfo{eid}{A8} (\bibinfo{year}{2011}),
  \eprint{1101.2024}.

\bibitem[{\citenamefont{Hasselfield et~al.}(2013)}]{Hasselfield:2013wf}
\bibinfo{author}{\bibfnamefont{M.}~\bibnamefont{Hasselfield}}
  \bibnamefont{et~al.}, \bibinfo{journal}{JCAP}
  \textbf{\bibinfo{volume}{1307}}, \bibinfo{pages}{008} (\bibinfo{year}{2013}),
  \eprint{1301.0816}.

\bibitem[{\citenamefont{{Reichardt} et~al.}(2013)\citenamefont{{Reichardt},
  {Stalder}, {Bleem}, {Montroy}, {Aird}, {Andersson}, {Armstrong}, {Ashby},
  {Bautz}, {Bayliss} et~al.}}]{2013ApJ...763..127R}
\bibinfo{author}{\bibfnamefont{C.~L.} \bibnamefont{{Reichardt}}},
  \bibinfo{author}{\bibfnamefont{B.}~\bibnamefont{{Stalder}}},
  \bibinfo{author}{\bibfnamefont{L.~E.} \bibnamefont{{Bleem}}},
  \bibinfo{author}{\bibfnamefont{T.~E.} \bibnamefont{{Montroy}}},
  \bibinfo{author}{\bibfnamefont{K.~A.} \bibnamefont{{Aird}}},
  \bibinfo{author}{\bibfnamefont{K.}~\bibnamefont{{Andersson}}},
  \bibinfo{author}{\bibfnamefont{R.}~\bibnamefont{{Armstrong}}},
  \bibinfo{author}{\bibfnamefont{M.~L.~N.} \bibnamefont{{Ashby}}},
  \bibinfo{author}{\bibfnamefont{M.}~\bibnamefont{{Bautz}}},
  \bibinfo{author}{\bibfnamefont{M.}~\bibnamefont{{Bayliss}}},
  \bibnamefont{et~al.}, \bibinfo{journal}{\apj} \textbf{\bibinfo{volume}{763}},
  \bibinfo{eid}{127} (\bibinfo{year}{2013}), \eprint{1203.5775}.

\bibitem[{\citenamefont{Aghanim et~al.}(2016)}]{Aghanim:2015eva}
\bibinfo{author}{\bibfnamefont{N.}~\bibnamefont{Aghanim}} \bibnamefont{et~al.}
  (\bibinfo{collaboration}{Planck}), \bibinfo{journal}{Astron. Astrophys.}
  \textbf{\bibinfo{volume}{594}}, \bibinfo{pages}{A22} (\bibinfo{year}{2016}),
  \eprint{1502.01596}.

\bibitem[{\citenamefont{{Sunyaev} and {Zeldovich}}(1970)}]{1970Ap&SS...7....3S}
\bibinfo{author}{\bibfnamefont{R.~A.} \bibnamefont{{Sunyaev}}}
  \bibnamefont{and} \bibinfo{author}{\bibfnamefont{Y.~B.}
  \bibnamefont{{Zeldovich}}}, \bibinfo{journal}{APSS}
  \textbf{\bibinfo{volume}{7}}, \bibinfo{pages}{3} (\bibinfo{year}{1970}).

\bibitem[{\citenamefont{{Hand} et~al.}(2012)\citenamefont{{Hand}, {Addison},
  {Aubourg}, {Battaglia}, {Battistelli}, {Bizyaev}, {Bond}, {Brewington},
  {Brinkmann}, {Brown} et~al.}}]{Hand12}
\bibinfo{author}{\bibfnamefont{N.}~\bibnamefont{{Hand}}},
  \bibinfo{author}{\bibfnamefont{G.~E.} \bibnamefont{{Addison}}},
  \bibinfo{author}{\bibfnamefont{E.}~\bibnamefont{{Aubourg}}},
  \bibinfo{author}{\bibfnamefont{N.}~\bibnamefont{{Battaglia}}},
  \bibinfo{author}{\bibfnamefont{E.~S.} \bibnamefont{{Battistelli}}},
  \bibinfo{author}{\bibfnamefont{D.}~\bibnamefont{{Bizyaev}}},
  \bibinfo{author}{\bibfnamefont{J.~R.} \bibnamefont{{Bond}}},
  \bibinfo{author}{\bibfnamefont{H.}~\bibnamefont{{Brewington}}},
  \bibinfo{author}{\bibfnamefont{J.}~\bibnamefont{{Brinkmann}}},
  \bibinfo{author}{\bibfnamefont{B.~R.} \bibnamefont{{Brown}}},
  \bibnamefont{et~al.}, \bibinfo{journal}{Physical Review Letters}
  \textbf{\bibinfo{volume}{109}}, \bibinfo{eid}{041101} (\bibinfo{year}{2012}),
  \eprint{1203.4219}.

\bibitem[{\citenamefont{De~Bernardis et~al.}(2016)}]{DeBernardis:2016pdv}
\bibinfo{author}{\bibfnamefont{F.}~\bibnamefont{De~Bernardis}}
  \bibnamefont{et~al.} (\bibinfo{year}{2016}), \eprint{1607.02139}.

\bibitem[{\citenamefont{Soergel et~al.}(2016)}]{Soergel:2016mce}
\bibinfo{author}{\bibfnamefont{B.}~\bibnamefont{Soergel}} \bibnamefont{et~al.}
  (\bibinfo{collaboration}{DES, SPT}), \bibinfo{journal}{Mon. Not. Roy. Astron.
  Soc.}  (\bibinfo{year}{2016}), \eprint{1603.03904}.

\bibitem[{\citenamefont{{Planck Collaboration}
  et~al.}(2016)\citenamefont{{Planck Collaboration}, {Ade}, {Aghanim},
  {Arnaud}, {Ashdown}, {Aubourg}, {Aumont}, {Baccigalupi}, {Banday}, {Barreiro}
  et~al.}}]{2016A&A...586A.140P}
\bibinfo{author}{\bibnamefont{{Planck Collaboration}}},
  \bibinfo{author}{\bibfnamefont{P.~A.~R.} \bibnamefont{{Ade}}},
  \bibinfo{author}{\bibfnamefont{N.}~\bibnamefont{{Aghanim}}},
  \bibinfo{author}{\bibfnamefont{M.}~\bibnamefont{{Arnaud}}},
  \bibinfo{author}{\bibfnamefont{M.}~\bibnamefont{{Ashdown}}},
  \bibinfo{author}{\bibfnamefont{E.}~\bibnamefont{{Aubourg}}},
  \bibinfo{author}{\bibfnamefont{J.}~\bibnamefont{{Aumont}}},
  \bibinfo{author}{\bibfnamefont{C.}~\bibnamefont{{Baccigalupi}}},
  \bibinfo{author}{\bibfnamefont{A.~J.} \bibnamefont{{Banday}}},
  \bibinfo{author}{\bibfnamefont{R.~B.} \bibnamefont{{Barreiro}}},
  \bibnamefont{et~al.}, \bibinfo{journal}{Astronomy \& Astrophysics}
  \textbf{\bibinfo{volume}{586}}, \bibinfo{eid}{A140} (\bibinfo{year}{2016}),
  \eprint{1504.03339}.

\bibitem[{\citenamefont{{Schaan} et~al.}(2016)\citenamefont{{Schaan},
  {Ferraro}, {Vargas-Maga{\~n}a}, {Smith}, {Ho}, {Aiola}, {Battaglia}, {Bond},
  {De Bernardis}, {Calabrese} et~al.}}]{2016PhRvD..93h2002S}
\bibinfo{author}{\bibfnamefont{E.}~\bibnamefont{{Schaan}}},
  \bibinfo{author}{\bibfnamefont{S.}~\bibnamefont{{Ferraro}}},
  \bibinfo{author}{\bibfnamefont{M.}~\bibnamefont{{Vargas-Maga{\~n}a}}},
  \bibinfo{author}{\bibfnamefont{K.~M.} \bibnamefont{{Smith}}},
  \bibinfo{author}{\bibfnamefont{S.}~\bibnamefont{{Ho}}},
  \bibinfo{author}{\bibfnamefont{S.}~\bibnamefont{{Aiola}}},
  \bibinfo{author}{\bibfnamefont{N.}~\bibnamefont{{Battaglia}}},
  \bibinfo{author}{\bibfnamefont{J.~R.} \bibnamefont{{Bond}}},
  \bibinfo{author}{\bibfnamefont{F.}~\bibnamefont{{De Bernardis}}},
  \bibinfo{author}{\bibfnamefont{E.}~\bibnamefont{{Calabrese}}},
  \bibnamefont{et~al.}, \bibinfo{journal}{Phys. Rev. D}
  \textbf{\bibinfo{volume}{93}}, \bibinfo{eid}{082002} (\bibinfo{year}{2016}).

\bibitem[{\citenamefont{{George} et~al.}(2014)\citenamefont{{George},
  {Reichardt}, {Aird}, {Benson}, {Bleem}, {Carlstrom}, {Chang}, {Cho},
  {Crawford}, {Crites} et~al.}}]{George14}
\bibinfo{author}{\bibfnamefont{E.~M.} \bibnamefont{{George}}},
  \bibinfo{author}{\bibfnamefont{C.~L.} \bibnamefont{{Reichardt}}},
  \bibinfo{author}{\bibfnamefont{K.~A.} \bibnamefont{{Aird}}},
  \bibinfo{author}{\bibfnamefont{B.~A.} \bibnamefont{{Benson}}},
  \bibinfo{author}{\bibfnamefont{L.~E.} \bibnamefont{{Bleem}}},
  \bibinfo{author}{\bibfnamefont{J.~E.} \bibnamefont{{Carlstrom}}},
  \bibinfo{author}{\bibfnamefont{C.~L.} \bibnamefont{{Chang}}},
  \bibinfo{author}{\bibfnamefont{H.}~\bibnamefont{{Cho}}},
  \bibinfo{author}{\bibfnamefont{T.~M.} \bibnamefont{{Crawford}}},
  \bibinfo{author}{\bibfnamefont{A.~T.} \bibnamefont{{Crites}}},
  \bibnamefont{et~al.}, \bibinfo{journal}{ArXiv e-prints}
  (\bibinfo{year}{2014}), \eprint{1408.3161}.

\bibitem[{\citenamefont{{Sunyaev} and {Zeldovich}}(1980)}]{SZ80}
\bibinfo{author}{\bibfnamefont{R.~A.} \bibnamefont{{Sunyaev}}}
  \bibnamefont{and} \bibinfo{author}{\bibfnamefont{I.~B.}
  \bibnamefont{{Zeldovich}}}, \bibinfo{journal}{MNRAS}
  \textbf{\bibinfo{volume}{190}}, \bibinfo{pages}{413} (\bibinfo{year}{1980}).

\bibitem[{\citenamefont{Abazajian et~al.}(2016)\citenamefont{Abazajian,
  Adshead, Ahmed, Allen, Alonso, Arnold, Baccigalupi, Bartlett, Battaglia,
  Benson et~al.}}]{CMBS42016}
\bibinfo{author}{\bibfnamefont{K.~N.} \bibnamefont{Abazajian}},
  \bibinfo{author}{\bibfnamefont{P.}~\bibnamefont{Adshead}},
  \bibinfo{author}{\bibfnamefont{Z.}~\bibnamefont{Ahmed}},
  \bibinfo{author}{\bibfnamefont{S.~W.} \bibnamefont{Allen}},
  \bibinfo{author}{\bibfnamefont{D.}~\bibnamefont{Alonso}},
  \bibinfo{author}{\bibfnamefont{K.~S.} \bibnamefont{Arnold}},
  \bibinfo{author}{\bibfnamefont{C.}~\bibnamefont{Baccigalupi}},
  \bibinfo{author}{\bibfnamefont{J.~G.} \bibnamefont{Bartlett}},
  \bibinfo{author}{\bibfnamefont{N.}~\bibnamefont{Battaglia}},
  \bibinfo{author}{\bibfnamefont{B.~A.} \bibnamefont{Benson}},
  \bibnamefont{et~al.} (\bibinfo{year}{2016}), \eprint{1610.02743}.

\bibitem[{\citenamefont{{LSST Science Collaboration}
  et~al.}(2009)\citenamefont{{LSST Science Collaboration}, Abell, Allison,
  Anderson, Andrew, Angel, Armus, Arnett, Asztalos, Axelrod
  et~al.}}]{LSSTScienceCollaboration2009}
\bibinfo{author}{\bibnamefont{{LSST Science Collaboration}}},
  \bibinfo{author}{\bibfnamefont{P.~A.} \bibnamefont{Abell}},
  \bibinfo{author}{\bibfnamefont{J.}~\bibnamefont{Allison}},
  \bibinfo{author}{\bibfnamefont{S.~F.} \bibnamefont{Anderson}},
  \bibinfo{author}{\bibfnamefont{J.~R.} \bibnamefont{Andrew}},
  \bibinfo{author}{\bibfnamefont{J.~R.~P.} \bibnamefont{Angel}},
  \bibinfo{author}{\bibfnamefont{L.}~\bibnamefont{Armus}},
  \bibinfo{author}{\bibfnamefont{D.}~\bibnamefont{Arnett}},
  \bibinfo{author}{\bibfnamefont{S.~J.} \bibnamefont{Asztalos}},
  \bibinfo{author}{\bibfnamefont{T.~S.} \bibnamefont{Axelrod}},
  \bibnamefont{et~al.}, \bibinfo{journal}{Science} p. \bibinfo{pages}{596}
  (\bibinfo{year}{2009}), \eprint{0912.0201}.

\bibitem[{\citenamefont{{Ho} et~al.}(2009)\citenamefont{{Ho}, {Dedeo}, and
  {Spergel}}}]{Ho09}
\bibinfo{author}{\bibfnamefont{S.}~\bibnamefont{{Ho}}},
  \bibinfo{author}{\bibfnamefont{S.}~\bibnamefont{{Dedeo}}}, \bibnamefont{and}
  \bibinfo{author}{\bibfnamefont{D.}~\bibnamefont{{Spergel}}},
  \bibinfo{journal}{ArXiv e-prints}  (\bibinfo{year}{2009}),
  \eprint{0903.2845}.

\bibitem[{\citenamefont{{Shao} et~al.}(2011)\citenamefont{{Shao}, {Zhang},
  {Lin}, {Jing}, and {Pan}}}]{Shao11b}
\bibinfo{author}{\bibfnamefont{J.}~\bibnamefont{{Shao}}},
  \bibinfo{author}{\bibfnamefont{P.}~\bibnamefont{{Zhang}}},
  \bibinfo{author}{\bibfnamefont{W.}~\bibnamefont{{Lin}}},
  \bibinfo{author}{\bibfnamefont{Y.}~\bibnamefont{{Jing}}}, \bibnamefont{and}
  \bibinfo{author}{\bibfnamefont{J.}~\bibnamefont{{Pan}}},
  \bibinfo{journal}{MNRAS} \textbf{\bibinfo{volume}{413}}, \bibinfo{pages}{628}
  (\bibinfo{year}{2011}), \eprint{1004.1301}.

\bibitem[{\citenamefont{{Zhang} and {Pen}}(2001)}]{Zhang01}
\bibinfo{author}{\bibfnamefont{P.}~\bibnamefont{{Zhang}}} \bibnamefont{and}
  \bibinfo{author}{\bibfnamefont{U.-L.} \bibnamefont{{Pen}}},
  \bibinfo{journal}{Astrophys. J.} \textbf{\bibinfo{volume}{549}},
  \bibinfo{pages}{18} (\bibinfo{year}{2001}), \eprint{astro-ph/0007462}.

\bibitem[{\citenamefont{Munshi et~al.}(2015)\citenamefont{Munshi, Iliev, Dixon,
  and Coles}}]{Munshi:2015anr}
\bibinfo{author}{\bibfnamefont{D.}~\bibnamefont{Munshi}},
  \bibinfo{author}{\bibfnamefont{I.~T.} \bibnamefont{Iliev}},
  \bibinfo{author}{\bibfnamefont{K.~L.} \bibnamefont{Dixon}}, \bibnamefont{and}
  \bibinfo{author}{\bibfnamefont{P.}~\bibnamefont{Coles}}
  (\bibinfo{year}{2015}), \eprint{1511.03449}.

\bibitem[{\citenamefont{Ferraro et~al.}(2016)\citenamefont{Ferraro, Hill,
  Battaglia, Liu, and Spergel}}]{Ferraro:2016ymw}
\bibinfo{author}{\bibfnamefont{S.}~\bibnamefont{Ferraro}},
  \bibinfo{author}{\bibfnamefont{J.~C.} \bibnamefont{Hill}},
  \bibinfo{author}{\bibfnamefont{N.}~\bibnamefont{Battaglia}},
  \bibinfo{author}{\bibfnamefont{J.}~\bibnamefont{Liu}}, \bibnamefont{and}
  \bibinfo{author}{\bibfnamefont{D.~N.} \bibnamefont{Spergel}}
  (\bibinfo{year}{2016}), \eprint{1605.02722}.

\bibitem[{\citenamefont{Hill et~al.}(2016)\citenamefont{Hill, Ferraro,
  Battaglia, Liu, and Spergel}}]{Hill:2016dta}
\bibinfo{author}{\bibfnamefont{J.~C.} \bibnamefont{Hill}},
  \bibinfo{author}{\bibfnamefont{S.}~\bibnamefont{Ferraro}},
  \bibinfo{author}{\bibfnamefont{N.}~\bibnamefont{Battaglia}},
  \bibinfo{author}{\bibfnamefont{J.}~\bibnamefont{Liu}}, \bibnamefont{and}
  \bibinfo{author}{\bibfnamefont{D.~N.} \bibnamefont{Spergel}},
  \bibinfo{journal}{Phys. Rev. Lett.} \textbf{\bibinfo{volume}{117}},
  \bibinfo{pages}{051301} (\bibinfo{year}{2016}), \eprint{1603.01608}.

\bibitem[{\citenamefont{{Zhang}}(2010)}]{Zhang10d}
\bibinfo{author}{\bibfnamefont{P.}~\bibnamefont{{Zhang}}},
  \bibinfo{journal}{MNRAS} \textbf{\bibinfo{volume}{407}}, \bibinfo{pages}{L36}
  (\bibinfo{year}{2010}), \eprint{1004.0990}.

\bibitem[{\citenamefont{Zhang and Johnson}(2015)}]{Zhang:2015uta}
\bibinfo{author}{\bibfnamefont{P.}~\bibnamefont{Zhang}} \bibnamefont{and}
  \bibinfo{author}{\bibfnamefont{M.~C.} \bibnamefont{Johnson}},
  \bibinfo{journal}{JCAP} \textbf{\bibinfo{volume}{1506}}, \bibinfo{pages}{046}
  (\bibinfo{year}{2015}), \eprint{1501.00511}.

\bibitem[{\citenamefont{Terrana et~al.}(2017)\citenamefont{Terrana, Harris, and
  Johnson}}]{Terrana2016}
\bibinfo{author}{\bibfnamefont{A.}~\bibnamefont{Terrana}},
  \bibinfo{author}{\bibfnamefont{M.-J.} \bibnamefont{Harris}},
  \bibnamefont{and} \bibinfo{author}{\bibfnamefont{M.~C.}
  \bibnamefont{Johnson}}, \bibinfo{journal}{Journal of Cosmology and
  Astroparticle Physics} \textbf{\bibinfo{volume}{2017}}, \bibinfo{pages}{040}
  (\bibinfo{year}{2017}), \eprint{1610.06919}.

\bibitem[{\citenamefont{{Alizadeh} and {Hirata}}(2012)}]{2012PhRvD..85l3540A}
\bibinfo{author}{\bibfnamefont{E.}~\bibnamefont{{Alizadeh}}} \bibnamefont{and}
  \bibinfo{author}{\bibfnamefont{C.~M.} \bibnamefont{{Hirata}}},
  \bibinfo{journal}{Physical Review D} \textbf{\bibinfo{volume}{85}},
  \bibinfo{eid}{123540} (\bibinfo{year}{2012}), \eprint{1201.5374}.

\bibitem[{\citenamefont{Deutsch et~al.}(2017)\citenamefont{Deutsch, Johnson,
  M{\"u}nchmeyer, and Terrana}}]{Deutsch:2017cja}
\bibinfo{author}{\bibfnamefont{A.-S.} \bibnamefont{Deutsch}},
  \bibinfo{author}{\bibfnamefont{M.~C.} \bibnamefont{Johnson}},
  \bibinfo{author}{\bibfnamefont{M.}~\bibnamefont{M{\"u}nchmeyer}},
  \bibnamefont{and} \bibinfo{author}{\bibfnamefont{A.}~\bibnamefont{Terrana}}
  (\bibinfo{year}{2017}), \eprint{1705.08907}.

\bibitem[{\citenamefont{Portsmouth}(2004)}]{Portsmouth2004}
\bibinfo{author}{\bibfnamefont{J.}~\bibnamefont{Portsmouth}},
  \bibinfo{journal}{Physical Review D} \textbf{\bibinfo{volume}{70}},
  \bibinfo{pages}{063504} (\bibinfo{year}{2004}), \eprint{0402173}.

\bibitem[{\citenamefont{Seto and Sasaki}(2000)}]{Seto:2000uc}
\bibinfo{author}{\bibfnamefont{N.}~\bibnamefont{Seto}} \bibnamefont{and}
  \bibinfo{author}{\bibfnamefont{M.}~\bibnamefont{Sasaki}},
  \bibinfo{journal}{Phys. Rev.} \textbf{\bibinfo{volume}{D62}},
  \bibinfo{pages}{123004} (\bibinfo{year}{2000}), \eprint{astro-ph/0009222}.

\bibitem[{\citenamefont{Seto and Pierpaoli}(2005)}]{Seto2005}
\bibinfo{author}{\bibfnamefont{N.}~\bibnamefont{Seto}} \bibnamefont{and}
  \bibinfo{author}{\bibfnamefont{E.}~\bibnamefont{Pierpaoli}},
  \bibinfo{journal}{Physical Review Letters} \textbf{\bibinfo{volume}{95}},
  \bibinfo{pages}{1} (\bibinfo{year}{2005}), \eprint{0502564}.

\bibitem[{\citenamefont{Bunn}(2006)}]{Bunn2006}
\bibinfo{author}{\bibfnamefont{E.~F.} \bibnamefont{Bunn}},
  \bibinfo{journal}{Physical Review D} \textbf{\bibinfo{volume}{73}},
  \bibinfo{pages}{123517} (\bibinfo{year}{2006}), \eprint{0603271}.

\bibitem[{\citenamefont{Abramo and Xavier}(2007)}]{Abramo:2006gp}
\bibinfo{author}{\bibfnamefont{L.~R.} \bibnamefont{Abramo}} \bibnamefont{and}
  \bibinfo{author}{\bibfnamefont{H.~S.} \bibnamefont{Xavier}},
  \bibinfo{journal}{Phys. Rev.} \textbf{\bibinfo{volume}{D75}},
  \bibinfo{pages}{101302} (\bibinfo{year}{2007}), \eprint{astro-ph/0612193}.

\bibitem[{\citenamefont{Liu et~al.}(2016)\citenamefont{Liu, Ichiki, Tashiro,
  and Sugiyama}}]{Liu2016}
\bibinfo{author}{\bibfnamefont{G.-C.} \bibnamefont{Liu}},
  \bibinfo{author}{\bibfnamefont{K.}~\bibnamefont{Ichiki}},
  \bibinfo{author}{\bibfnamefont{H.}~\bibnamefont{Tashiro}}, \bibnamefont{and}
  \bibinfo{author}{\bibfnamefont{N.}~\bibnamefont{Sugiyama}},
  \bibinfo{journal}{Monthly Notices of the Royal Astronomical Society: Letters}
  \textbf{\bibinfo{volume}{460}}, \bibinfo{pages}{L104} (\bibinfo{year}{2016}),
  \eprint{1603.06166}.

\bibitem[{\citenamefont{Louis et~al.}(2017)\citenamefont{Louis, Bunn, Wandelt,
  and Silk}}]{Louis:2017hoh}
\bibinfo{author}{\bibfnamefont{T.}~\bibnamefont{Louis}},
  \bibinfo{author}{\bibfnamefont{E.~F.} \bibnamefont{Bunn}},
  \bibinfo{author}{\bibfnamefont{B.}~\bibnamefont{Wandelt}}, \bibnamefont{and}
  \bibinfo{author}{\bibfnamefont{J.}~\bibnamefont{Silk}}
  (\bibinfo{year}{2017}), \eprint{1707.04102}.

\bibitem[{\citenamefont{Hu and Okamoto}(2002)}]{Hu:2001kj}
\bibinfo{author}{\bibfnamefont{W.}~\bibnamefont{Hu}} \bibnamefont{and}
  \bibinfo{author}{\bibfnamefont{T.}~\bibnamefont{Okamoto}},
  \bibinfo{journal}{Astrophys. J.} \textbf{\bibinfo{volume}{574}},
  \bibinfo{pages}{566} (\bibinfo{year}{2002}), \eprint{astro-ph/0111606}.

\bibitem[{\citenamefont{Okamoto and Hu}(2002)}]{Okamoto:2002ik}
\bibinfo{author}{\bibfnamefont{T.}~\bibnamefont{Okamoto}} \bibnamefont{and}
  \bibinfo{author}{\bibfnamefont{W.}~\bibnamefont{Hu}}, \bibinfo{journal}{Phys.
  Rev.} \textbf{\bibinfo{volume}{D66}}, \bibinfo{pages}{063008}
  (\bibinfo{year}{2002}), \eprint{astro-ph/0206155}.

\bibitem[{\citenamefont{Okamoto and Hu}(2003)}]{Okamoto:2003zw}
\bibinfo{author}{\bibfnamefont{T.}~\bibnamefont{Okamoto}} \bibnamefont{and}
  \bibinfo{author}{\bibfnamefont{W.}~\bibnamefont{Hu}}, \bibinfo{journal}{Phys.
  Rev.} \textbf{\bibinfo{volume}{D67}}, \bibinfo{pages}{083002}
  (\bibinfo{year}{2003}), \eprint{astro-ph/0301031}.

\bibitem[{\citenamefont{Dvorkin and Smith}(2009)}]{Dvorkin:2008tf}
\bibinfo{author}{\bibfnamefont{C.}~\bibnamefont{Dvorkin}} \bibnamefont{and}
  \bibinfo{author}{\bibfnamefont{K.~M.} \bibnamefont{Smith}},
  \bibinfo{journal}{Phys. Rev.} \textbf{\bibinfo{volume}{D79}},
  \bibinfo{pages}{043003} (\bibinfo{year}{2009}), \eprint{0812.1566}.

\bibitem[{\citenamefont{Dvorkin et~al.}(2009)\citenamefont{Dvorkin, Hu, and
  Smith}}]{Dvorkin:2009ah}
\bibinfo{author}{\bibfnamefont{C.}~\bibnamefont{Dvorkin}},
  \bibinfo{author}{\bibfnamefont{W.}~\bibnamefont{Hu}}, \bibnamefont{and}
  \bibinfo{author}{\bibfnamefont{K.~M.} \bibnamefont{Smith}},
  \bibinfo{journal}{Phys. Rev.} \textbf{\bibinfo{volume}{D79}},
  \bibinfo{pages}{107302} (\bibinfo{year}{2009}), \eprint{0902.4413}.

\bibitem[{\citenamefont{Smith and Ferraro}(2016)}]{Smith:2016lnt}
\bibinfo{author}{\bibfnamefont{K.~M.} \bibnamefont{Smith}} \bibnamefont{and}
  \bibinfo{author}{\bibfnamefont{S.}~\bibnamefont{Ferraro}}
  (\bibinfo{year}{2016}), \eprint{1607.01769}.

\bibitem[{\citenamefont{Meerburg et~al.}(2017)\citenamefont{Meerburg, Meyers,
  and van Engelen}}]{Meerburg:2017xga}
\bibinfo{author}{\bibfnamefont{P.~D.} \bibnamefont{Meerburg}},
  \bibinfo{author}{\bibfnamefont{J.}~\bibnamefont{Meyers}}, \bibnamefont{and}
  \bibinfo{author}{\bibfnamefont{A.}~\bibnamefont{van Engelen}}
  (\bibinfo{year}{2017}), \eprint{1704.00718}.

\bibitem[{\citenamefont{Yasini and
  Pierpaoli}(2016{\natexlab{a}})}]{Yasini:2016pby}
\bibinfo{author}{\bibfnamefont{S.}~\bibnamefont{Yasini}} \bibnamefont{and}
  \bibinfo{author}{\bibfnamefont{E.}~\bibnamefont{Pierpaoli}},
  \bibinfo{journal}{Phys. Rev.} \textbf{\bibinfo{volume}{D94}},
  \bibinfo{pages}{023513} (\bibinfo{year}{2016}{\natexlab{a}}),
  \eprint{1605.02111}.

\bibitem[{\citenamefont{Yasini and
  Pierpaoli}(2016{\natexlab{b}})}]{Yasini:2016dnd}
\bibinfo{author}{\bibfnamefont{S.}~\bibnamefont{Yasini}} \bibnamefont{and}
  \bibinfo{author}{\bibfnamefont{E.}~\bibnamefont{Pierpaoli}}
  (\bibinfo{year}{2016}{\natexlab{b}}), \eprint{1610.00015}.

\bibitem[{web()}]{web:pla}
\urlprefix\url{http://pla.esac.esa.int/pla/\#cosmology}.

\bibitem[{\citenamefont{Ade et~al.}(2016{\natexlab{b}})}]{Ade:2015xua}
\bibinfo{author}{\bibfnamefont{P.~A.~R.} \bibnamefont{Ade}}
  \bibnamefont{et~al.} (\bibinfo{collaboration}{Planck}),
  \bibinfo{journal}{Astron. Astrophys.} \textbf{\bibinfo{volume}{594}},
  \bibinfo{pages}{A13} (\bibinfo{year}{2016}{\natexlab{b}}),
  \eprint{1502.01589}.

\bibitem[{\citenamefont{{Hinshaw} et~al.}(1996)\citenamefont{{Hinshaw},
  {Banday}, {Bennett}, {Gorski}, {Kogut}, {Smoot}, and
  {Wright}}}]{1996ApJ...464L..17H}
\bibinfo{author}{\bibfnamefont{G.}~\bibnamefont{{Hinshaw}}},
  \bibinfo{author}{\bibfnamefont{A.~J.} \bibnamefont{{Banday}}},
  \bibinfo{author}{\bibfnamefont{C.~L.} \bibnamefont{{Bennett}}},
  \bibinfo{author}{\bibfnamefont{K.~M.} \bibnamefont{{Gorski}}},
  \bibinfo{author}{\bibfnamefont{A.}~\bibnamefont{{Kogut}}},
  \bibinfo{author}{\bibfnamefont{G.~F.} \bibnamefont{{Smoot}}},
  \bibnamefont{and} \bibinfo{author}{\bibfnamefont{E.~L.}
  \bibnamefont{{Wright}}}, \bibinfo{journal}{ApjL}
  \textbf{\bibinfo{volume}{464}}, \bibinfo{pages}{L17} (\bibinfo{year}{1996}),
  \eprint{astro-ph/9601058}.

\bibitem[{\citenamefont{Efstathiou}(2003)}]{Efstathiou:2003wr}
\bibinfo{author}{\bibfnamefont{G.}~\bibnamefont{Efstathiou}},
  \bibinfo{journal}{Mon. Not. Roy. Astron. Soc.}
  \textbf{\bibinfo{volume}{346}}, \bibinfo{pages}{L26} (\bibinfo{year}{2003}),
  \eprint{astro-ph/0306431}.

\bibitem[{\citenamefont{Yadav and Wandelt}(2005)}]{Yadav:2005tf}
\bibinfo{author}{\bibfnamefont{A.~P.~S.} \bibnamefont{Yadav}} \bibnamefont{and}
  \bibinfo{author}{\bibfnamefont{B.~D.} \bibnamefont{Wandelt}},
  \bibinfo{journal}{Phys. Rev.} \textbf{\bibinfo{volume}{D71}},
  \bibinfo{pages}{123004} (\bibinfo{year}{2005}), \eprint{astro-ph/0505386}.

\bibitem[{\citenamefont{Bardeen et~al.}(1986)\citenamefont{Bardeen, Bond,
  Kaiser, and Szalay}}]{Bardeen1986}
\bibinfo{author}{\bibfnamefont{J.~M.} \bibnamefont{Bardeen}},
  \bibinfo{author}{\bibfnamefont{J.~R.} \bibnamefont{Bond}},
  \bibinfo{author}{\bibfnamefont{N.}~\bibnamefont{Kaiser}}, \bibnamefont{and}
  \bibinfo{author}{\bibfnamefont{a.~S.} \bibnamefont{Szalay}},
  \bibinfo{journal}{The Astrophysical Journal} \textbf{\bibinfo{volume}{304}},
  \bibinfo{pages}{15} (\bibinfo{year}{1986}).

\end{thebibliography}

\end{document}